\documentclass[12pt]{article}

\usepackage{subfigure}

\usepackage[numbers,sort&compress]{natbib}
\bibpunct[, ]{[}{]}{,}{n}{,}{,}

\usepackage{hyperref}
\hypersetup{
colorlinks=true, 
breaklinks=true, 
urlcolor= blue, 
}

\newtheorem{theorem}{Theorem}[section]

\newtheorem{definition}[theorem]{Definition}

\usepackage[top=3.5cm, bottom=3cm, left=3cm , right=2.5cm]{geometry}
\usepackage{amsmath}
\DeclareMathOperator{\vect}{vec}

\usepackage[final]{graphicx}

\begin{document}


\title{Multiscale Bayesian State Space Model for Granger Causality Analysis of Brain Signal}


  \author{Sezen Cekic
   \hspace{.2cm}\\
    Methodology and Data Analysis, Department of Psychology, \\
    University of Geneva,\\
    Didier Grandjean\\
    Neuroscience of Emotion and Affective Dynamics Lab, \\Department of Psychology, \\ University of Geneva,\\
	and \\
    Olivier Renaud \\
    Methodology and Data Analysis, Department of Psychology, \\ University of Geneva}


\maketitle

\thispagestyle{empty}

\begin{abstract}
 Modelling time-varying and
  frequency-specific relationships between two brain signals is becoming an essential methodological tool to answer theoretical questions in experimental neuroscience. In this article, we propose to estimate a frequency Granger
  causality statistic that may vary in time in order to evaluate the functional
  connections between two brain regions during a task. We use for that purpose an adaptive Kalman filter type of estimator of
  a linear Gaussian vector autoregressive model with coefficients
  evolving over time. The estimation procedure is achieved through
  variational Bayesian approximation and is extended for multiple
  trials. This Bayesian State Space (BSS) model provides a dynamical
  Granger-causality statistic that is quite natural. We propose to
  extend the BSS model to include the \textit{\`{a} trous} Haar
  decomposition. This wavelet-based forecasting method is based on a
  multiscale resolution decomposition of the signal using the redundant
  \textit{\`{a} trous} wavelet transform and allows us to capture short- and
  long-range dependencies between signals. Equally importantly it allows us to
  derive the desired dynamical and frequency-specific
  Granger-causality statistic. The application of these models to
  intracranial local field potential data recorded during a psychological
  experimental task shows the complex frequency based cross-talk
  between amygdala and medial orbito-frontal cortex.  

  keywords: \textit{\`{A} trous} Haar wavelets; Multiple trials; Neuroscience data; Nonstationarity; Time-frequency; Variational methods

  The published version of this article is

  Cekic, S., Grandjean, D., Renaud, O. (2018). Multiscale Bayesian state-space model for Granger causality analysis of brain signal. \textit{Journal of Applied Statistics}. \url{https://doi.org/10.1080/02664763.2018.1455814}
  
\end{abstract}

\section{Introduction}
\label{sec:intro}

In many neuroscientific experiments, data are recorded in an experimental situation where
stimuli are presented at fixed times and are expected to induce a reaction. For
psychologists and neuroscientists, being able to model and explain the dynamics of the functional and effective links between neural and behavioural signals recorded during the experiment is of primary interest. They often have strong prior hypotheses about these causal links and therefore need reliable statistical tools to draw valid conclusions.

\subsection{Granger causality}

The question of how to operationally formalize and test for causality is a fundamental and philosophical problem. A mathematical solution, which relies on the causal nature of predictability, was provided in the 60's by the economist Clive Granger and was latter coined
``Granger causality''. According to \citet{granger_investigating_1969}, if a signal $X$ ``Granger-causes'' a signal $Y$, then the history of $X$ should contains information that helps to predict $Y$ above and beyond the information contained in the history of $Y$ alone. The axiomatic imposition of a temporal ordering is the crucial element that enables us to interpret such dependence as causal.
The presence of this relation between $X$ and $Y$ will be referred to ``Granger causality'' throughout the text.
 In the 1960s, \citet{granger_investigating_1969} adapts the
 definition of causality proposed by \citet{e_theory_1956} into a
 practical form and since that time Granger causality has been widely
 used in economics and econometrics. It is however only since last few
 years that it became popular in neuroscience, see \citet{cekic_time_2018} for a recent review.  

\subsection{Existing methods and limits}

In the context of linear Gaussian autoregressive models, for which the
restrictions are Gaussianity and linearity, which imply
stationarity in most cases, a common way to test for  Granger causality between two series is to estimate a vector
autoregressive model (VAR) and then test the significance of the off diagonal coefficients of interest \citep{hamilton_time_1994,lutkepohl2005new}.

However, in neuroscience the data are usually nonstationary and this
characteristic is moreover of interest. In the simplest case, the data are
stationnary up to a particular point where the Gaussian process has
been perturbed away from its stationary distribution (perhaps by some
external intervention). We therefore want to derive a
causality-statistic that allows us to capture the causal structure
differentially for each time.

Basic causality statistic 
 has therefore to be extended
to the nonstationary case in order to be suitably applied in a
neuroscience context, which can be achieved by letting the VAR model evolve in time. 

Time-varying VAR model estimate implies three challenges: o\-ver-\-pa\-ra\-me\-tri\-za\-tion, model order selection and multiple trials.

The two widely used approaches allowing us to deal with the nonstationarity are the windowing approach, based on the locally-stationary assumption and the adaptive estimation approach, based on the slowly-varying assumption of the parameters (see \citet{ding_short-window_2000}, \citet{schlogl_electroencephalogram_2000} and \citep{chen_vector_2012}).
 
The windowing approach consists in estimating VAR models in short temporal sliding windows where the underlying process is assumed to be (locally) stationary (see \citet{ding_short-window_2000} for a methodological tutorial of windowing estimation approach in neuroscience).

The windows size is a trade-off between the accuracy of the parameter estimates and the resolution in time. The choice of the model order is a delicate issue, and depends on the choice of the segment length. Some criteria have been proposed in order to optimized simultaneously the windows length and the model order \citep{lin_dynamic_2009}.

The usual approach with multiple trials is to average the estimation or do a global optimization to get an overall estimation. 

In \citet{cekic_lien_2010} we found that this windowing methodology presents several limits. First, the improvement of the time resolution implies short time-windows and so few residuals for assessing the quality of the fit. In addition the size of the temporal windows is subjective (even if it depends on a criterion) as is the overlap between the time-windows. The order of the model in turns depends on the size of the windows and so the quality of the estimation strongly relies on several subjective parameters.

The adaptive estimation approach consists on estimating a different model at each time, where the observations at time $t$ are expressed as a linear combination of the past with coefficients evolving slowly over time. The differences between the methods consist on the way the transition and the update from coefficients at time $t$ to those at time $t+1$ are processed \citep[see][]{schlogl_electroencephalogram_2000}.

All these adaptive estimation methodologies depend on a
quantity that acts as a tuning parameter and defines the relative
influence of past values and innovation noise on the recursive estimation. Generally this free tuning parameter determines the speed of adaptation as well as the smoothness of the time-varying VAR parameter estimates. 
The algorithms are very sensitive to this tuning parameter \citep{schlogl_electroencephalogram_2000} and therefore the estimation quality strongly depends on it. The ``ad-hoc'' nature of this tuning parameter is obviously a problem in term of statistical inference and this issue was not raised in the development of these algorithms coming from the engineering field.
The model order and the tuning parameters are usually optimized together by a Mean Square Error criterion \citep{schlogl_criterion_2000}.
 
Kalman filtering algorithm \citep{kalman_new_1960} can be used in order to estimate time-varying VAR models when we express it in a state space form \citep{arnold_adaptive_1998,barnett_granger_2015}.

If the transition matrix and variance-covariance matrices of the observed and state equation are known, the Kalman smoother algorithm gives the best linear unbiased estimator for the state vector \citep{kalman_new_1960}, which in this specific case contains the time-varying VAR coefficients.
In the engineering and neuroscience literature, these matrices are systematically set to fixed values or estimated through some ``ad-hoc'' estimation procedure (see \citet{schlogl_electroencephalogram_2000} and \citet{hesse_use_2003,arnold_adaptive_1998} for applications in neuroscience). There is moreover the very important issue of model order selection which becomes very tricky with model complexity and the plurality of the trials.

\subsection{Neuroscience data specificities} 
\label{Neuroscience data specificities}

Our model has to be applied to experimental neuroscience data,
whose intrinsic specificities must be taken into account in its
development.
 Therefore, in order to derive a suitable dynamical causal statistic, we need a model that allows us to get a reliable estimate of the dynamical VAR coefficients based on several trials and, last but not least, that also allows us to capture short- and long-range causal dependencies between signals due to specific frequency characteristics of the data.

\subsection{Proposal}
Faced with data with a time-varying structure (like neuroscience data), none of the above methods relies on a tailored statistical model that provides satisfying estimation and inference procedures and proposes a solution to deal with short and long range causal dependencies potentially present in the data. 

We propose a new methodology for suitably modelling multivariate nonstationary time series in order to get a reliable Granger-type dynamic causal statistic. 
    It is based on a linear Gaussian vector autoregressive (VAR) model with coefficients evolving over time according to a linear dynamical system. Given that this model is strongly over-parametrized, we propose to place it in a Bayesian framework and to use the variational method to estimate all the densities.
This variational Bayesian methodology \citep{beal_variational_2003}
estimates all the necessary quantities. 
The Bayesian nature of the model moreover offers a natural criterion for model order selection. In Section \ref{The Bayesian State Space Model}, we describe our Bayesian state space (BSS) model and discuss its technical specificities and the estimation procedure. We extend it to deal with multiple trials  (or epochs), in a proper
manner for the estimation and the inference procedure (section \ref{multipletrials}).
In Section \ref{The Variational Bayesian State Space Multiscale
  Model}, we propose an additionnal extension of the BSS model, called the multiscale Bayesian state space (MSBSS) model, which is based on the \textit{\`{a} trous} multiscale wavelet transform. The latter approach has never been used in this context of time-varying VAR coefficient estimate and we will show that it allows better estimate of short- and long-range specific dependencies between recorded signals and offers a very simple way to deal with time-frequency uncertainty bounds. In Section \ref{Bayesian Granger-Causality statistic}, we will present a Bayesian dynamical Granger-causality statistic based on the time-varying estimated VAR coefficients and in Section \ref{Assessment of accuracy}, we present simulation studies to assess our proposed methodology. Finally, we present in Section \ref{Application} the results of the application of the method to intracranial local field potential recorded during a psychological experiment in specific brain areas, namely in the amygdala and orbitofrontal cortex.

\section{The Bayesian State Space Model} 
\label{The Bayesian State Space Model}

We propose to write the dynamic VAR model in a state space form with an observation equation in which the dynamic VAR coefficients are driven by the state equation \citep[this modelling proposal was previously made by][]{cassidy_bayesian_2002}. This leads to the following system of equations:
 \begin{equation}
 \left\{\begin{aligned}
 	\varphi_{t+1}  =  A\varphi_t+w_t\quad &w_t\sim{\mathcal{N}_k(0,Q)} \\
    Z_t  =  C_t \varphi_t+v_t \quad &v_t\sim{\mathcal{N}_d(0,R)}\\ 
       \end{aligned}
 \right.
 \quad \text{with}\quad
 \begin{cases}
    \varphi_t&=\vect[\vartheta_{1(t)} ,\vartheta_{2(t)},..,\vartheta_{p(t)}]', \\
    Z_t&=(Y_t\; X_t)^{'},\\
    C_t \varphi_t&=\sum\limits_{j=1}^p \vartheta_{j(t)} (Y_{t-j}\;  X_{t-j})^{'},
    \end{cases}
 \label{kalman.def}
\end{equation}
where $Z_t=(Y_t\; X_t)^{'}$ is the value of the $d=2$ signals at time
$t$,  $\vartheta_{j(t)}$ are the time-varying VAR coefficients (up to
order $p$)
 and the vector $\varphi_t$ of size $k=pd^{2}$ contains all the time-varying VAR coefficients that have to be estimated for the time $t$. The matrix $A$ is the transition matrix of the state vector $\varphi_t$, $Q$ is the $k \times k$ variance-covariance matrix of the state equation, and $R$ is the $d \times d$ variance-covariance matrix of the observation equation. We will use the notation $\varphi_{1}^T$ and $Z_1^T$ to denote the entire set of values from $t=1$ to $t=T$.
Although the state equation seems to be only a first-order
autoregression, note that the vectors  $\varphi_t$ and $\varphi_{t-1}$
contain all the coefficients up to order $p$, and therefore the direct
dependency of $\varphi_t$ on past values is actually unlimited (and
driven by the choice of the order $p$). This formulation is actually
similar to the state-space representation of AR(p) or ARMA(p,q) models (see e.g. \cite{barnett_granger_2015} examples 12.1.4-5)

 With a slight abuse of notation, we can write $p(Z_t|C_t, \varphi_t,R)=\mathcal{N}_d(C_t \varphi_t,R), \\
p(\varphi_{t} |A,\varphi_{t-1},Q) =\mathcal{N}_k(A\varphi_{t-1},Q)$ and $p(\varphi_{1} )=\mathcal{N}_k(\mu_1,\Sigma_1)$, which are respectively the observation, the state and the initial state densities.

Given the Bayesian framework, we are interested in obtaining the posterior distribution of the unknowns of this model. 
 In simple cases, these distributions can be obtained analytically but here, the curse of dimensionality makes the computation intractable and forces us to rely on approximation techniques.

\subsection{Variational Bayes}
\label{ElementsVB}

Variational approximation was applied to the linear Gaussian state space model in \citet{ghahramani_propagation_2001} and \citet{cassidy_bayesian_2002}, and more recently in \citet{luessi_variational_2014}. This approximation methodology has been extensively explored and used during the past years; see for example \citet{titterington_bayesian_2004}, \citet{beal_variational_2003}, \citet{ormerod_explaining_2010} and \citet{fox_tutorial_2011}.
It is not widely known within the statistical community dominated by Monte Carlo and Laplace approximation methods but is however much faster than Monte Carlo, especially for large models, and allows us to deal with models containing a very large amount of parameters (see \citet{friston_variational_2007} for a comparison of variational and Laplace approximations).

 For sake of clarity, we will define the set of unknown parameters as $\Omega_1^b$, so the full set of unknowns for model \eqref{kalman.def} is $\{\varphi_1^{T};\Omega_1^b\}$.

The target quantity is the posterior distribution of the parameters which is an intractable integral of very high dimension. The variational approach allows us to approximate this posterior density $p(\varphi_{1}^T,\Omega_1^b|Z_1^T )$ by a variational posterior density $q(\varphi_{1}^T,\Omega_1^b|Z_1^T)$
that will be selected to be optimal according to the Kullback--Leibler distance dissimilarity criterion \citep{beal_variational_2003} (see Appendix A for further details on variational Bayesian approximation).

                          \subsection{Mean-field approximation}
\label{Factorization assumption}
Variational Bayesian methodology
 allow the
approximating density $q(\varphi_{1}^T,\Omega_1^b|Z_1^T)$ to
factorize over groups of parameters. The researcher has the
  choice here, but the explicit link between the $\varphi$'s in
  equation~\ref{kalman.def} urge not to factorize $\varphi_1^T$. The
  less intricate links between  $\varphi_1^T$, $A$, $Q$ and $R$
  allow for the following factorization (see Figure~A.1 from the
  supplementary material)
 \begin{equation} 
q(\varphi_{1}^T,\Omega_1^b|Z_1^T) = q(\varphi_1^T|Z_1^T)\prod_{j=1}^b q(\Omega_j|Z_1^T).
\label{mean field approx}
\end{equation}
       Note that it may lead to serious degradation in the resulting inference if unsatisfied \citep{ormerod_explaining_2010,titterington_bayesian_2004,beal_variational_2003}.

\subsection{The variational evidence lower bound}
\label{The Variational Evidence Lower Bound}

As presented in details in Appendix B, 
   the variational Bayesian methodology provides a quantity $\text{F}$ that is a lower bound for the evidence of the model and that can be computed efficiently. This leads to a natural criterion for model order selection which is crucial to estimate our strongly over-parametrized time-varying VAR model. We can therefore perform model selection by comparing the $\text{F}$ quantities computed for each model order $m_p$ and select the model that exhibits the highest $\text{F}_{m_p}$. 
                   The specific analytic form of $\text{F}_{m_p}$ for our model is derived in  Appendix B. 

\section{Model specification}
\label{Model specification}

We will now describe the fully hierarchical Bayesian model that we propose. 

        The optimal form for $q^*(\varphi_1^T|Z_1^T)$ and
$q^*(\Omega_1^b|Z_1^T)$ of course depends on the choice of the prior
distributions $p(\varphi_1^T)$ and $p(\Omega_m)$. However, the analytical form of $q^*(.|Z_1^T)$ will be of the same distributional form as the prior distributions $p(.)$ if the complete-data likelihood $p(\varphi_1^T,Z_1^T | \Omega_1^b)$ is part of the exponential family and if the hidden and parameter prior distributions $p(\varphi_1^T)$ and $p(\Omega_1^b)$ are conjugate to this complete-data likelihood (this condition is known as ``conjugate exponential'', see \citet{beal_variational_2003}).
  
Recalling the model \eqref{kalman.def}, the complete data likelihood  $p(\varphi_1^T,Z_1^T | \Omega_1^b)$
 is Gaussian. We will now describe the model by defining conjugate priors over the model parameters. 
\subsection{Prior distributions}
\label{Prior distributions}

\subsubsection{Prior for $\varphi_1^T$}
\label{Prior for phi}
The state space structure of the model \eqref{kalman.def} 
 leads to the conditional prior
distributions for the hidden state $\varphi_1^T$ defined in Section
\ref{The Bayesian State Space Model} and the
 prior mean $\mu_1$ and variance $\Sigma_1$ for $\varphi_1$ are set to $1_{k} \times 0$ and $I_{k} \times 0.1$ respectively.

\subsubsection{Prior for $\Omega_1=\{A,\mathbf{\alpha},\mathbf{\delta}\}$}
\label{Prior for delta,alf and A}

 We propose a diagonal structure for the $A$ matrix, meaning that a specific causal coefficient at time $t$ will only depend on its own past value plus a white noise and not on the past values of other VAR coefficients. 
This assumption seems reasonable in terms of brain connectivity, where the dynamic of a particular causal input can be assumed to be driven by its own trajectory and not by that of other causal or auto-causal inputs. 
To satisfy the conjugacy condition explained in details in Appendix A, 
We place a Gaussian prior on each diagonal entry of the $A$ matrix:
\begin{equation}
p(A|\mathbf{\alpha}) = \prod_{i=1}^k \mathcal{N}_1(A_{ii}|m_{A_{ii}},\alpha_i),
\label{p(A1)}
\end{equation}
where $\alpha_i$ is the hyperparameter variance of the specific diagonal element $A_{ii}$ whose distribution will be discussed below.
We choose to impose a conservative prior for $A$ by setting each location hyperparameter $\{m_{A_{ii}}\}_{i=1}^{k}$ to $0.9$.

As we have no subjective input for the variances $\alpha_i$, it is desirable to have priors that exhibit very little information. 
   The main recommendation in \citet{gelman2006prior} is to use half-t priors on standard deviation
parameters to achieve arbitrarily high noninformativeness. 
The definition of the half-t distribution can be found e.g.\ in \citet{wand2011}.
    They showed that the half-t distribution can be written as a scaled mixture of inverse-gamma distributions. So the half-t prior is obtained for each $\alpha_i^{0.5}$ element through the auxiliary variable construction
\begin{equation}
\begin{aligned}
\begin{split}
p(\alpha_i|\delta_i) = \mathcal{IG}(c_{p_i},b_{p_i}) = \mathcal{IG}(\frac{1}{2},\frac{1}{\delta_i}), \\ 
p(\delta_i) = \mathcal{IG}(\kappa_{p_i},\beta_{p_i})= \mathcal{IG}(\frac{1}{2},\frac{1}{D_i^2}), \quad i=1,\dots,k,
\end{split}
\end{aligned}
\label{p(alf)}
\end{equation}
which ensures that $p(\alpha_i^{0.5}) = \text{Half-t}(1,D_i)$.
The hierarchical representation in equation \eqref{p(alf)} respects the conditional conjugacy condition
 due to the conjugacy properties of the inverse-gamma distribution. 
 
By its noninformativeness property, this half-t prior choice will let each variance element $\{\alpha_{i}\}_{i=1}^{k}$ get an appropriately high or low value, thereby allowing them to play the role of shrinkage parameters for the distribution of $A_{ii}$. 
 In our specific case, it will allow us to be conservative and to tend to avoid an erroneous causality assessment.

\subsubsection{Prior for $\Omega_2=\{Q,a_q\}$}
\label{Prior for Q and aq}

The accuracy of the Markovian conditional distribution followed by the VAR coefficients $\varphi_t$ is established through the variance-covariance matrix $Q$. Throughout this work we assume that this matrix is diagonal (for the same theoretical reason discussed for the $A$ matrix) and that all its elements are equal. In fact there is no reason to assume that the variability of one VAR coefficient is different from that of another. This choice is also motivated by the concern to limit the number of parameters to estimate.
 
In order to satisfy the conjugacy conditions and for the same theoretical reason mentioned earlier, we set the same weakly-informative half-t prior distribution for the single standard deviation parameter $Q_{ii}^{0.5}$ of the diagonal $Q$ matrix: 
\begin{equation}
\begin{aligned}
\begin{split}
p(Q_{ii}|a_q) = \mathcal{IG}(n_p,d_p) = \mathcal{IG}(\frac{1}{2},\frac{1}{a_q}), \quad
p(a_q) = \mathcal{IG}(a_{qp},b_{qp}) = \mathcal{IG}(\frac{1}{2},\frac{1}{A_q^2}), 
\end{split}
\end{aligned}
\label{p(Q)}
\end{equation}
which again ensures that $p(Q_{ii}^{0.5}) = \text{Half-t}(1,A_q)$.

\subsubsection{Prior for $\Omega_3=\{R,a_r\}$}
\label{Prior for R and ar}

The variance-covariance matrix $R$ of the observed equation is not supposed to be diagonal due to the theoretical interdependence between brain signals modelled in the system.

We impose for $R$ a generalisation of the multivariate case of the Half-t prior used for $\alpha$ and $Q_{ii}$. \citet{huang2013simple} derived this prior and explained that with a suitable hyperparameter choice, it induces an Half-t distribution for each standard
deviation term corresponding to the diagonal of $R$, as well as a marginal uniform distribution for all correlations. 
Explicitly we have
\begin{equation}
\begin{aligned}
\begin{split}
p(R|a_{r_1},\dots,a_{r_d}) &= \mathcal{IW}_d\big(r_p,B_p) = \mathcal{IW}_d\big(\nu+d-1,2\,\nu\, \text{diag}[\frac{1}{a_{r_1}},\dots,\frac{1}{a_{r_d}}]\big),\\
 p(a_{r_i}) &= \mathcal{IG}(a_{pr},b_{pr}) =\mathcal{IG}(\frac{1}{2},\frac{1}{A_R^2}), \qquad i=1,\dots,d,
\label{p(R)}
\end{split}
\end{aligned}
\end{equation}
where $\text{diag}()$ denotes a diagonal matrix.
This hierarchical structure ensures that each diagonal element $R_{i,i}$ is distributed as $p(R_{i,i}^{0.5}) = \text{Half-t}(\nu,A_R)$, and the particular choice of $\nu = 2$ leads to marginal uniform
distributions over $[-1;1]$ for all correlation terms \citep{huang2013simple}.

Lastly, we note that the selected prior distributions impose the choice of hyperparameters $D_1,\dots,D_k$, $A_q$, $\nu$ and $A_R$. We will choose $\nu = 2$ for the uniform property to hold. For the remaining hyperparameters, the larger they are the less informative the priors are. We follow \citet{menictas2013variational} and set them to $10^5$.

\subsection{Update equations}
\label{Update equations}

The variational Bayesian iterative algorithm and
optimal posterior distributions are based on the results presented in Appendix A and factorization chosen in Section \ref{Factorization assumption}.
    The optimal variational posterior distribution for the hidden state sequence $\varphi_1^T$ (Variational E-Step) is multivariate Gaussian at each time $t$:
\begin{equation}
q^{*}(\varphi_t|Z_1^T) =  \mathcal{N}_k(\varphi_t|\mu_{t},\Sigma_{t}),
\label{POST_H}
\end{equation} 
where the sufficient statistics $\{\mu_{t};\Sigma_{t}\}_{t=1}^T$, as well as the cross-moments \\ 
$\{\mu_{t}\mu_{t-1};\Sigma_{t,t-1}\}_{t=2}^T$, are obtained  by the
Kalman--Rauch--Tung--Striebel (KRTS) smoother algorithm
\citep{kalman_new_1960,kalman_new_1961}, applied to an augmented
system of equations, see Appendices C, D and E. 
  All the derivations for the Variational M-steps can be found in Appendices F, G, H, I, J, K and L.

\subsection{Multiple trials}
\label{multipletrials}

One important contribution of the present article is to show that our model can be modified to deal with $N$ conditionally independent sequences $\{{Z_1^T}_{(j)}\}_{j=1}^{N}$ which are supposed to have the same hidden state. This reflects the case that arises during an event-related experimental paradigm, where many trials on the same condition are measured. 

In \citet{beal_variational_2003} and \citet{cassidy_bayesian_2002}, this extension is treated by first estimating the necessary sufficient statistics for each sequence independently in the E-step, and then by averaging these statistics to get only one set of sufficient statistics, which is representative of the entire set of independent sequences before performing the M-step. However this approach does not take into account the complex dependence of the variational posterior $q(\varphi_{1}^T|\{{Z_1^T}_{(j)}\}_{j=1}^{N})$ on the whole dataset $\{{Z_1^T}_{(j)}\}_{j=1}^{N}$. 
In fact, all the computations done so far can be adapted to multiple trials. A detailed derivation of the model and related variational posteriors distributions in a multiple trials setting can be found in Appendix M.

\section{The Multiscale Bayesian State Space Model} 
\label{The Variational Bayesian State Space Multiscale Model}

In a neuroscience context, an important limitation of many models,
including the BSS model, is the inability to capture both the short-
and long-range possible causal dependencies between signals.
Causal interactions in a neuroscientific experiment context may not be instantaneous, but delayed over a certain time interval $(\upsilon)$ that must be subjectively chosen depending on the research hypothesis. Another important subjective parameter is the time-lag ($\tau$), that determines the interval between two data points
As shown in \citet{barnett_detectability_2017} and \citet{solo_state-space_2016}
these choices strongly influence
classical Granger statistics in their ability to detect 
causalities,
 whereas the result should be as invariant as possible to arbitrary choices
like the chosen sampling frequency or added time-lag in the prediction model.

Based on these considerations, and because Granger causality is based on predictive ability, if the auto-causal information contained in the history of the predicted signal $Y_t$ in model \eqref{kalman.def} is not well represented in $C_t$, 
 the Granger-causality evaluation, which is based on the predictability improvement of $Y_t$ by adding the information contained in the history of the second variable $X$, may be spuriously assessed as significant. On the other hand, if the predictive ability of the history of the causal signal $X$ is not informative enough, we can miss some crucial information about causal interdependencies between signals $Y$ and $X$.

We propose a new solution that has the ability to appropriately select the short- and long-term causal histories of $Y$ and $X$ by combining the BSS model with the \textit{\`{a} trous} multiscale decomposition methodology and that remains within the conjugate exponential framework \citep{renaud_prediction_2003}. 

\subsection{The \textit{\`{a} trous} Haar wavelets transform} 
\label{The a trous Haar wavelets transform}

For that purpose, we will perform the \textit{\`{a} trous} multiscale decomposition of signals $Y_t$ and $X_t$ contained in $Z_t$ in model~\eqref{kalman.def}, in order to use these quantities as predictive histories in the matrices $\{C_t\}_{t=1}^T$. The reader is  referred to \citet{renaud_prediction_2003} and references therein for a complete overview of the method. We define $w^y_{j,t}$ as the \textit{\`{a} trous} wavelet coefficient of the signal $Y$ at time $t$ for scale $j$. Since we will use the \textit{\`{a} trous} wavelet coefficients for prediction, they should not be based on future values and the only family that satisfies this constraint is the \textit{\`{a} trous} Haar wavelet transform. Then $S_{j+1}(t)=[S_j(t)+S_j(t-2^j)]/2, \quad w_{j+1}(t)=S_j(t)-S_{j+1}(t), \quad t=1, \ldots, T, \quad j=1, \ldots,J$, where the finest scale is the original series $S_0(t)=Y_t$.

\subsection{The multiscale Bayesian state space model} 

Based on the derivation in Section \ref{The a trous Haar wavelets transform}, we can modify the VAR model in equation \eqref{kalman.def}. We keep the quantity to predict $Y_{t}$ in the time domain, but the histories of the series $Y$ and $X$ will equal the \textit{\`{a} trous} Haar wavelet transforms of series $Y_t$ and $X_t$ respectively. 
We can therefore define the set $C_t^w$ that contains the decompositions of the histories of series $Y_t$ and $X_t$
as
\begin{equation}
\begin{aligned}
C_t^w = &{\{w^y_{j,t-1-2^j(k-1)}\}_{j=1,\ldots,J,k=1,\ldots,p_j}}, \{s^y_{J,t-1-2^J(k-1)}\}_{k=1,\ldots,p_{J+1}},\\
&\{w^x_{j,t-1-2^j(k-1)}\}_{j=1,\ldots,J,k=1,\ldots,p_j},\{s^x_{J,t-1-2^J(k-1)}\}_{k=1,\ldots,p_{J+1}},
\end{aligned}
\label{C_t^w}
\end{equation} 
and therefore adapt equation \eqref{kalman.def} replacing $C_t$ with $C_t^w$.

In particular, we underline that for any $J$, for a suitable choice of $p_1 ,\dots, \\
p_{J+1}$, $C_t^w$ is an orthogonal transform of $C_t$
and therefore in this specific case BSS and MSBSS models are
equivalent. For any choice of the $p$'s, all results in Section
\ref{The Bayesian State Space Model} and all the estimation procedures
in Section \ref{Model specification}
can be applied. The only
difference is that the matrix $C_t$ is replaced by $C_t^w$, and that
the dimension of $\varphi_t$ is changed accordingly.  
   The model order $p_j$ for the different scales as well as the number of scales $J$ to be taken in the model must now be selected. 
   As argued in \citet{renaud_prediction_2003}, the relative
non-overlapping frequencies used in each scale motivate an independent selection of the model order $p_j$ for each scale. Defining a maximal scale decomposition $J_{\text{max}}$ and a maximum model order for each scale $p_{\text{max}}$, the model order selection procedure set in Section \ref{The Variational Evidence Lower Bound} is iteratively applied to select $p_j$ in a stepwise manner for each scale from 1 to $J_{\text{max}}$. 
 The free-energy quantities related to the $J_{\text{max}}$ models are then compared and the $J$ that exhibits the highest free energy is selected. The model order for the smooth $p_{J+1}$ is finally selected.

This \textit{\`{a} trous} extension can thus be viewed as a
generalisation of the BSS model that contains information relative to
the frequencies. Each wavelet scale indeed is directly related to a specific frequency band and the resulting dynamic Granger causality statistic is therefore directly interpretable in terms of frequencies (see Section \ref{Application}). 
	
\section{Bayesian Granger-Causality Statistic} 
\label{Bayesian Granger-Causality statistic}

Based on the model proposed in this article, a necessary and sufficient condition for $X_t$ not to be Granger-causal for $Y_t$ at a given time $t$, is that each element of the subset $\widetilde{\varphi_{t}}$ of $\varphi_{t}$, that contains all the causal coefficients of interest, equals zero. 

The most appropriate approach to evaluate the compatibility of this
type of hypothesis with the data would be to compute for each time $t$
a Bayes factor between the VAR model under
the restriction $\widetilde{\varphi_{t}}=0$ for just one value of $t$ ($M_1^t$) and the VAR
model without restrictions ($M_2^t$). For each $t$, this would
requires the computation of the evidence of model $M_1^t$,
which seems untractable: one would need a Markovian process for $\varphi_{t}$ that is
conditionned (sort of bridge) on
the restriction that $\widetilde{\varphi_{t}}=0$ for just one given $t$. At the very least, one would need to estimate a different (conditional) model for each $t$ and compute its free energy.
 Additionnaly, we have no guarantee that the free energy approximation
is of the same magniture for all these models.

We will use a simpler approach that rely only on the posterior density of the (unconditional) model. The use of highest posterior density (HPD) regions for Bayesian testing was introduced in \citet{box2011bayesian} and used in the Bayesian literature, as for example in \citet{kim1994bayesian} and \citet[p.~280]{harrison1999bayesian}. 
  For a given time $t$, consider the sub-vector $\widetilde{\varphi_{t}}$ and let $c$ be its dimension. A suitable partition of the posterior parameters $\mathbf{\mu_{t}}$ and $\Sigma_{t}$ gives the marginal posterior density $q(\widetilde{\varphi_{t}}|Z_1^T)=\mathcal{N}_c(\widetilde{\mu_{t}},\widetilde{\Sigma_{t}})$. The contribution of each element of $\widetilde{\varphi_{t}}$ to the prediction of $Y_t$ may be assessed by considering the compatibility in this marginal posterior with the value $\widetilde{\varphi_{t}}=\widetilde{\varphi_{0_{t}}}$, where $\widetilde{\varphi_{0_{t}}}$ is a zero vector of dimension $c$. The question is therefore to know whether the parameter point $\widetilde{\varphi_{t}}=\widetilde{\varphi_{0_{t}}}$ is included in the highest posterior density region (HPD) of size $1-\alpha$. 
 This happens if and only if 
$(\widetilde{\varphi_{0_{t}}}-\widetilde{\mu_{t}})' \widetilde{\Sigma_{t}}^{-1}(\widetilde{\varphi_{0_{t}}}-\widetilde{\mu_{t}})<k$,
where $k$ is the $1-\alpha$ quantile of the standard $\chi^2$ distribution with $c$ degrees of freedom. 
As stated in \citet[p.~125]{box2011bayesian}, it follows that the parameter point $\widetilde{\varphi_{0_{t}}}$ is covered by the HPD region of content $1-\alpha$ if and only if $\mathrm{Pr} \{p(\widetilde{\varphi_{t}}|Z_1^T)> p(\widetilde{\varphi_{0_{t}}}|Z_1^T)|Z_1^T \} \leq 1-\alpha$. 
Equivalently, we can search the region of minimum coverage that contains $\widetilde{\varphi_{0_{t}}}$.

The above approach gives only pointwise evaluations (i.e., for a given time $t$). When jointly testing a set of values for a complete time, frequency or time-frequency connectivity map, it is important to suitably correct the significance threshold for multiple comparisons. We do not correct it in Section \ref{Assessment of accuracy} as we are interested in the separate evaluation for each time but we do for the application in Section \ref{Application}.

\section{Assessment of Accuracy}
\label{Assessment of accuracy}
We now turn our attention to the accuracy of variational Bayesian inference for our model \eqref{kalman.def} under the mean-field assumption described in Section \ref{Factorization assumption} and priors discussed in Section \ref{Model specification}. 
 We provide here a study of the quality of variational Bayesian estimate for model order selection and Granger causality detection. The first simulation study in Section \ref{Granger causality detection} evaluates the ability of the BSS and MSBSS models to detect Granger causalities between two signals and in Section \ref{Windowing comparison}, we present a systematic comparison between the proposed BSS and MSBSS models and the windowing approach in term of Granger causality detection ability.
 Note that a Monte-Carlo type of simulation is not feasible due to the very large amount of unknown parameters of our model and so direct comparison of the obtained variational posterior with true posterior density was unfortunately not feasible.   

\subsection{Practical implementation}
\label{Practical Implementation}
To initialize the BSS and the MSBSS model algorithms, we run 10 iterations of the simple EM algorithm \citep{shumway_approach_1982}. We thereby obtain reliable starting values for $\{\varphi_1,\Sigma_1,\Omega_1^b\}$ under initial conditions defined in Section \ref{Model specification}. We then iteratively update the parameters of the variational approximate posteriors through the 
 Variational Bayes EM until the relative free-energy criterion described in Section \ref{The Variational Evidence Lower Bound} between two consecutive iterations changes less than a tolerance value that we choose here to be equal $10^{-2}$ for the model order selection study and to $10^{-4}$ for the Granger causality detection study following \citet{menictas2013variational}.

\subsection{Granger-causality detection}
\label{Granger causality detection}

We will now assess the ability of the proposed models BSS and MSBSS to detect Granger causality. We simulated signals with parameters that vary slowly in time, which is a reasonable simulation of neuroscientific data. It will show how the method performs with data that are not generated according to the model. The simulation consists in $50$ replications of a bivariate BSS model but with hidden variables $\varphi_1^T$ that evolve slowly and deterministically through time. Based on these deterministic $\varphi_1^T$, data are generated from the observed equation in model \eqref{kalman.def}. The simulations were carried out for model orders $\{1; 2; 4; 8\}$, series lengths $\{500;1000;2000\}$, number of trials $\{1;10\}$ and also for different values of the causal parameter $\{1;0.8;0.6;0.4;0.2\}$ in order to see the limits of the methods. The simulations for the causal parameter values $\{0.8;0.6;0.4;0.2\}$ were carried out for all model orders and number of trials, but only for a series length of $500$ time points (which is the case where the method will break first).

The $R$ matrix was fixed to the $I_d \times 0.1$ values and the slowly-varying parameters are all set to zero except for the entries related to the causal parameters $\{\varphi_{2 \rightarrow 1}\}_{t=1}^T$ and $\{\varphi_{1 \rightarrow 1}\}_{t=1}^T$ for order $p$.  The values of the simulated parameters can be seen on the two panels (which are identical) on the top of Figures in Appendix N.
Signals  were simulated with normal and non-normal errors. Additional
simulation settings and additional results are available in \cite{cekic_time-frequency_2015}

\paragraph{Data generated with slowly-varying parameters and normal errors}
\label{Slowly varying normal}

Concerning the MSBSS model, for each of the $50$ simulation, the model order $p_j$ for the different scales as well as the number of scales $J$ to be taken in the model are selected as described in Section \ref{The Variational Bayesian State Space Multiscale Model}. The maximum number of scale was set at $4$ and the maximal model order per scale at $5$. In multiple trial scenarios, the number of scales as well as the model order selection procedure was carried on one single trial only (for computational simplicity). Estimation procedure is always conditioned on the same number of time points, allowing us to compare the models.
The Granger-causality detection capability of a method is quantified by the true negative rate (TNR) and the true positive rate (TPR). For each time, TNR is the percentage of the $50$ simulations for which the $95$\% HPD region defined in Section \ref{Bayesian Granger-Causality statistic} contains the causal statistic when it must actually contain it, and TPR is the percentage of the $50$ simulations for which the $95$\% HPD region does not contain the causal statistic when it should not. Secondly, the BSS model is estimated with the true model order (oracle order) and the relative TNR and TPR are computed for that model. Finally, for the BSS model, the model order $p$ is selected with the free-energy criterion as described in Section \ref{The Variational Evidence Lower Bound} and the model is estimated based on this selected order $p$. The relative TNR and TPR are calculated.
 
The causality detection ability for the BSS model with estimated model order is not shown in the results, because the model order was selected correctly by the free-energy criterion for each simulation (the causality detection ability is thus exactly the same as this for the BSS model with oracle order). 

\begin{figure}[!tb]
\centering
\hspace*{-1in}
\includegraphics[width=5in]{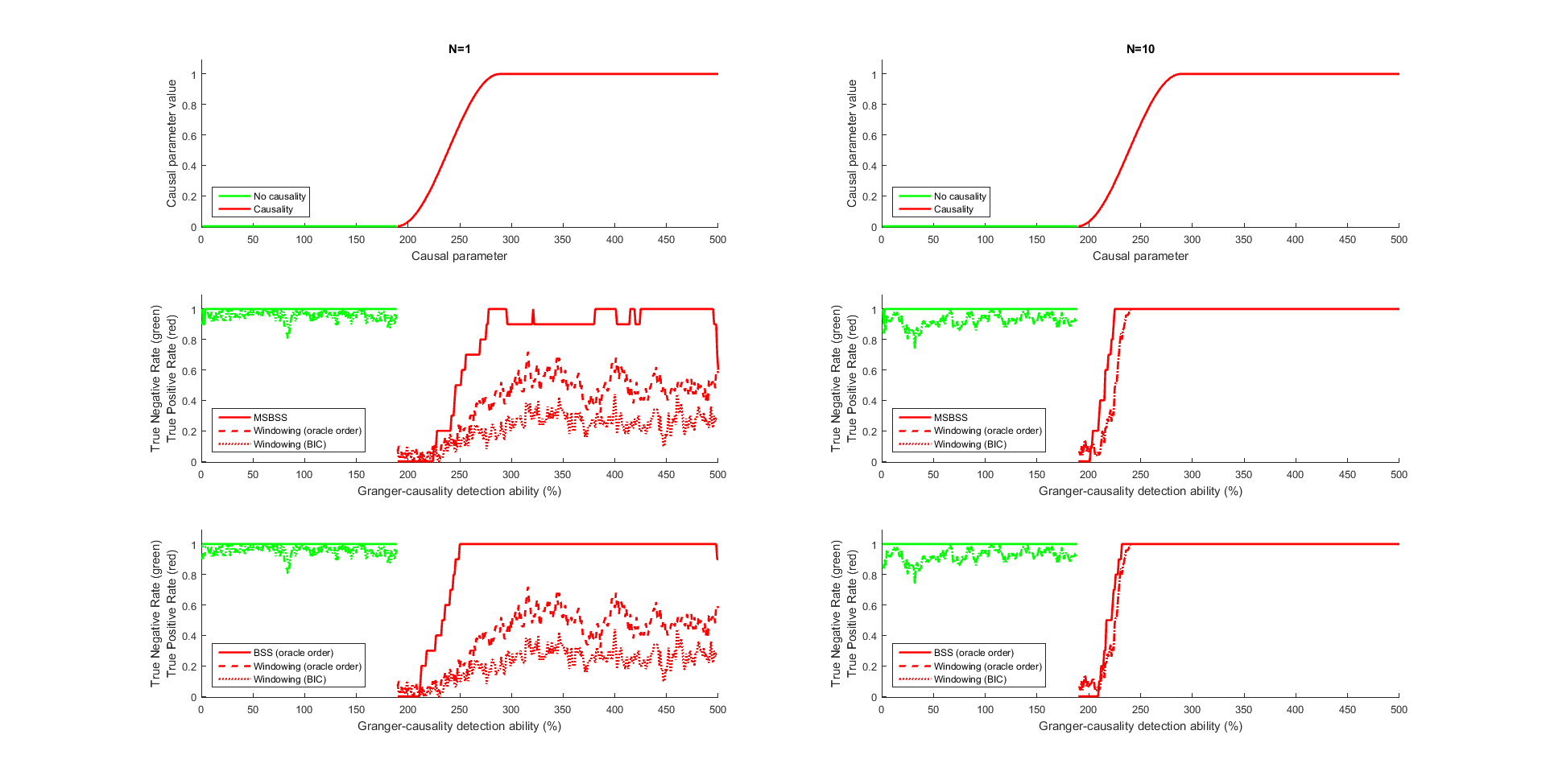}
\caption{Granger-causality detection ability for order $4$, series length $500$ and causal parameter $1$. Top graphs show the value of the true parameters. Middle graphs display the true positive and true negative rates for the MSBSS model estimation, the windowing estimation with the true model order estimate (oracle) and the windowing estimation with the model order selected based on the BIC criterion. Bottom graphs display the true positive and true negative rates for the BSS model estimation, the windowing estimation with the true model order estimate (oracle) and the windowing estimation with the model order selected based on the BIC criterion.}
\label{AOAS_3_1}
\end{figure}

\begin{figure}[!tb]
\centering
\hspace*{-1in}
\includegraphics[width=5in]{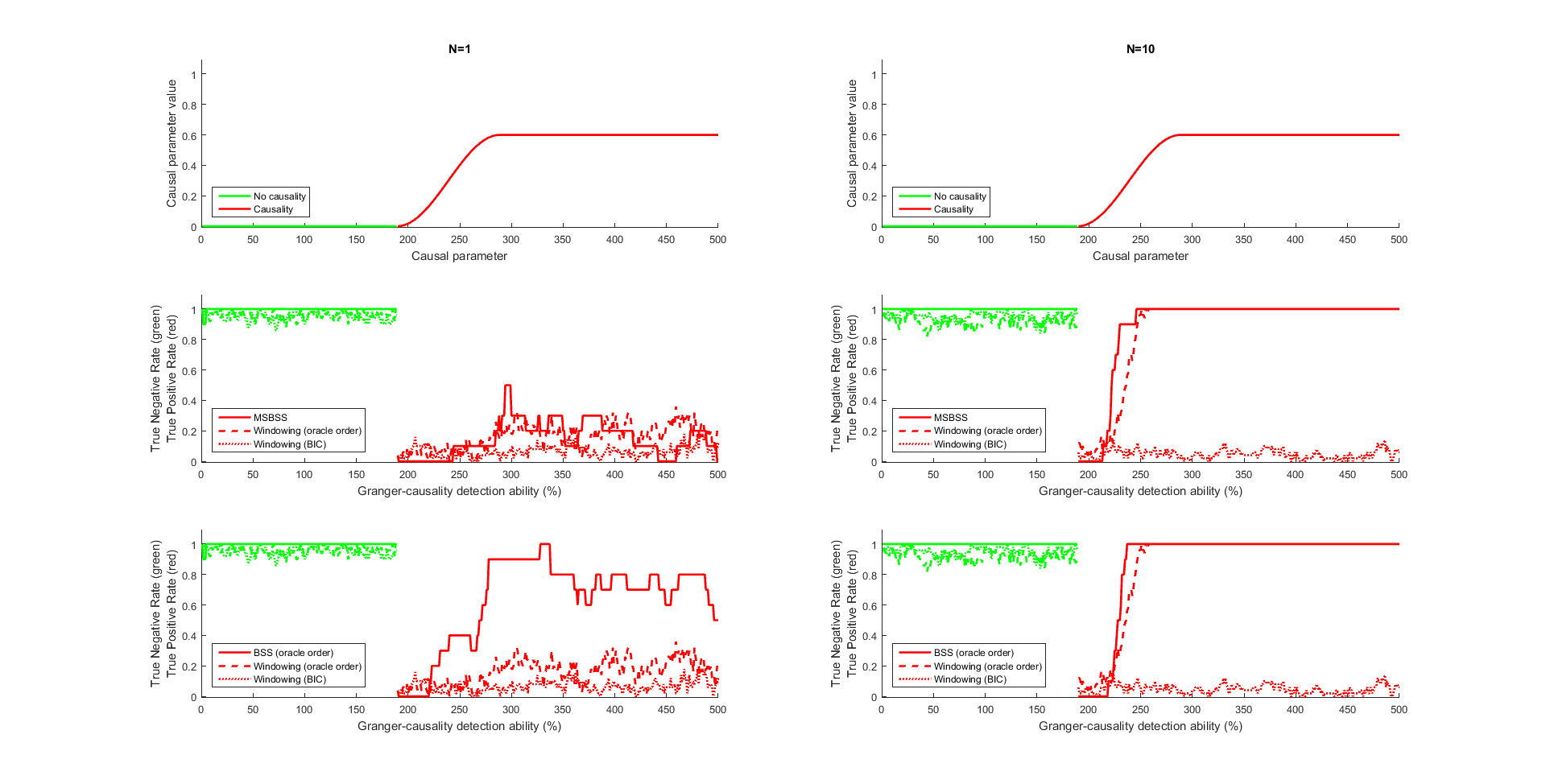}
\caption{Results for the same settings as Figure~\ref{AOAS_3_1} except
  for the causal parameter which is set to $0.6$}
\label{AOAS_3_3}
\end{figure}

Figures \ref{AOAS_3_1} and~\ref{AOAS_3_3} show Granger-causality
detection accuracy for a model order $4$, series length of $500$ and
causal parameter equal to $1$ and $0.6$ respectively. By construction,
the causality arises between times $190-500$, as shown in the two
graphs on the top of the figures. For Figure~\ref{AOAS_3_1}, the BSS model performs slightly better than the MSBSS model with $1$ trial and both methods yield very similar results, correctly recovering 
the underlying directional influences with $10$ trials. One can
observe that the causality detection evolves as the causal parameter
changes from zero to one. The TPR is indeed gradually increasing as
the parameter changes from zero to one, slightly faster for the BSS
than for the MSBSS model for $N=1$, and identically for $N=10$. For
Figure~\ref{AOAS_3_3}, with such a low causal parameter, the TPR related to the MSBSS model is worse than the one related to the BSS model for $1$ trial, but both method perform very well with $10$ trials although the causal parameter is low.

Figures for Granger-causality detection accuracy related to all other model orders, series lengths and causal parameter values for normal errors are given in Appendix N. 
Globally, for a causal parameter that equals $1$, the MSBSS and the BSS models show very good Granger-causality detection accuracy in terms of TPR and TNR for $1$ and $10$ trials. The only cases which display poorer results are the causality detection for the MSBSS model with $1$ trial for model order $2$ and series length $500$, $1000$ and $2000$. 
When the causal parameter value decreases, the causality detection ability decreases as well. Poorer causality detection results are globally observed for a causal parameter from $0.6$ to $0.2$ for $1$ trial and for a causal parameter value of $0.2$ for $10$ trials, especially for the MSBSS model.

\paragraph{Data generated with slowly-varying parameters and non-normal errors}
\label{Slowly varying non-normal}

We also simulated data with the same settings as above, except for the
observation equation errors which are here multivariate t-distributed with
parameters $\nu=5$ and $R=I_d \times 0.1$. All graphical results can
be found in Appendix N.  Globally, the detection accuracy remains
satisfactory with data generated with non-normal errors.  The
Granger-causality detection ability of the proposed methods is
therefore robust to this model assumption departure.

\paragraph{Comparison with the windowing approach.}
\label{Windowing comparison}

We will now compare our proposed models (BSS and MSBSS) with the windowing approach. The comparison is performed on data generated with slowly-varying parameters with normal errors as explained in Section \ref{Slowly varying normal}. Simulations are performed for model orders $\{1;2;4\}$, series length $500$, number of trials $\{1;10\}$ and causal parameter values $\{1;0.8;0.6;0.4;0.2\}$. Data are fitted using the sliding window methodology proposed in \citet{ding_short-window_2000}. The overlap parameter between the time-windows is chosen to be equal to $1$ time point and the subjective windows size is chosen to be equal to $15$ time points. The model estimation procedure is performed using the Viera--Morf algorithm implemented in BSMART \citep{cui_bsmart:_2008} and GCCA toolboxes \citep{seth2010matlab}. A first estimate was obtained with the true model order (oracle). 
For a second estimate, the model order selection is performed by considering the mode of the optimal model order calculated in each temporal window based on the Bayesian Information Criterion (BIC) as proposed in the BSMART toolbox \citep{cui_bsmart:_2008}, in the GCCA toolbox \citep{seth2010matlab} and in the SIFT toolbox \citep{mullen2010electrophysiological}. Based on the estimated models, a time-domain Granger-causal $F$ statistic is computed in each temporal window and its significance is assessed through the asymptotic $F$ distribution (see \citep{cui_bsmart:_2008} and \citep{seth2010matlab} for further details).

In Figures \ref{AOAS_3_1} and~\ref{AOAS_3_3}, the dash and the dot lines
represent the results for the windowing approach. For Figure \ref{AOAS_3_1}, the MSBSS and the BSS models perform much better than the windowing approach with the true model order and with the model order selected based on the BIC criterion for $1$ trial. MSBSS and the BSS models detect the causality faster than the windowing approach with $10$ trials.
The TPR is worse for the windowing estimation than for the MSBSS and the BSS models for both $1$ and $10$ trials. 
For the case with a low causal parameter (Figure \ref{AOAS_3_3}), the windowing estimation with oracle order performs almost identically as the MSBSS model and performs much worse than the BSS model for $1$ trial. MSBSS and the BSS models detect the causality faster than the windowing approach with oracle order with $10$ trials. The windowing estimation with the model order selected based on the BIC criterion display very poor detection accuracy for $1$ and for $10$ trials.

All other graphical results related to the windowing estimation procedure can be found in Appendix N.
Globally, for the simulations with $1$ trial, the model order selection procedure based on the BIC performs well for order $1$ and is inaccurate for orders $2$ and $4$ and the related Granger-causality detection fails. One can observe that the causality detection evolves as the causal parameter changes from zero to the maximal value of the causal coefficient. For the simulations with $1$ trial, the windowing estimate detects the causality pattern globally slower than the MSBSS and the BSS models, whereas the TNR remains almost identical. When the value of the causal parameter decreases, the TPR for the MSBSS model decreases as well and becomes worse than the TPR for the windowing estimate in some cases (e.g., for the causal parameter $0.6$ and model order $1$). The TPR for the BSS model estimate decreases with the value of the causal parameter as well, but its TPR is always better than the TPR for the windowing estimation procedure. When the causal parameter equals $0.2$, however, the MSBSS and BSS model estimates do not detect anything whereas the windowing approach displays a TPR around $5-10$\%.

For the simulations with $10$ trials, the model order selection procedure based on the Bayesian Information Criterion performs well overall. Windowing estimate (with model order selected by the BIC and oracle model order) globally detects the causality pattern slower than the MSBSS and the BSS models, whereas the TNR is almost identical everywhere. When the value of the causal parameter decreases, the model order selection procedure based on the BIC performs less well and its TPR degrades even more than the one for the MSBSS and the BSS models (e.g., for order $2$ and causal coefficient $0.4$ and $0.2$ and for order $4$ and causal coefficient $0.6$, $0.4$ and $0.2$).

\section{Application}
\label{Application}

We will now apply the proposed MSBSS model to real iEEG (intracranial electroencephalogram) data recorded during a psychological experimental situation. Brain recordings
 are localized within the amygdala (AMY) and medial orbito-frontal cortex (mOFC) regions in order to study the dynamics of neuronal processes between these regions in response to emotional prosody exposure. It is known in the literature that the emotional content of the stimulus induces the presence of causal links AMY $\rightarrow$ mOFC and mOFC $\rightarrow$ AMY (\citet{grandjean_voices_2005},\citet{grandjean_intonation_2006},\citet{grandjean_unpacking_2008}). 

\subsection{Results}
\label{Results} 
Here we present the results for one patient and the experimental
condition anger. The patient was exposed to short pseudowords pronounced with an angry prosody (\citet{grandjean_voices_2005}). The data were acquired with a high
resolution of $512$Hz during $2.25$sec. There are $28$ trials
available. As high frequency behaviour of data was not of interest, we
downsampled the data by a factor of four with an exponential smoother
(it is of course vital to process with a smoother that is not based on
future values). A MSBSS model was estimated following the procedure
described in Section \ref{The Variational Bayesian State Space
  Multiscale Model}. As researchers were already aware about frequency
bands where they expect a causal relationship, the number of
scales $J_{\text{max}}$ was fixed to $4$ (on the downsampled data),
corresponding to four frequency bands respectively around $64$, $32$,
$16$ and $8$Hz, plus a smooth that represents the frequency content
below $8$Hz with a maximal order per scale $p_j$ of $[5,5,3,3,1,1]$. These particular choices for $p_{\text{max}}$ are motivated by the occurrence of the onset of the stimulus at $250$ms. Indeed, due to the use of past
values in the model, the estimation procedure starts at time
$\max p_{\text{max},j}2^{j}+1$ and we do not want to miss the onset of the stimulus in the model estimation. The model order per scale $p_j$ is selected on the whole set of trials.  

Based on the MSBSS model, a significant Granger causality from signal 1 to signal 2 at frequency $f$ and at time $t$ means that the energy of signal 1 at frequency $f$ significantly improves the prediction of the value of signal 2 at time $t$. 

The testing procedure was performed as follows: we first tested the overall set of causal VAR parameters of interest and then the scale-specific causal VAR parameters ($\widetilde{\varphi_{t}}$) by considering the compatibility in their marginal posterior with the value $\widetilde{\varphi_{t}}=\widetilde{\varphi_{0_{t}}}$, where $\widetilde{\varphi_{0_{t}}}$ is a zero vector of dimension $c$. We computed the coverage of the smallest highest posterior density (HPD) region that contains $\widetilde{\varphi_{0_{t}}}$. We will call the complementary probability of this coverage the significance level. 

The results are reported for the two directional causalities AMY $\rightarrow$ mOFC and mOFC $\rightarrow$ AMY. 
The first line of each graph in Figure \ref{CLUSTERS_SIGN} represents the results related to the overall statistic and each following line represents the results related to each scale corresponding to the frequency bands respectively around $64$, $32$, $16$ and $8$Hz.

The number of tests provided by the proposed method is basically
proportional to the number of time points. To circumvent the multiple testing problem,
a solution that seems suitable is the cluster mass
test which consists in defining clusters of neighbouring time regions using a permutation scheme to assess its significance. We applied it for the overall testing for each frequency with a threshold corresponding to a level of $.2$ and this procedure controls for the family wise error rate (FWER) \citep{maris2007nonparametric}.

\subsubsection{Testing the scale specific causality}
\label{Testing the scale specific causality}

\begin{figure}[!tb]
\centering
\hspace*{-1in}
\includegraphics[width=6.4in]{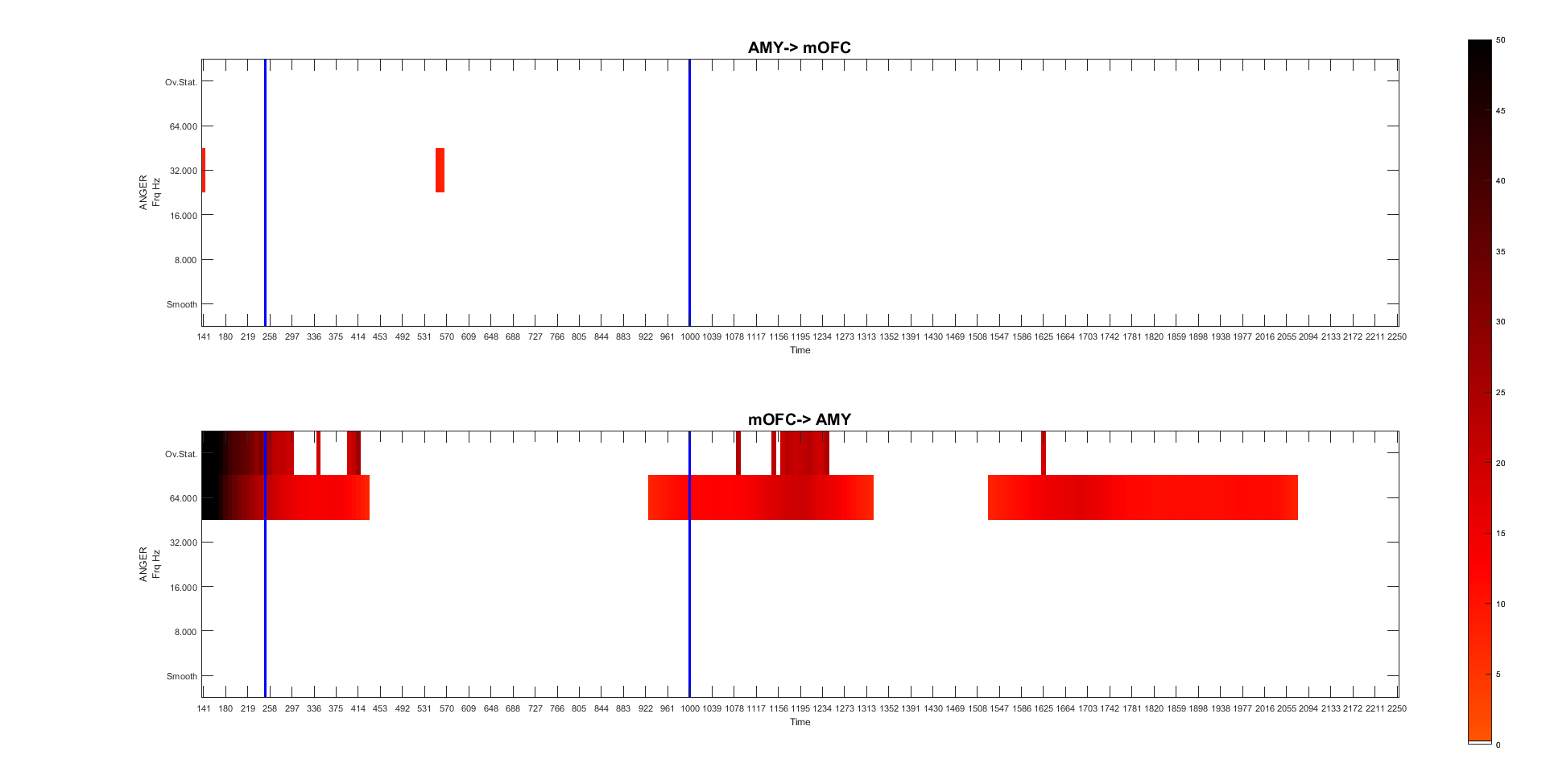}
\caption{Results of the estimated Granger causality for the two causal links of interest AMY $\rightarrow$ mOFC and mOFC $\rightarrow$ AMY for the overall causality statistic 
and the scale specific causality statistics. Intensity represents the
value of the individual statistic. Only the clusters assessed
significant by the cluster mass test are displayed. Vertical lines
represent the onset and the offset of the stimulus which occur
respectively at $250$ms and $1000$ms. They are displayed for the
interpretation of the results but were not provided to the model.}
\label{CLUSTERS_SIGN}
\end{figure}

Figure \ref{CLUSTERS_SIGN} shows the results of the estimated
Granger causality for the two causal links of interest AMY $\rightarrow$ mOFC
and mOFC $\rightarrow$ AMY. Vertical lines represent the onset and the offset of the stimulus which occur respectively at $250$ms and $1000$ms. They are displayed for the
interpretation of the results but were not provided  to the model. The estimated model order for each scale
plus the smooth is [5 5 3 1 1]. Only the clusters assessed significant
by the cluster mass test are displayed and the colors represent the
value of the individual statistic. As mentioned in Section \ref{The
  Variational Bayesian State Space Multiscale Model}, the smooth
represents the frequency content of the series from the largest scale to the lowest frequency in the signal.

These results give a partial answer to the question of how AMY and
mOFC regions are functionally and causally related during the exposure
of auditory emotional stimuli. Actually, an initial Granger causality
event from mOFC $\rightarrow$ AMY is observed in gamma range ($64$Hz) just
before the stimuli onset and during stimulus exposure followed by an
AMY $\rightarrow$ mOFC in beta range ($32$Hz) during this same period. At the
offset of the stimuli the Granger causality is again from mOFC
$\rightarrow$ AMY in gamma range ($64$Hz), this directional Granger causality
is sustained during the period after the stimuli presentation. These
results are compatible with a known complex cross-talk between these two
brain regions during emotional stimulus exposure and the related
meaning of such event for the organism. Of course such results should
be extended to several patients with similar brain recordings in
targeted brain regions and compared to the processing of other emotional
stimuli.

\section{Conclusion}
\label{Conclusion}

We derived a time-varying Granger-causality statistic through a
Bayesian nonstationary multivariate time series model with dynamic
coefficients. While similar models have been proposed
\citep{cassidy_bayesian_2002}, one of the main contributions of this
article is to provide an assessment of accuracy of the method, and
especially an extension to the \textit{\`{a} trous} Haar wavelets
transform \citep{renaud_prediction_2003}. This very flexible
\textit{\`{a} trous} Haar procedure enables us to capture short- and
long-range dependencies between signals with only few parameters to be
estimated and allows us to be specific for the frequency in the assessment of causality, which is a main point of interest in the neuroscience community. The central finding of this article is that variational
Bayesian estimate of time-varying VAR models enables us to achieve
good (if not excellent) accuracy in terms of Granger-causality
detection for normal, but also for non-normal errors and that the method performs much better than the commonly used windowing methodology in terms of Granger-causality detection accuracy. This model thus provides a very powerful tool for dynamical spectral causality analysis in a neuroscientific context.

     Further points to highlight are the suitability of the model to deal with multiple trial in a fully correct way and the potential extensibility of the model to deal with more than two signals, or two sites of interest.

The choice of the sampling frequency (e.g. for EEG recordings) and the preprocessing steps commonly perform by researchers in neuroscience are two delicate issues for subsequent Granger-causality analysis, because different choices may lead to different Granger-causality results. Due to the multiscale decomposition, the MSBSS model is probably much more robust in this regard, and this would be an interesting topic for further research.

\section*{Toolbox and Appendix} 
    
An open matlab toolbox called MSGranger is available at the following
url:
\url{https://www.unige.ch/fapse/mad/services/matlab/msgranger/}. Single
and multiple trials are implemented and both the Bayesian State Space
(BSS) and the multiscale Bayesian state space (MSBSS) models can be used
to obtain  dynamical and frequency-specific
  Granger-causalities.

Appendices with all the derivations, Figures referenced in the text,
as well as the cluster mass test and data analysed in Section
\ref{Application} are available in the appendix.

\section*{Acknowledgements}
The authors gratefully acknowledge Center for advanced modelling
science, Swiss National Science Foundation under Grant 100014\_156493 and Lifebrain H2020-SC1-2016-201 under Grant 732592.

\bibliographystyle{abbrvnat}
\bibliography{stat_OR}

\appendix

\clearpage
\begin{center}
{\large\bf SUPPLEMENTARY MATERIAL}
\end{center}

\renewcommand\contentsname{}
\tableofcontents

\clearpage

\section{Elements of variational Bayes}
\label{Appendix VB}

The most common type of variational Bayesian methodology, known as
mean-field approximation uses the Kullback--Leibler distance (KL distance) between $q(.|Z_1^T)$ and $p(.|Z_1^T)$ as a
dissimilarity function. 

\subsubsection{Learning rules}
By simple algebra, we will decompose the evidence of the model $p(Z_1^T)$ by inserting the variational density $q$. Using the fact that $p(\varphi_{1}^T,\Omega_1^b,Z_1^T)=p(\varphi_{1}^T,\Omega_1^b|Z_1^T)p(Z_1^T)$, we have
\begin{equation}
\begin{aligned}
\mathrm{KL}\big(q(\varphi_{1}^T,\Omega_1^b|Z_1^T)\| p(\varphi_{1}^T,\Omega_1^b|Z_1^T)\big) = \left< \log \frac{q(\varphi_{1}^T,\Omega_1^b|Z_1^T)}{p(\varphi_{1}^T,\Omega_1^b,Z_1^T)}\right>_{q(\varphi_{1}^T,\Omega_1^b|Z_1^T)} +\log p(Z_1^T),
\end{aligned}
\label{KL}
\end{equation}
and therefore
\begin{equation}
\begin{aligned}
\log p(Z_1^T) &= \mathrm{KL}\big(q(\varphi_{1}^T,\Omega_1^b|Z_1^T)\| p(\varphi_{1}^T,\Omega_1^b|Z_1^T)\big)-\left<\log \frac{q(\varphi_{1}^T,\Omega_1^b|Z_1^T)}{p(\varphi_{1}^T,\Omega_1^b,Z_1^T)}\right>_{q(\varphi_{1}^T,\Omega_1^b|Z_1^T)}, \\
\label{VB_freeNRJ}
\end{aligned}
\end{equation}
where $\left<.\right>$ denotes expectation and its subscript denotes the density used for this expectation.
Equation \eqref{VB_freeNRJ} is the fundamental equation of variational Bayesian
methodology. By necessary positiveness of the KL distance, we have obtained a lower bound for the logarithm of the evidence
\begin{equation}
\begin{aligned}
\log p(Z_1^T) &\geq -\left<\log \frac{q(\varphi_{1}^T,\Omega_1^b|Z_1^T)}{p(\varphi_{1}^T,\Omega_1^b,Z_1^T)}\right>_{q(\varphi_{1}^T,\Omega_1^b|Z_1^T)} := \text{F}\big (q(\varphi_{1}^T,\Omega_1^b|Z_1^T)\big),
\label{VB_freeNRJ2}
\end{aligned}
\end{equation}
where
$\text{F}\big(q(\varphi_{1}^T,\Omega_1^b|Z_1^T)\big) $ is called the
negative free energy. By minimization of the KL distance between $q$
and $p$, we maximize
$\text{F}\big(q(\varphi_{1}^T,\Omega_1^b|Z_1^T)\big)$. Furthermore,
since the KL distance is equal to zero if and only if the two
densities $q(\varphi_{1}^T,\Omega_1^b|Z_1^T)$ and
$p(\varphi_{1}^T,\Omega_1^b|Z_1^T )$ are identical, the functional
quantity $\text{F}\Big(q(\varphi_{1}^T,\Omega_1^b|Z_1^T)\Big)$ equals
the model evidence if and only if $q(\varphi_{1}^T,\Omega_1^b|Z_1^T)$
equals the true target posterior $p(\varphi_{1}^T,\Omega_1^b|Z_1^T
)$. 
The aim is thus to find a density $q(\varphi_{1}^T,\Omega_1^b|Z_1^T)$  for which the  integrals in $\text{F}\Big (q(\varphi_{1}^T,\Omega_1^b|Z_1^T)\Big)$ are tractable and which is close to $p(\varphi_{1}^T,\Omega_1^b|Z_1^T)$.

\subsubsection{Mean-field approximation}
\label{Factorization assumption}
The choice underlying the variational Bayesian methodology, known
as mean-field approximation in physics, is to allow the
approximating density $q(\varphi_{1}^T,\Omega_1^b|Z_1^T)$ to
factorize over groups of parameters \citep[see][for a comparison between Laplace and variational Bayesian assumptions]{friston_variational_2007}. We will suppose here that the approximating density factorizes as
\begin{equation}
q(\varphi_{1}^T,\Omega_1^b|Z_1^T) = q(\varphi_1^T|Z_1^T)\prod_{j=1}^b q(\Omega_j|Z_1^T),
\label{mean field approx}
\end{equation}
and the lower bound for the model evidence can be rewritten based on this factorization as
\begin{equation}
\text{F}\Big (q(\varphi_1^T|Z_1^T),q(\Omega_1|Z_1^T), \dots,q(\Omega_b|Z_1^T)\Big).
\end{equation}
Depending on the model at hand, mean field approximation may have minor to major impacts on the resulting inference. For example, if $p(\varphi_{1}^T,\Omega_1^b|Z_1^T)$ is such that $\varphi_{1}^T$ and $\Omega_1^b$ have a high degree of dependence, then the restriction $q(\varphi_{1}^T,\Omega_1^b|Z_1^T)$ =
$q(\varphi_{1}^T|Z_1^T)$ $q(\Omega_1^b|Z_1^T)$  will lead to serious degradation in the resulting inference \citep{ormerod_explaining_2010,titterington_bayesian_2004,beal_variational_2003}. The factorization in equation \eqref{mean field approx} is obviously not unique. For instance, some authors factorize also $\varphi_1^T$ into $[\varphi_1, \dots, \varphi_T]$ \citep{wang2004lack}.

\subsubsection{Variational EM algorithm}
\label{Variational EM algorithm}

With the use of the calculus of variations (hence the name variational Bayes), it can
be shown that under assumption \eqref{mean field approx}, the variational distributions $q^*(\varphi_1^T|Z_1^T)$ and $q^*(\Omega_j|Z_1^T)$,
that maximize the functional $\text{F}\Big (q(\varphi_1^T|Z_1^T),$ $q(\Omega_1|Z_1^T),\dots, q(\Omega_b|Z_1^T)\Big)$, can be expressed and therefore maximized in an iterative way \citep{beal_variational_2003,fox_tutorial_2011}. First,
\begin{equation}
q^*(\varphi_1^T|Z_1^T)^{(l+1)}  \propto\exp \left< \log p(\varphi_1^T|\Omega_1^k,Z_1^T)\right> _{q(\Omega_1^b|Z_1^T)}^{(l)},
\label{optimal q (x)2}
\end{equation}
where superscript $(l)$ denotes the iteration number. The other steps for $m=1 ,\dots, b$ are
\begin{equation}
q^*(\Omega_m|Z_1^T)^{(l+1)} \propto \exp \left< \log p(\Omega_1^k|\varphi_1^T,Z_1^T)\right> _{-\Omega_m}^{(l)},
\label{optimal q Omega 2}
\end{equation}
where $\left< .\right> _{-\Omega_m}^{(l)}$ is the expectation over all the distributions at iteration ${l}$ except $q(\Omega_m|Z_1^T)^{(l)}$. See \citet{beal_variational_2003} and \citet{ostwald2014tutorial} for all proofs. 

The distributions $\exp \left< \log p(\varphi_1^T|\Omega_1^k,Z_1^T)\right> _{-\varphi_1^T}^{(l)}$ and $\exp \left< \log p(\Omega_1^k|\varphi_1^T,Z_1^T)\right> _{-\Omega_m}^{(l)}$ are known as full conditionals
in the MCMC literature. The mutual dependence of the optimal variational posterior densities in equations \eqref{optimal q (x)2} and \eqref{optimal q Omega 2} suggest a similarity with Gibbs sampling \citep{casella1992explaining} which involves successive draws from the full conditionals. Mean field approximation indeed leads to tractable solutions in situations where Gibbs sampling is also
applicable.

Thus, if we set $q(\varphi_1^T|Z_1^T)$ equal to $q^*(\varphi_1^T|Z_1^T)$ and $q(\Omega_m|Z_1^T)$ equal to $q^*(\Omega_m|Z_1^T)$, we have maximized the lower bound for the model evidence $\text{F}$ under the \eqref{mean field approx} constraint.

The form of equations \eqref{optimal q (x)2} and \eqref{optimal q Omega 2} define a circular dependence which explains the use of an iterative algorithm whose convergence can be assessed by monitoring the relative increase of $\text{F}$. This iterative algorithm is known as the variational Bayesian expectation-maximisation algorithm \citep{beal_variational_2003}.
The result is that the formulas for the sufficient statistics of each
unknown distribution $q^*(.|Z_1^T)$  can be expressed as a series of
equations with mutual dependencies. 
As discussed in \citet{beal_variational_2003} and \citet{cassidy_bayesian_2002}, this variational Bayesian EM algorithm reduces to the ordinary frequentist EM algorithm for maximum likelihood estimate \citep{dempster_maximum_1977} if the parameter priors are flat.

\subsubsection{Directed acyclic graphs and Markov blanket theory} 

\begin{figure}
\begin{center}
\includegraphics[width=5in]{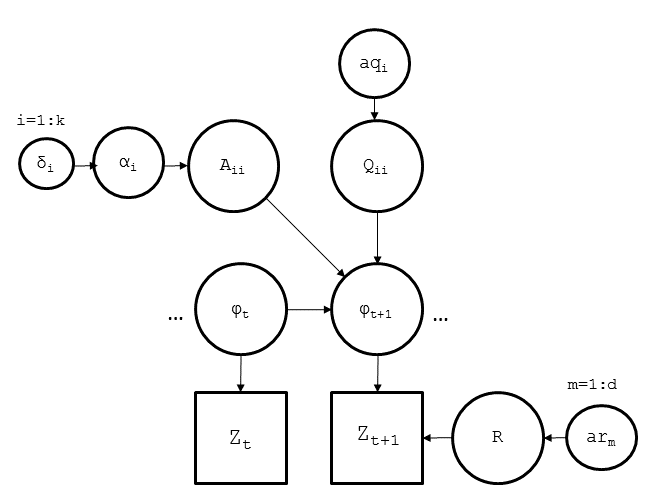}
\end{center}
\caption{DAG representation of the full model given in Section 3.}
\label{DAG_Figure2}
\end{figure}

The model proposed in Section 2 can be viewed as a
hierarchical Bayesian model and hence can be represented as a
probabilistic directed acyclic graph (DAG). This DAG representation is
very useful for visualising the relationships between hidden variables
($\varphi_1^T$), parameters ($\Omega_1^b$) and observations ($Z_1^T$),
each of them being represented as nodes. Typically, square nodes
indicate observed variables, round nodes indicate latent random variables and arrows act for conditional dependence. Figure \ref{DAG_Figure2} contains the DAG for the full model considered in the present article. Moreover for models having such a structure, the variational Bayesian algorithm benefits from a graphical-related concept hailing from machine learning theory, known as variational message passing \citep{winn_variational_2005}. More specifically, the benefits for variational Bayesian models are directly related to the concept of Markov blanket that we will first define.

\begin{definition} 
The Markov blanket of a node $x_i$ in a DAG ($mb(x_i)$) is defined as the set of its parents, $pa(x_i)$, children, $ch(x_i)$, and co-parents, $cop(x_i)$. Two nodes are defined co-parents if they have at least one child node in common \citep{winn_variational_2005}.
\label{MB}
\end{definition}

The point of particular interest for variational Bayesian theory is that the variational sequential update equation for a node $\Omega_m$ only depends on expectations over variables in its Markov blanket.   
It directly follows that equation \eqref{optimal q Omega 2} can be rewritten as
\begin{equation}
q^*(\Omega_m|Z_1^T)^{(l+1)} \propto \exp \left<\log p(\Omega_1^b|mb(\Omega_m),Z_1^T)\right>_{mb(\Omega_m)}^{(l)}.
\label{VMP_OMEGA}
\end{equation} 
\citet{fox_tutorial_2011} show that equation \eqref{VMP_OMEGA} can be rewritten in an even simpler form as
\begin{equation}
\begin{aligned}
\ln  q^*(\Omega_m|Z_1^T)^{(l+1)} = \qquad & \left< \ln  p(\Omega_m|pa(\Omega_m),Z_1^T)\right>_{pa(\Omega_m)}^{(l)} + \\
& \sum\limits_{ch_i \in ch}\left< \ln  p(ch_i|\Omega_m,cop(\Omega_m);ch_i,Z_1^T)\right>_{ch_i,cop(\Omega_m;ch_i)}^{(l)}+ \text{ct}.
\end{aligned}
\label{VMP_OMEGA2}
\end{equation} 
Equation \eqref{VMP_OMEGA} is much simpler than equation \eqref{optimal q Omega 2}.
Similar simplifications can be obtained for equation \eqref{optimal q (x)2}.

\subsubsection{Conjugate exponential}
\label{Conjugate Exponential}

The optimal form for $q^*(\varphi_1^T|Z_1^T)$ and $q^*(\Omega_1^b|Z_1^T)$ of course depends on the choice of the prior distributions $p(\varphi_1^T)$ and $p(\Omega_m)$. The analytical form of $q^*(.|Z_1^T)$ can be assessed via the following theorem \citep{beal_variational_2003}.

\begin{theorem}
For models with observed variables $Z_1^T$, hidden variables $\varphi_1^T$ and 
parameters $\Omega_1^b$, the mean field variational Bayesian approximation has the following characteristic: 
if the complete-data likelihood $p(\varphi_1^T,Z_1^T | \Omega_1^b)$ is part of the exponential family (in its ``natural form'', meaning parametrized by its natural parameter), and if the hidden and parameter prior distributions $p(\varphi_1^T)$ and $p(\Omega_1^b)$ are conjugate to this complete-data likelihood, then the corresponding variational approximate posterior distributions that maximize $\text{F}$, $q^*(\varphi_1^T|Z_1^T)$ and $q^*(\Omega_1^b|Z_1^T)$, are of the same distributional form as the prior distributions $p(\varphi_1^T)$ and $p(\Omega_1^b)$ respectively.
\label{Thm conj exponential}
\end{theorem}
As stated in \citet[p.~160]{beal_variational_2003}, with suitable priors, the state space model with unknown parameters is in the conjugate-exponential family, but the ``natural form'' pa\-ra\-me\-tri\-za\-tion required by Theorem \ref{Thm conj exponential} presents a parameter-to-natural parameter mapping that is non-invertible. 
We can however use the fact that all the nodes of the model
defined in Section 2 are conditionally conjugate, implying that the
optimal variational approximate posterior distributions
$q^*(\varphi_1^T | Z_1^T)$ and $q^*(\Omega_1^b|Z_1^T)$ will be of the same distributional form as, respectively, the prior distributions $p(\varphi_1^T)$ and $p(\Omega_1^b)$ (a node is said to be conditionally conjugate when its conditional distribution given its Markov blanket (see Definition \ref{MB}) is in the same family as its conditional distribution given its parents). 

Theorem \ref{Thm conj exponential} moreover ensures that the analytical form of the variational distributions $q^*(\Omega_m|Z_1^T)^{(l)}$ and $q^*(\varphi_1^T|Z_1^T)^{(l)}$  does not change during iterations. Since this property does not hold for general equations \eqref{optimal q (x)2} and \eqref{optimal q Omega 2}, variational posterior distributions become quickly unmanageable outside the conjugate exponential framework. 
All proofs of Theorem \ref{Thm conj exponential} and related properties can be found in \citet{beal_variational_2003}.

\subsection{The variational evidence lower bound}
\label{The Variational Evidence Lower Bound}

In Bayesian analysis, the evidence (or marginal likelihood) provides a natural criterion for model selection by comparing the evidences obtained for the models to be compared. Well-known criteria based on evidence comparison are the Bayesian Information Criterion (BIC) and the Bayes Factor.

As observed in equation \eqref{VB_freeNRJ2}, the variational Bayesian methodology has the advantage of providing a quantity, the free energy, that is a lower bound for the marginal likelihood of the model and that can be computed efficiently. This leads to a natural criterion for model order selection, which is crucial to estimating our time-varying VAR model, where we have many more
variables to estimate than available observations. We can therefore perform model selection by comparing the free-energy quantities computed for each model order $m_p$ and select the model that exhibits the highest $\text{F}_{m_p}$. This latter comparison supposes that we have placed uniform priors over each model structure $m_p$, thereby considering them as equiprobable.

Recalling equations \eqref{VB_freeNRJ2}, \eqref{mean field approx} and letting $m_p$ be the model estimated for a specific order $p$, the free-energy quantity can be re-expressed as
\begin{equation}
\begin{aligned}
\text{F}_{m_p} & =\left<\log \frac{p(\varphi_{1}^T,Z_1^T,\Omega_1^b|m_p)}{q(\varphi_{1}^T,Z_1^T|m_p)}\right>_{q(\varphi_{1}^T,\Omega_1^b|m_p)}-\left<\log \frac{q(\Omega_1^b|Z_1^T,m_p)}{p(\Omega_1^b|m_p)}\right>_{q(\Omega_1^b|Z_1^T,m_p)}, 
\label{VB_freeNRJ4}
\end{aligned}
\end{equation}
where the first right-hand side term is the average log-likelihood over the entire set of parameters and  hidden states $[\Omega_1^b;\varphi_1^T]$ that therefore acts as an accuracy term, and the second right-hand side term is the Kullback--Leibler distance between the prior and the variational posterior distributions for the entire set of parameters $\Omega_1^b$. Since the Kullback--Leibler distance increases with the number of parameters, this second term acts as a penalty.
 
The choice of $\text{F}_{m_p}$ (instead of $ p(Z_1^T)$) as a criterion for model order selection implicitly assumes that the free energy lies at the same distance  to the evidence whatever the model order $p$. This seems reasonable, given that the dataset $Z_1^T$ is the same for all models. A similar procedure for model order selection can be found in \citet{corduneanu2001variational} and \citet{roberts2002variational}.

It has been shown that in the large sample limit, the free energy becomes equivalent to
the Bayesian information criterion (BIC) \citep{Attias:1999:IPS:2073796.2073799}. This popular model order selection criterion can therefore be seen as a limiting case of the
variational Bayesian framework \citep{beal_variational_2003,penny2012comparing}. 

The specific analytic form of $\text{F}_{m_p}$ for our model is derived in Appendix \ref{Appendix F}.

\section{Computation of the Free Energy}
\label{Appendix F}

It is straightforward to re-express the free-energy quantity as
\begin{equation}
\begin{aligned}
\begin{split}
\mathbf{F}
=&-\left<\log \frac{q(\Omega_1^b|Z_1^T)}{p(\Omega_1^b)}\right>_{q(\Omega_1^b|Z_1^T)} - \left<\log q(\varphi_{1}^T|Z_1^T)\right>_{q(\varphi_{1}^T|Z_1^T)} \\
&+
\left<\log p(\varphi_{1}^T,Z_1^T|\Omega_1^b,m_p)\right>_{q(\varphi_{1}^T,\Omega_1^b|Z_1^T)},  \\
\end{split}
\end{aligned}
\label{AppF1}
\end{equation}
where we omit the conditional dependence to the model $m_p$ for notational simplicity.
The first term on the r.h.s. of equation \eqref{AppF1} represents the Kullback--Leibler divergence between the prior and the variational posterior for the entire set of parameters $\Omega_1^b$, the second r.h.s. term represents the entropy of the variational posterior distribution of the hidden variables $\varphi_1^T$ and the third r.h.s. term is the average log-likelihood of the data and hidden state parameters taken over the entire set of parameters $\Omega_1^b$ and hidden state $\varphi_1^T$.

Recalling the mean-field factorization assumed in our model as well as the priors distributions set in Section 4, the free-energy quantity takes the explicit form:
\begin{equation}
\begin{aligned}
\begin{split}
\mathbf{F}=&-\left<\log\frac{q(A)}{p(A|\mathbf{\alpha)}}\right>_{q(A)}
-\left<\log\frac{q(\mathbf{\alpha}|\mathbf{\delta})}{p(\mathbf{\alpha}|\mathbf{\delta}}\right>_{q(\mathbf{\alpha}|\mathbf{\delta})}
-\left<\log \frac{q(\mathbf{\delta})}{p(\mathbf{\delta})}\right>_{q(\mathbf{\delta})}\\
&-\left<\log\frac{q(Q|a_q)}{p(Q|a_q)}\right>_{q(Q|a_q)} 
-\left<\log\frac{q(a_q)}{p(a_q)}\right>_{q(a_q)}\\
&\underbrace{- \left<\log \frac{q(R|\{a_{ri}\}_{i=1}^d)}{p(R|\{a_{ri}\}_{i=1}^d)}\right>_{q(R|\{a_{ri}\}_{i=1}^d)}
-\left<\log \frac{q(\{a_{ri}\}_{i=1}^d)}{p(\{a_{ri}\}_{i=1}^d)}\right>_{q(\{a_{ri}\}_{i=1}^d)}}_{\text{term 1}}\\
&\underbrace{-\left<\log q(\varphi_{1}^T)\right>_{q(\varphi_{1}^T)}
}_{\text{term 2}}
\underbrace{+\left<\log p(\varphi_1^T,Z_1^T|\Omega_1^b)\right>_{q(\varphi_{1}^T)q(\Omega_1^b)}
}_{\text{term 3}},
\end{split}
\end{aligned}
\label{F2}
\end{equation}
where the conditional dependance of the variational posterior distributions $q(.)$ to the data $Z_t$ is now omitted for notational simplicity.

The KL divergences of term 1 have closed form for conjugate exponential distributions and are therefore straightforward to obtain despite their intensive computation. Let us focus on the entropy term that appears in term 2. Following \cite{beal_variational_2003}, this can be rewriten as
\begin{equation}
\begin{aligned}
\text{term 2}&=-\left<\log q(\varphi_{1}^T)\right>_{q(\varphi_{1}^T)}
=-\left<  - \log \Upsilon+\underbrace{\left<\log p(\varphi_{1}^T,Z_{1}^T|\Omega_1^b)\right>_{q(\Omega_1^b)}}_{\text{term 3}} \right>_{q(\varphi_{1}^T)}.
\label{Phi_Entropy}
\end{aligned}
\end{equation} 
Term 3 therefore disappears in equation \eqref{F2} and the quantity $\Upsilon$ becomes
\begin{equation}
\begin{aligned}
\Upsilon&=\left<   \exp  \left< \log p(\varphi_{1}^T,Z_{1}^T|\Omega_1^b)\right>_{q(\Omega_1^b)} \right>_{q(\varphi_{1}^T)}\\
&= \underbrace{\left<   \exp  \left< \log p(\varphi_{1}^T|Z_{1}^T,\Omega_1^b)\right>_{q(\Omega_1^b)} \right>_{q(\varphi_{1}^T)}}_{=1} + \left<\exp \left< \log p(Z_{1}^T|\Omega_1^b)\right>_{q(\Omega_1^b)}  \right>_{q(\varphi_{1}^T)}.
\label{Z1}
\end{aligned}
\end{equation} 
Assuming now that the parameters $\Omega_1^b$ have a point mass density rather than their variational posterior distribution $q(.)$, the $\Upsilon$ quantity becomes
\begin{equation}
\begin{aligned}
\Upsilon&=\left<\exp \left< \log p(Z_{1}^T|\Omega_1^b)\right>_{q(\Omega_1^b)}  \right>_{q(\varphi_{1}^T)}
=p(Z_{1}^T|\bar{R})=p(Z_1) \prod_{t=2}^{T} p(Z_t|Z_1^{t-1}).
\label{Z2}
\end{aligned}
\end{equation} 
The quantity $\Upsilon$ can be obtained just after the forward recursion step and therefore amounts to 
\begin{equation}
\begin{aligned}
\Upsilon&= p(Z_1) \prod_{t=2}^{T} p(Z_t|Z_1^{t-1})= \prod_{t=1}^{T} \mathcal{N}_d(C_t\mu_t^{t-1},C_t\Sigma_t^{t-1}C_t^{'}+R),
\label{Z4}
\end{aligned}
\end{equation} 
where the quantities $\mu_t^{t-1}$ and $\Sigma_t^{t-1}$ are defined in Appendix D \citep{sarkka2013bayesian}.

\section{Mean and Fluctuation Theorem}
\label{Mean and Fluctuation Theorem}
The \textit{Mean and Fluctuation Theorem} is a decomposition proved by  \cite{chiappa2007unified}. Using our notation, based on the model and the Conditional distributions defined in Section 2 and the 
 set of unknown parameters $\Omega_1^b$, the following decomposition of the density $\left< \log p(\varphi_1^T,Z_1^T|\Omega_1^b)\right> _{q(\Omega_1^b|Z_1^T)}$ holds
\begin{equation}
\begin{aligned}
\begin{split}
\left< \log p(\varphi_1^T,Z_1^T|\Omega_1^b)\right> _{q(\Omega_1^b|Z_1^T)}= &\log p(\varphi_1^T,Z_1^T|\left<\Omega_1^b\right>)+F_{A,Q}+F_{C_t,R}, \quad \text{where}
\\
\log p(\varphi_1^T,Z_1^T|\left<\Omega_1^b\right>)\propto&\frac{1}{2}(\varphi_1-\mu_1)\Sigma_1^{-1}(\varphi_1-\mu_1)^{'}\\
&-\frac{1}{2}\sum_{t=2}^{T}(\varphi_t-\left<A\right>\varphi_{t-1})Q^{-1}(\varphi_t-\left<A\right>\varphi_{t-1})^{'}
\\
&-\frac{1}{2}\sum\limits_{t=1}^T (Z_t-C_t\varphi_t) R^{-1} (Z_t-C_t\varphi_t)^{'}, 
\\
F_{A,Q}= &-\frac{1}{2}\sum_{t=1}^{T-1}\varphi_t^{'}\big(\left<AQ^{-1}A\right>-\left<A\right>\left<Q\right>^{-1}\left<A\right>\big)\varphi_t,\\
F_{C_t,R}= &-\frac{1}{2}\sum_{t=1}^{T}\varphi_t^{'}\big(\left<C_tR^{-1}C_t\right>-\left<C_t\right>\left<R\right>^{-1}\left<C_t\right>\big)\varphi_t.
\end{split}
\end{aligned}
\end{equation} 
\label{MFTheorem}
Note that the formulation in \cite{chiappa2007unified} considers the matrix $C_t$ as time-invariant.

\section{Unified Inference Theorem}
\label{Unified inference Theorem}

For the BSS model, recall equation \eqref{optimal q (x)2}, which is a key quantity we want to evaluate
\begin{equation*}
q^*(\varphi_1^T)^{(t+1)} \propto \exp \left< \log p(\varphi_1^T,\Omega_1^b,Z_1^T)\right>_{q(\Omega_1^b)}^{(t)}\propto \exp \left< \log p(\varphi_1^T|\Omega_1^b,Z_1^T)\right>_{q(\Omega_1^b)}^{(t)}.
\end{equation*} 
In the situation where $\Omega_1^b$ are random parameters rather than fixed values, \citet{chiappa2007unified} propose an elegant solution based on a suitably augmented system of equations that allows to infer $q(\varphi_1^T)$ through classical state space model inference algorithms. The theorem states that the above density can be written as
\begin{equation}
\exp \left< \log p(\varphi_1^T|\Omega_1^b,Z_1^T)\right>_{q(\Omega_1^b)} = 
p(\varphi_1^T|\widetilde{\Omega_1^b},\widetilde{Z}_1^T),
\label{UIT2}
\end{equation} 
where the augmented elements are
\begin{equation*}
\begin{aligned}
\begin{split}
\widetilde{\Omega}_1^b=\{\widetilde{A};\widetilde{Q};\widetilde{R}\}, \qquad
\widetilde{A}=\left<A\right>,\qquad
\widetilde{Q}=\left<Q\right>,\qquad
\widetilde{R}=\begin{pmatrix}\left<R\right> &0 & 0\\0 & I_k & 0 \\0 & 0 &I_d \end{pmatrix},\\
\widetilde{Z}_t=\begin{pmatrix}Z_t\\0_k\\0_d\end{pmatrix},\qquad
\widetilde{C}_t=\begin{pmatrix}C_t\\U_A\\U_{C_t}\end{pmatrix}, \qquad
\end{split}
\end{aligned}
\end{equation*} 
and where $U_A$ and $U_{C_t}$ are defined as the Cholesky decompositions of
\begin{equation*}
\begin{aligned}
\begin{split}
U_A^{'}U_A &= \left<AQ^{-1}A\right>-\left<A\right>\left<Q\right>^{-1}\left<A\right>,
\\
U_{C_t}^{'}U_{C_t} &= \left<C_tR^{-1}C_t\right>-\left<C_t\right>\left<R\right>^{-1}\left<C_t\right>.
\end{split}
\end{aligned}
\label{UIT4}
\end{equation*} 
In our specific case, we have that $U_{C_t}^{'}U_{C_t}=0$, due to the non-randomness of the matrix $C_t$, and so the unique quantity to define is $U_A^{'}U_A$. In the case where $A$ and $Q$ are diagonal as defined in Section 3, we have that
\begin{equation}
\begin{aligned}
\begin{split}
U_A^{'}U_A =& \left<AQ^{-1}A\right>-\left<A\right>\left<Q\right>^{-1}\left<A\right>\\
= & \begin{pmatrix}\left<A_{1,1}q_1^{-1}A_{1,1}\right> &0  & 0 \\ \vdots  & \ddots  &\vdots  \\ 0  & 0& \left<A_{k,k}q_k^{-1}A_{k,k}\right> \end{pmatrix}\\
&-
\begin{pmatrix}\left<A_{1,1}\right>\left<q_1^{-1}\right>\left<A_{1,1}\right> &0  & 0 \\ \vdots  & \ddots  &\vdots  \\ 0  & 0& \left<A_{k,k}\right>\left<q_k^{-1}\right>\left<A_{k,k}\right>\end{pmatrix},\\
= &\begin{pmatrix}\left<A_{1,1}^{2}\right>\left<q_1^{-1}\right> &0  & 0 \\ \vdots  & \ddots  &\vdots  \\ 0  & 0& \left<A_{k,k}^{2}\right>\left<q_k^{-1}\right> \end{pmatrix}-
\begin{pmatrix}\left<A_{1,1}\right>^{2}\left<q_1^{-1}\right> &0  & 0 \\ \vdots  & \ddots  &\vdots  \\ 0  & 0& \left<A_{k,k}\right>^{2}\left<q_k^{-1}\right> \end{pmatrix},\\
= &\begin{pmatrix}\sigma_{A_{1,1}}^{2}\left<q_1^{-1}\right>&0  & 0 \\ \vdots  & \ddots  &\vdots  \\ 0  & 0& \sigma_{A_{k,k}}^{2}\left<q_k^{-1}\right>\end{pmatrix},
\end{split}
\end{aligned}
\label{UIT5}
\end{equation}
where $\sigma_{A_{i,i}}^{2}$ is the variational posterior variance of the $i$-th
entry of the $A$ matrix defined in Section \ref{A diagonal} and $\left<q_i^{-1}\right>$ is the variational posterior mean of the $i$-th entry of the inverse variance-covariance matrix $Q^{-1}$ defined in Section \ref{Q diagonal with 1 element} that can be straightforwardly obtained due to the properties of the inverse-gamma and gamma distributions. 

For our model, the augmented system of equations then gives
\begin{equation}
\begin{aligned}
\begin{split}
\widetilde{Z}_t=\begin{pmatrix}Z_t\\0_k\end{pmatrix},\qquad
\widetilde{A}=\left<A\right>,\qquad
\widetilde{Q}=\left<Q\right>,\qquad
\widetilde{C}_t=\begin{pmatrix}C_t\\U_A\end{pmatrix}, \qquad
\widetilde{R}=\begin{pmatrix}\left<R\right> &0 \\0 &I_k \end{pmatrix},
\end{split}
\end{aligned}
\label{UIT6}
\end{equation} 
where $U_A$ is defined in equation \eqref{UIT5}. 
Complete proofs can be found in \cite{chiappa2007unified} and \cite{ostwald2014tutorial}.

\section{E-step: Computation of the Distribution of $\varphi_1^T$}
\label{Appendix x}

As discussed in Section~\ref{Appendix VB}, variational Bayesian algorithms lead to EM-like iterative equations for the optimal variational densities $q^{*}(.|Z_1^T)$.
The variational update equation form for the hidden variables $\varphi_1^T$ was state in equation \eqref{optimal q (x)2}. The derivation of equation \eqref{optimal q (x)2} can be found in Appendix \ref{Appendix x}. Due to the conjugacy condition discussed in Section \ref{Appendix VB} and equation the Conditional distributions stated in Section 2, the variational posterior distribution $q^*(\varphi_1^T|Z_1^T)$ is multivariate Gaussian of dimension $k \times T$. As explained in \citet{beal_variational_2003} and \citet{cassidy_bayesian_2002}, if the parameters $\Omega_1^b$ were, as they called, point estimated, equation \eqref{optimal q (x)2} would be straightforwardly resolved with classical state space model tools.

However in our variational Bayesian scenario, parameters $\Omega_1^b$ are random variables governed by a specific variational distribution $q(.|Z_1^T)$, and so $\exp \left< \log p(\varphi_1^T|\Omega_1^b,Z_1^T)\right>_{q(\Omega_1^b|Z_1^T)}$ has to be computed for each variable with respect to its variational posterior distribution $q(\Omega_1^b|Z_1^T)$, sequentially for all parameters in the set $\Omega_1^b$. As described in \citet{beal_variational_2003} and \citet{chiappa2007unified} and fully explained in \citet{ostwald2014tutorial}, the simplification in equation \eqref{optimal q (x)2} no longer holds in this random parameter scenario because 
\begin{equation}
\exp \left< \log p(\varphi_1^T|\Omega_1^b,Z_1^T)\right>_{q(\Omega_1^b|Z_1^T)}\neq p(\varphi_1^T|\bar{\Omega}_1^b,Z_1^T).
\label{q (x)_3}
\end{equation}
The difference between the two terms in equation \eqref{q (x)_3} is evaluated in \citet{chiappa2007unified} and its decomposition is known as the mean and fluctuation theorem, reproduced in Appendix \ref{Mean and Fluctuation Theorem}. To nevertheless use standard state space model algorithms to solve $\exp \left< \log p(\varphi_1^T|\Omega_1^b,Z_1^T)\right>_{q(\Omega_1^b|Z_1^T)}$ in the variational Bayesian framework, and therefore to capitalize on the vast literature and results that exist on the topic, \citet{chiappa2007unified} prove the so-called unified inference theorem recalled in Appendix C. The idea is to reformulate $\exp \left< \log p(\varphi_1^T,\Omega_1^b,Z_1^T)\right>_{q(\Omega_1^b|Z_1^T)}$ as a standard density $ \widetilde{p}(\varphi_1^T,\widetilde{\Omega}_1^b,\widetilde{Z}_1^T)$ with known parameters $\widetilde{\Omega}_1^b$, by suitably augmenting the first equation in the system defined in Section 2. The classical KRTS smoother algorithms may then be suitably applied to the new density $ \widetilde{p}(\varphi_1^T,\widetilde{\Omega}_1^b,\widetilde{Z}_1^T)$, and allows us to derive the target quantity $ \widetilde{p}(\varphi_1^T|\widetilde{\Omega}_1^b,\widetilde{Z}_1^T)$. This finally corresponds to
$\exp \left< \log p(\varphi_1^T|\Omega_1^b,Z_1^T)\right>_{q(\Omega_1^b|Z_1^T)}$, and by extension to $q^*(\varphi_1^T|Z_1^T)^{(t+1)}$ by equation \eqref{optimal q (x)2}.

The optimal variational posterior distribution for the hidden state sequence $\varphi_1^T$ is therefore multivariate Gaussian at each time $t$:
\begin{equation}
q^{*}(\varphi_t|Z_1^T) =  \mathcal{N}_k(\varphi_t|\mu_{t},\Sigma_{t}),
\label{POST_H}
\end{equation} 
where the sufficient statistics $\{\mu_{t};\Sigma_{t}\}_{t=1}^T$, as well as the cross-moments $\{\mu_{t}\mu_{t-1};\Sigma_{t,t-1}\}_{t=2}^T$, are obtained through the KRTS smoother recursive equations applied to the suitable augmented system of equations discussed above.

\paragraph{Forward recursions.} 
This step implements the recursive equation for $q(\varphi_t|Z_1^t)^{(l+1)}$. Let $\mu_t^t = \mathrm{E}_{q}\{\varphi_t|Z_1^t\}$ , $\Sigma_t^t = \mathrm{VAR}_{q}\{\varphi_t|Z_1^t\}$, $\mu_t^{t-1} = \mathrm{E}_{q}\{\varphi_t|Z_1^{t-1}\}$ , $\Sigma_t^{t-1} = \mathrm{VAR}_{q}\{\varphi_t|Z_1^{t-1}\}$ and $\left<.\right>$ denotes the expectation with respect to the suitable variational distribution $q(.|Z_1^T)^{(l)}$. These quantities are obtained recursively as
\begin{equation}
\begin{aligned}
\mu_t^{t-1}  &= \left<A\right> \mu_{t-1}^{t-1},\\
\Sigma_t^{t-1}&=\left<A\right>\Sigma_{t-1}^{t-1}\left<A\right>^T+\left<Q\right>,\\
\mu_t^{t} &= \mu_t^{t-1}+K_t(Z_t-C_t\mu_t^{t-1}),\\
\Sigma_t^{t}&=\Sigma_t^{t-1}-K_tC_t\Sigma_t^{t-1},
\end{aligned}
\end{equation}
where the Kalman gain $K_t$ is given by
\begin{equation}
\begin{aligned}
K_t&=\Sigma_t^{t-1}C_t^{'}(C_t\Sigma_t^{t-1}C_t^{'}+\left<R\right>)^{-1}.
\end{aligned}
\label{Kalman_GAIN}
\end{equation}

\paragraph{Backward recursions.} 
The backward recursions implement the recursive equations for $q(\varphi_t|Z_1^T)^{(l+1)}$. Let now $\mu_t^T = \mathrm{E}_{q}\{\varphi_t|Z_1^T\}$ and $\Sigma_t^T = \mathrm{VAR}_{q}\{\varphi_t|Z_1^T\}$. These quantities are obtained recursively as
\begin{equation}
\begin{aligned}
J_t&=\Sigma_t^t\left<A\right>^{'}{(\Sigma_{t+1}^t)}^{-1},\\
\mu_t^T &=\mu_t^t+J_t(\mu_{t+1}^T -\left<A\right>\mu_t^t),\\
\Sigma_t^T&=\Sigma_t^t+J_t(\Sigma_t^t-\Sigma_{t+1}^t)J_t{'}.\\
\end{aligned}
\end{equation}
We also define the cross-time quantities:
\begin{equation}
\begin{aligned}
M_1&=\sum\limits_{t=2}^T \Sigma_t^T+\mu_t^T{\mu_t^T}^{'},\\
M_2&=\sum\limits_{t=1}^{T-1} \Sigma_t^T+\mu_t^T{\mu_t^T}^{'},\\
M_3 &=\sum\limits_{t=2}^T \Sigma_{t,t-1}^T+\mu_t^T{\mu_{t-1}^T}^{'}, \qquad \text{where}\\
\Sigma_{t,t-1}^T&=\Sigma_{t}^tJ_{t-1}^T+(\Sigma_{t+1,t}^T-\left<A\right>\Sigma_{t}^t)J_{t-1}^T.
\end{aligned}
\label{M_123}
\end{equation}
This last element represents $\text{cov}_q(\varphi_t,\varphi_{t-1}|Z_1^T)$.

\section{M-step: Computation of the Distribution of $\delta$}
\label{Appendix delta}

We will derive here the update equation for $\mathbf{\delta}$ in the case where $\{\alpha_i\}_{i=1}^k$ is different for each $\{A_{i,i}\}_{i=1}^k$ diagonal entry of $A$. Extension to the situation where $\delta$ is a unique parameter related to a unique variance parameter $\alpha$ is straightforward. 

Recalling the prior form for $\mathbf{\delta}$ defined in Section 3 as well as equation \eqref{optimal q Omega 2}, the optimal form for the variational posterior $q^*(\mathbf{\delta}|Z_1^T)$ becomes:
\begin{equation}
\begin{split}
\begin{aligned}
\log  q^*(\mathbf{\delta}|Z_1^T)^{(l+1)}=\sum_{i=1}^k \log  q^*(\delta_i|Z_1^T)^{(l+1)} = \underbrace{ \sum_{i=1}^k \log  p(\delta_i)}_{\text{term 1}} + \underbrace{\sum_{i=1}^k\left<\log  p(\alpha_i|\delta_i)\right>_{q(\alpha_i)^{l}}}_{\text{term 2}}+\text{ct},
\end{aligned}
\end{split}
\label{UPDATE_delta1}
\end{equation}
where throughout this Supplementary Material, ct will denote the normalization constant for the given density or log-density.
Recalling equation $p(\alpha_i|\delta_i)$, we have that
\begin{equation}
\begin{split}
\begin{aligned}
\text{term 1}  & = \sum_{i=1}^k (-\kappa_{p_{i}}-1)\log\delta_i - \delta_i^{-1}\beta_{p_i} +\text{ct},
\end{aligned}
\end{split}
\label{UPDATE_delta2}
\end{equation}
and 
\begin{equation}
\begin{split}
\begin{aligned}
\text{term 2}  & = \sum_{i=1}^k -c_{p_{i}} \log \,\delta_i-\delta_i^{-1}\left<\alpha_i^{-1}\right>+\text{ct}.
\end{aligned}
\end{split}
\label{UPDATE_delta3}
\end{equation}
The variational posterior for each element $\{\delta_i\}_{i=1}^k$ is therefore an inverse-gamma distribution with shape and scale parameters $\{\kappa_i;\beta_i\} $ defined as:
\begin{equation}
\begin{split}
\begin{aligned}
\kappa_i&=\kappa_{p_{i}}+c_{p_{i}},\\
\beta_i&=\beta_{p_i}+\left<\alpha_i^{-1}\right>.
\end{aligned}
\end{split}
\label{UPDATE_delta4}
\end{equation}

\section{M-step: Computation of the Distribution of $\alpha$}
\label{Alpha different for each A element}

We will derive here the update equation for $\mathbf{\alpha}$ in the situation where $A$ is diagonal. Extension to the situation where $A$ is full is straightforward. 

Recalling the prior form for each $\{\alpha_i\}_{i=1}^k$ defined in Section 3 as well as equation \eqref{optimal q Omega 2}, the optimal form for the variational posterior $q^*(\alpha_i|\delta_i,Z_1^T)$ becomes:
\begin{equation}
\begin{split}
\begin{aligned}
\log  q^*(\alpha|\delta,Z_1^T)^{(l+1)}=&\sum_{i=1}^k \log  q^*(\alpha_i|\delta_i,Z_1^T)^{(l+1)} \\ = &\underbrace{\sum_{i=1}^k \left< \log  p(\alpha_i|\delta_i)\right>_{q(\delta_i)^{(l)}}}_{\text{term 1}} + \underbrace{\sum_{i=1}^k \left<\log  p(a_i|\alpha_i)\right>_{q(a_i)^{(l)}}}_{\text{term 2}}+\text{ct},
\end{aligned}
\end{split}
\label{UPDATE_alpha1}
\end{equation}
where all terms not depending on $\mathbf{\alpha}$ are put in the constant term ct. Let us consider the two terms separately. Under the prior specified for $p(\alpha_i|\delta_i)$, term 1 simply equals
\begin{equation}
\begin{split}
\begin{aligned}
\text{term 1}  =\sum_{i=1}^k (-c_{p_{i}}-1)\log\alpha_i - \alpha_i^{-1}\left<\delta_{i}^{-1}\right>+\text{ct}.
\end{aligned}
\end{split}
\label{UPDATE_alpha_term1}
\end{equation}
Recalling equation for $p(\alpha_i|\delta_i)$, term 2 can be rewritten as
\begin{equation}
\begin{split}
\begin{aligned}
\text{term 2}  & = -\sum\limits_{i=1}^k\frac{1}{2}\log\alpha_i  - \frac{\alpha_i^{-1}}{2} \sum\limits_{i=1}^k \left<(A_{ii}-m_{A_{ii}})^{2}\right> +\text{ct}\\
&=-\sum\limits_{i=1}^k\frac{1}{2}\log\alpha_i-\frac{\alpha_i^{-1} }{2} \sum\limits_{i=1}^k (\left<A_{ii}^2\right>+m_{A_{ii}}^2-2m_{A_{ii}}\left<A_{ii}\right>)^{2}+\text{ct}\\
&=-\sum\limits_{i=1}^k\frac{1}{2}\log\alpha_i-\frac{\alpha_i^{-1} }{2} \sum\limits_{i=1}^k (\theta_i+(\psi_i\theta_i)^2+m_{A_{ii}}^2-2m_{A_{ii}}\psi_i\theta_i)+\text{ct}.
\end{aligned}
\end{split}
\label{UPDATE_alpha_term2}
\end{equation}
The whole expression for $q^*(\mathbf{\alpha}|\mathbf{\delta},Z_1^T)^{(l+1)}$ then becomes 
\begin{equation}
\begin{split}
\begin{aligned}
q^*(\mathbf{\alpha}|\mathbf{\delta},Z_1^T)^{(l+1)} =&\sum_{i=1}^k \log  q^*(\alpha_i|\delta_i,Z_1^T)  \\
 =& \sum_{i=1}^k (-c_{p_{i}}-1)\log\alpha_i - \alpha_i^{-1}\left<\delta_{i}^{-1}\right>
-\frac{1}{2}\log\alpha_i\\
&-\alpha_i^{-1}  \frac{(\theta_i+(\psi_i\theta_i)^2+m_{A_{ii}}^2-2m_{A_{ii}}\psi_i\theta_i)}{2}\big]+\text{ct},
\end{aligned}
\end{split}
\label{UPDATE_alpha_full}
\end{equation}
and so the variational posterior for each element $\{\alpha_i\}_{i=1}^k$ is an inverse-gamma distribution with respectively $\{c_i;b_i\} $ shape and scale parameters defined as:
\begin{equation}
\begin{split}
\begin{aligned}
c_i&=c_{p_{i}}+\frac{1}{2},\\
b_i&=\left<\delta_{i}^{-1}\right>+\frac{(\theta_i+(\psi_i\theta_i)^2+m_{A_{ii}}^2-2m_{A_{ii}}\psi_i\theta_i)}{2}.
\end{aligned}
\end{split}
\label{UPDATE_alpha_PARAMETERS}
\end{equation}

\section{M-step: Computation of the Distribution of A}
\label{A diagonal}

The optimal form for the variational posterior $q^*(A|\alpha,\delta,Z_1^T)$ is:
\begin{equation}
\begin{split}
\begin{aligned}
\log  q^*(A|\alpha,\delta,Z_1^T)^{(l+1)} =  & \underbrace{\left<\log  p(A|\alpha)\right>_{q(\varphi_1^T)^{(l)},q(Q)^{(l)},q(\alpha)^{(l)}}}_{\text{term 1}} 
\\
 &+\underbrace{\sum\limits_{t=2}^T \left<\log  p(\varphi_t|\varphi_{t-1},A,Q)\right>_{q(\varphi_1^T)^{(l)},q(Q)^{(l)},q(\alpha)^{(l)}}}_{\text{term 2}}+ \text{ct},
\end{aligned}
\end{split}
\label{UPDATE_Aa}
\end{equation}
where all terms not depending on $A$ are stacked in the constant term $\text{ct}$ and the conditional dependence on the data $Z_1^T$ is dropped for sake of brevity. Let us consider the two terms separately.
The development of term 2 is the same whatever the form for $A$ (full, diagonal or proportional to identity). It becomes
\begin{equation}
\begin{split}
\begin{aligned}
\text{term 2} & =  \sum\limits_{t=2}^T \left<\log  p(\varphi_t|\varphi_{t-1},Q,\alpha)\right>_{{q(\varphi_1^T)^{(l)},q(Q)^{(l)}}} +\text{ct}\\
&=\sum\limits_{t=2}^T\left<\underbrace{\frac{1}{2}\log |Q|}_{\longrightarrow \text{ct}}-\frac{1}{2}\mathrm{Tr}[(\varphi_t-A\varphi_{t-1})(\varphi_t-A\varphi_{t-1})^{'}Q]\right>_{q(\varphi_1^T)^{(l)},q(Q)^{(l)}}+\text{ct}\\
&=\sum\limits_{t=2}^T \left< -\frac{1}{2}\mathrm{Tr}\big[(\varphi_t\varphi_t^{'}-2A\varphi_{t-1}\varphi_{t}^{'}+A\varphi_{t-1}\varphi_{t-1}^{'}A^{'}) Q\big]\right>_{q(\varphi_1^T)^{(l)},q(Q)^{(l)}}+\text{ct}\\
&=-\frac{1}{2}\mathrm{Tr}\left<\big[\underbrace{\sum\limits_{t=2}^T \left<\varphi_t\varphi_t^{'}\right>}_{\longrightarrow \text{ct}} -2A  \underbrace{\sum\limits_{t=2}^T\left<\varphi_{t-1}\varphi_{t}^{'}\right>}_{= M_3 \eqref{M_123}} + A \underbrace{\sum\limits_{t=2}^T\left<\varphi_{t-1}\varphi_{t-1}^{'}\right>}_{=M_2 \eqref{M_123}} A^{'}) Q\big]\right>_{q(Q)^{(l)}}+\text{ct}.
\end{aligned}
\end{split}
\label{UpdateA_term2}
\end{equation}
We can now rewrite \eqref{UpdateA_term2} in term 2 of equation \eqref{UPDATE_Aa}:
\begin{equation}
\begin{split}
\begin{aligned}
\text{term 2} & = -\frac{1}{2} \left<\mathrm{Tr}\big[-2A M_3Q + A M_2 A^{'} Q\big]\right>_{q(Q)^{(l)}}+\text{ct}\\
& = -\frac{1}{2}\mathrm{Tr}\big[-2A M_3 \left<Q\right> + A M_2 A^{'} \left<Q\right>\big] + \text{ct}.
\end{aligned}
\end{split}
\label{UpdateA_term2b}
\end{equation}

If we suppose the matrix $A$ diagonal, as implied by the prior form $p(A|\mathbf{\alpha})$ as well as equation \eqref{optimal q Omega 2}, term 1 in equation \eqref{UPDATE_Aa} is straightforward:
\begin{equation}
\begin{split}
\begin{aligned}
\text{term 1} & = \sum\limits_{i=1}^k \left<\log  p(A_{ii}|\alpha_i)\right>_{q(\alpha_i)^{(l)}}
& =  - \frac{\left<\alpha_i^{-1}\right> }{2} \sum\limits_{i=1}^k (A_{ii}-m_{A_{ii}})^{2} + \text{ct}.
\end{aligned}
\end{split}
\label{UpdateA_term1}
\end{equation}
The full expression then gives:
\begin{equation}
\begin{split}
\begin{aligned}
\log  q^*(A|\alpha,\delta,Z_1^T)^{(l+1)} =&   - \frac{1}{2}\big[ \sum\limits_{i=1}^k \left<\alpha_i^{-1}\right>(A_{ii}-m_{A_{ii}})^{2} \\
&+\mathrm{Tr}\{-A M_3 \left<Q\right> - M_3 A^{'}\left<Q\right> + A M_2 A^{'} \left<Q\right>\}\big] + \text{ct}\\
= &- \frac{1}{2} \sum\limits_{i=1}^k \big[\left<\alpha_i^{-1}\right>(A_{ii}^2+m_{A_{ii}}^2-2A_{ii}m_{A_{ii}}) -2A_{ii} M_{3_{i,i}} \left<Q_{ii}\right> \\
&+ A_{ii} M_{2_{i,i}} A_{ii} \left<Q_{ii}\right>\big] + \text{ct}\\
= &- \frac{1}{2} \sum\limits_{i=1}^k \big[-2A_{ii}\underbrace{(\left<\alpha_i^{-1}\right> m_{A_{ii}}+M_{3_{i,i}}\left<q_i\right>)}_{:=\psi_i} \\
&+ A_{ii}^2\underbrace{(\left<Q_{ii}\right>M_{2_{i,i}}+\left<\alpha_i^{-1}\right>)}_{:={\theta_i}^{-1}}+
 \underbrace{\left<\alpha_i^{-1}\right> m_{A_{ii}}^2}_{\longrightarrow\text{ct}}] + \text{ct}\\
=& - \frac{1}{2} \sum\limits_{i=1}^k \big[-2A_{ii}\psi_i{\theta_i}{\theta_i}^{-1} + A_{ii}^2{\theta_i}^{-1}\big] + \text{ct}.
\end{aligned}
\end{split}
\label{UPDATE_Ab}
\end{equation}
Therefore
\begin{equation}
\begin{split}
\begin{aligned}
\log  q^*(A|\alpha,\delta,Z_1^T)&=\sum\limits_{i=1}^{k} \log  q^*(A_{ii}|\alpha,\delta,Z_1^T),
\\
 \text{with} \qquad & q^*(A_{ii}|Z_1^T) =\mathcal{N}_1(A_{ii}|\psi_i{\theta_i},{\theta_i}), \\ 
\text{where} \qquad &\theta_i=(\left<Q_{ii}\right>M_{2_{i,i}}+\left<\alpha_i^{-1}\right>)^{-1} \qquad \text{and} \\
\qquad &\psi_i=\left<\alpha_i^{-1}\right> m_{a}+M_{3_{i,i}}\left<Q_{ii}\right>.
\end{aligned}
\end{split}
\label{UPDATE_Ac}
\end{equation}

\section{M-step: Computation of the Distribution of $a_q$}
\label{Appendix aq}

We will derive here the update equation for $a_q$ in the situation where the parameter $Q_{ii}$ is unique. Extension to the situation where $\{a_{q_i}\}_{i=1}^k$ is related to different $\{Q_{ii}\}_{i=1}^k$ is straightforward. 
Recalling the prior form for $a_q$ in Section 2 as well as equation \eqref{optimal q Omega 2}, the optimal form for the variational posterior $q^*(a_q|Z_1^T)$ becomes:
\begin{equation}
\begin{split}
\begin{aligned}
\log  q^*(a_q|Z_1^T)^{(l+1)}= \underbrace{\log  p(a_q)}_{\text{term 1}} + \underbrace{\left<\log  p(Q_{ii}|a_q)\right>_{q(Q_{ii})^{(l+1)}}}_{\text{term 2}}+\text{ct}.
\end{aligned}
\end{split}
\label{UPDATE_aq1}
\end{equation}
Recalling the equation for $p(\alpha_i|\delta_i)$, we have that
\begin{equation}
\begin{split}
\begin{aligned}
\text{term 1}  & = (-a_{qp}-1)\log \,a_q - a_q^{-1} b_{qp} + \text{ct}
\end{aligned}
\end{split}
\label{UPDATE_aq2}
\end{equation}
and 
\begin{equation}
\begin{split}
\begin{aligned}
\text{term 2}  &=  -n_{p} \log a_q - a_q^{-1}\left<q^{-1}\right> + \text{ct}.
\end{aligned}
\end{split}
\label{UPDATE_aq3}
\end{equation}
The variational posterior $\log  q^*(a_q|Z_1^T)$ is therefore an inverse-gamma distribution with  shape and scale parameters $\{a_{qq};b_{qq}\}$
\begin{equation}
\begin{split}
\begin{aligned}
a_{qq}&=a_{qp}+n_{p},
\\
b_{qq}&=b_{qp}+\left<q^{-1}\right>.
\end{aligned}
\end{split}
\label{UPDATE_aq4}
\end{equation}

\section{M-step: Computation of the Distribution of $Q$}
\label{Q diagonal with 1 element}
If the matrix $Q$ is proportional to identity, i.e. with only one element $Q_{ii}$, the optimal variational form for the posterior $q^*(Q|a_q)$ becomes:
\begin{equation}
\begin{aligned}
\log  q^*(Q|a_q)^{(l+1)} =   \underbrace{\log p(Q_{ii}|a_q)}_{\text{term 1}} + \underbrace{\sum\limits_{t=2}^T \left<\log p(\varphi_t|A\varphi_{t-1})\right>_{q(A)^{(l)},q(\varphi_1^T)^{(l)}}}_{\text{term 2}}+\text{ct},
\end{aligned}
\label{UPDATE_QDD1a}
\end{equation}
where all terms not depending on $Q$ are included in the constant term. Recalling the equation of $p(Q_{ii}|a_q)$, we have that
\begin{equation}
\begin{split}
\begin{aligned}
\text{term 1}   =(-n_{p}-1)\log Q_{ii}  - Q_{ii}^{-1}\left<a_q^{-1}\right> + \text{ct},
\label{UpdateQDD1_term1}
\end{aligned}
\end{split}
\end{equation}
and
\begin{equation}
\begin{split}
\begin{aligned}
\text{term 2}  =& -\frac{T-1}{2} \log |Q|-\frac{1}{2} \sum\limits_{t=2}^T \left< \mathrm{Tr}\big[(\varphi_t-A\varphi_{t-1})(\varphi_t-A\varphi_{t-1})^{'}Q^{-1}\big]\right>_{q(\varphi_t,\varphi_{t-1})^{(l)},q(A)^{(l)}} \\& + \text{ct}
\\
 = &-\frac{T-1}{2} \log |Q| \\&-\frac{1}{2} \left< \mathrm{Tr}\big[\sum\limits_{t=2}^T\varphi_t\varphi_t^{'}Q^{-1}-2\sum\limits_{t=2}^T\varphi_t\varphi_{t-1}^{'}A^{'}Q^{-1}+A\sum\limits_{t=2}^T\varphi_{t-1}\varphi_{t-1}^{'}A^{'}Q^{-1}\big] \right>_{q(\varphi_t,\varphi_{t-1})^{(l)},q(A)^{(l)}} 
 \\&+ \text{ct}
 \\
 =& -\frac{T-1}{2} \log |Q|\\
 &-\frac{1}{2} \left< \mathrm{Tr}\big[\left<\sum\limits_{t=2}^T\varphi_t\varphi_t^{'}\right>Q^{-1}-2\left<\sum\limits_{t=2}^T\varphi_t\varphi_{t-1}^{'}\right>A^{'}Q^{-1}+A\left<\sum\limits_{t=2}^T\varphi_{t-1}\varphi_{t-1}^{'}\right>A^{'}Q^{-1}\big] \right>_{q(A)^{(l)}} 
 \\&+ \text{ct}\\
= & -k\frac{T-1}{2}\log q-\frac{1}{2} \sum\limits_{i=1}^k \big[M_{1_{i,i}}-2M_{3_{i,i}}\left<a_i\right>+M_{2_{i,i}}\left<a_i^2\right>\big]Q_{ii}^{-1} + \text{ct}
\\
= & -k\frac{T-1}{2}\log Q_{ii}-\frac{1}{2} \underbrace{\sum\limits_{i=1}^k \big[M_{1_{i,i}}-2M_{3_{i,i}}\psi_i\theta_i+M_{2_{i,i}}[(\psi_i\theta_i)^{2}+\theta_i^2]\big]}_{:=\Gamma}Q_{ii}^{-1} + \text{ct}.
\end{aligned}
\end{split}
\label{UpdateQDD1_term2}
\end{equation}
The whole expression for $q^*(Q_{ii}|a_q)$ can the be rewritten as
\begin{equation}
\begin{split}
\begin{aligned}
\log  q^*(Q_{ii}|a_q) = (-n_{p}-1)\log Q_{ii}  - Q_{ii}^{-1}\left<a_q^{-1}\right>
  -k\frac{T-1}{2}\log Q_{ii}- Q_{ii}^{-1}\frac{\Gamma}{2} + \text{ct}.\\
\end{aligned}
\end{split}
\label{UPDATE_QDD1b}
\end{equation}
The variational posterior distribution $ q^*(Q_{ii}|a_q)$  is therefore gamma with shape and scale parameters
\begin{equation}
\begin{aligned}
\begin{split}
n&=n_{p}+k\frac{T-1}{2},\\
d&=\left<a_q^{-1}\right>+\frac{\Gamma}{2},
\end{split}
\end{aligned}
\label{UPDATE_QDD1parameters}
\end{equation}
where $\Gamma= \sum\limits_{i=1}^k \big[M_{1_{i,i}}-2M_{3_{i,i}}\psi_i\theta_i+M_{2_{i,i}}[(\psi_i\theta_i)^{2}+\theta_i^2]\big]$  and $\{\theta_i;\psi_i\}$ are defined in  equation~\eqref{UPDATE_Ac}.

\section{M-step: Computation of the Distribution of $a_r$}
\label{Appendix ar}

We will derive here the update equation for the parameters $\{a_{r_{i}}\}_{i=1}^d$. 
Recalling the prior form for $\{a_{r_{i}}\}_{i=1}^d$ defined in Section 3 and equation \eqref{optimal q Omega 2}, the optimal form for the variational posterior $q^*(\{a_{r_{i}}\}_{i=1}^d)$ becomes:
\begin{equation}
\begin{split}
\begin{aligned}
\log  q^*(\{a_{r_{i}}\}_{i=1}^d)^{(l+1)}=  \underbrace{\log  p(\{a_{r_{i}}\}_{i=1}^d)}_{\text{term 1}} + \underbrace{\left<\log  p(R|\{a_{r_{i}}\}_{i=1}^d)\right>_{q(R)^{(l)}}}_{\text{term 2}}+\text{ct}.
\end{aligned}
\end{split}
\label{UPDATE_ar1}
\end{equation}
Recalling equation for $p(R|a_{r_1},\dots,a_{r_d})$, we have that
\begin{equation}
\begin{split}
\begin{aligned}
\text{term 1}  & = \sum_{i=1}^d (-a_{pr}-1)\log a_{r_{i}} - \sum_{i=1}^d a_{r_{i}}^{-1} b_{pr}+\text{ct},
\end{aligned}
\end{split}
\label{UPDATE_ar2}
\end{equation}
and 
\begin{equation}
\begin{split}
\begin{aligned}
\text{term 2}  &= \frac{r_p}{2} \log |B_p| -\frac{1}{2}\mathrm{Tr}[B_p \left<R^{-1}\right>]+\text{ct}, \qquad \text{where} \quad B_p=2\nu\, \text{diag}[\frac{1}{a_{r_1}}...\frac{1}{a_{r_d}}]. 
\end{aligned}
\end{split}
\label{UPDATE_ar3}
\end{equation}
We therefore have

\begin{equation}
\begin{split}
\begin{aligned}
\text{term 2}  &= \frac{r_p}{2} \log |2\nu\, \text{diag}[\frac{1}{a_{r_1}}...\frac{1}{a_{r_d}}]| -\frac{1}{2}\mathrm{Tr} \,[2\nu\, \text{diag}[\frac{1}{a_{r_1}}...\frac{1}{a_{r_d}}] \left<R^{-1}\right> \,]+\text{ct}\\
&= -\frac{r_p}{2} \sum_{i=1}^d \log a_{r_i}-\nu\sum_{i=1}^d  a_{r_i}^{-1}\left<R^{-1}_{\{i,i\}}\right>+\text{ct}.
\end{aligned}
\end{split}
\label{UPDATE_ar4}
\end{equation}
And then regrouping term 1 and term 2 yield to the inverse-gamma distribution for the variational posterior $\log  q^*(a_q|Z_1^T)$ with  shape and scale parameters  $\{a_{qr};b_{qr}\} $ defined as
\begin{equation}
\begin{split}
\begin{aligned}
a_{qr}&=a_{pr}+\frac{r_{p}}{2},
 \\
b_{qr}&=b_{pr}+\nu\left<R^{-1}_{\{i,i\}}\right>.
\end{aligned}
\end{split}
\label{UPDATE_ar5}
\end{equation}

\section{M-step: Computation of the Distribution of $R$}
\label{Appendix R}

Recalling the prior form for $R$, $p(R|a_{r_1},\dots,a_{r_d})$, and equation \eqref{optimal q Omega 2}, the optimal form for the variational posterior $q^*(R|\{a_{r_{i}}\}_{i=1}^d,Z_1^T)$ is
\begin{equation}
\begin{split}
\begin{aligned}
\log  q^*(R|\{a_{r_{i}}\}_{i=1}^d)^{(l+1)} =  & \underbrace{\left<\log p(R|\{a_{r_{i}}\}_{i=1}^d)\right>_{q(a_r)}}_{\text{term 1}} + \underbrace{\sum\limits_{t=1}^T \left<\log p(Z_t|C_t\varphi_{t})\right>_{q(\varphi_1^T)^{(l)}}}_{\text{term 2}}+\text{ct},
\end{aligned}
\end{split}
\label{UPDATE_Ra}
\end{equation}
where all terms not depending on $R$ are included in the constant term $\text{ct}$. Let us consider first the term 1. As the prior for $R$ defined in Section 3 is inverse-Wishart, term 1 becomes
\begin{equation}
\begin{split}
\begin{aligned}
\text{term 1}    = & -\frac{r_p+d+1}{2} \log |R| -\frac{1}{2}\rm {Tr}(\left<B_p\right> R^{-1})+\text{ct}\\
=&-\frac{r_p+d+1}{2} \log |R| -\frac{1}{2}\rm {Tr}(2\nu\, \text{diag}[\frac{1}{\left<a_{r_1}\right>}...\frac{1}{\left<a_{r_d}\right>}] R^{-1})+\text{ct},
\end{aligned}
\end{split}
\label{UpdateR_term1}
\end{equation}
Let us now consider term 2:
\begin{equation}
\begin{split}
\begin{aligned}
\text{term 2} & = \sum\limits_{t=1}^T \left< \log p(Z_t|C_t\varphi_t)\right>_{q(\varphi_1^T)^{(l)}} +\text{ct}\\
& = -\frac{T}{2}\log |R|-\frac{1}{2}\sum\limits_{t=1}^T\mathrm{Tr}\big[\left< (Z_t-C_t\varphi_t){(Z_t-C_t\varphi_t)}^{'}\right>R^{-1}\big]+\text{ct}\\
& =- \frac{T}{2}\log |R|-\frac{1}{2}\sum\limits_{t=1}^T \mathrm{Tr}\big[\left<(Z_tZ_t^{'}-2C_t\varphi_tZ_t^{'}+C_tZ_tZ_t^{'}C_t^{'})\right> R^{-1}\big]+\text{ct}\\
& = -\frac{T}{2}\log |R|-\frac{1}{2}\mathrm{Tr}\sum\limits_{t=1}^T [(Z_t-C_t\mu_t){(Z_t-C_t\mu_t)}^{'}+C_t\Sigma_tC_t^{'}]R^{-1}+\text{ct}.\\
\end{aligned}
\end{split}
\label{UpdateR_term2}
\end{equation}
Assembling terms 1 and 2 in equations \eqref{UpdateR_term1} and \eqref{UpdateR_term2} yields to
\begin{equation}
\begin{aligned}
\begin{split}
\log  q^*(R|\{a_{r_{i}}\}_{i=1}^d,Z_1^T)^{(l+1)} = & -\frac{r_p+d+1}{2} \log |R| 
-\frac{1}{2}\rm {Tr}(\left<B_p\right> R^{-1})-{\frac{T}{2}} \log |R|\\
&-\frac{1}{2}\mathrm{Tr}\underbrace{\sum\limits_{t=1}^T [(Z_t-C_t\mu_t){(Z_t-C_t\mu_t)}^{'}+C_t\Sigma_tC_t^{'}]}_{:= B}R^{-1}+\text{ct},
\end{split}
\end{aligned}
\label{UPDATE_R1}
\end{equation}
and then 
\begin{equation}
\begin{split}
\begin{aligned}
\log  q^*(R|\{a_{r_{i}}\}_{i=1}^d,Z_1^T) =\mathcal{IW}_d(R|r_p+T,\left<B_p\right>+B).  
\end{aligned}
\end{split}
\label{UPDATE_R2}
\end{equation}

\section{Multiple trials}
\label{Multiple trials}

The model can be modified to deal with $N$ conditionally independent sequences $\{{Z_1^T}_{(j)}\}_{j=1}^{N}$ which are supposed to have the same hidden state. This reflects the case that arises during an event-related experimental paradigm, where many trials on the same condition are measured. 

In \citet{beal_variational_2003} and \citet{cassidy_bayesian_2002}, this extension is solved by first estimating the necessary sufficient statistics for each sequence independently in the E-step, and then by averaging these statistics to get only one set of sufficient statistics, which is representative of the entire set of independent sequences before performing the M-step. However this approach does not take into account the complex dependence of the variational posterior $q(\varphi_{1}^T|\{{Z_1^T}_{(j)}\}_{j=1}^{N})$ on the whole dataset $\{{Z_1^T}_{(j)}\}_{j=1}^{N}$. We can however write and solve the full evidence of all the data if we rewrite
 \begin{equation}
 \left\{\begin{aligned}
 	\varphi_{t+1}  =  A\varphi_t+w_t\qquad &w_t\sim{\mathcal{N}_k(0,Q)} \\
    \mathbf{Z}_t  =  \mathbf{C}_t \varphi_t+v_t \qquad &v_t\sim{\mathcal{N}_d(0,\mathbf{R})}\\ 
       \end{aligned}
 \right.
 \quad \text{where}
 \begin{cases}
    \varphi_t=&\vect{[\vartheta_{1(t)} ,\vartheta_{2(t)},..,\vartheta_{p(t)}]}', 
    \\
    \mathbf{Z}_t=&( Y_{t(1)}  \hdots  Y_{t(N)} X_{t(1)} \hdots X_{t(N)})^{'}
    \\
    \mathbf{C}_t \varphi_t=&\sum\limits_{j=1}^p \vartheta_{j(t)} (Y_{t-j(1)} \hdots Y_{t-j(N)} \\ & X_{t-j(1)} \hdots X_{t-j(N)})^{'},
    \\
    \mathbf{R}=&\text{diag}(R_1 \hdots R_N)^{'},
    \end{cases}
 \label{kalman2.def}
\end{equation}
where $\mathbf{R}$ is block diagonal of dimensions $d^2N \times d^2N$ with each diagonal element $R_j$ being identically distributed. The state equation remains the same as in model defined in Section 2. We can observe in model \eqref{kalman2.def} that the time-varying parameter vectors $\{\varphi_t\}_{t=1}^T$ are unique, taking into account the whole dataset $\{{Z_1^T}_{(j)}\}_{j=1}^{N}$. The evidence of the complete model can be rewritten as 
\begin{equation}
\begin{aligned}
\begin{split}
p(\{{Z_1^T}_{(j)}\}_{j=1}^{N},\varphi_{1}^T,\Omega_1^b)=&p({Z_1^T}_{(j)}\}_{j=1}^{N},\varphi_{1}^T,\{A_{ii}\}_{i=1}^k\{\alpha_{i}\}_{i=1}^k,\{\delta_i\}_{i=1}^k,Q_{i,i},a_q,\mathbf{R},\{a_{r_{i}}\}_{i=1}^d)
\\
=&\prod_{j=1}^{N}\prod_{i=1}^{T}p(Z_t(j)|C_t(j),\varphi_{t},R_j)p(\varphi_1)\prod_{i=2}^{T}p(\varphi_t|A,\varphi_{t-1},Q)\prod_{i=1}^{k} p(A_{ii}|\alpha_i)\\
&\times\prod_{i=1}^{k}p(\alpha_i|\delta_i)\prod_{i=1}^{k}
p(\delta_i)p(Q_{i,i}|a_{q})p(a_{q})p(R_j|\{a_{r_{i}}\}_{i=1}^d)\prod_{i=1}^{d}p(a_{r_i}).
\label{Full_distribution_TOT}
\end{split}
\end{aligned}
\end{equation}
All the computations done so far can easily be adapted to this new setting.
The variational posterior distributions relative to model \eqref{kalman2.def} are now conditional on the whole dataset $\{{Z_1^T}_{(j)}\}_{j=1}^{N}$. A detailed derivation of the model and related variational posterior distributions with multiple trials can be founded in Section \ref{Multiple trials}.

In the DAG depicted in Figure~\ref{DAG_Figure2}, we can see that conditional dependence on the observed variables $Z$ is present only for nodes $\{\varphi_1^T;Z_1^T;\mathbf{R};a_r\}$. 
For the hidden variables, the variational E-step KRTS algorithm can be run on the whole state space system \eqref{kalman2.def} and the resulting variational posterior density will therefore be conditional on the whole dataset $q(\varphi_1^T|\{{Z_1^T}_{(j)}\}_{j=1}^{N})$. In Appendix \ref{Appendix x}, the update equations remain the same, except for the Kalman gain in equation \eqref{Kalman_GAIN} which becomes 
\begin{equation}
\begin{aligned}
K_t=\Sigma_t^{t-1}C_t^{'}(C_t\Sigma_t^{t-1}C_t^{'}+\left<\mathbf{R}\right>)^{-1}.
\end{aligned}
\end{equation} 
Looking at the update equations for the $a_r$ parameter in Appendix \ref{Appendix ar}, we see that for multiple sequences equation \eqref{UPDATE_ar1} becomes 
\begin{equation}
\begin{split}
\begin{aligned}
\log  q^*(\{a_{r_{i}}\}_{i=1}^d)^{(l+1)}=  \underbrace{\log  p(\{a_{r_{i}}\}_{i=1}^d)}_{\text{term 1}} + \underbrace{\sum\limits_{j=1}^N \left<\log  p(R_j|\{a_{r_{i}}\}_{i=1}^d)\right>_{q(R_j)^{(l)}}}_{\text{term 2}}+\text{ct},
\end{aligned}
\end{split}
\end{equation}
where ct denotes the normalization constant for the given density. Therefore equation \eqref{UPDATE_ar4} can be rewritten as
\begin{equation}
\begin{split}
\begin{aligned}
\text{term 2}  &= -\frac{r_p}{2} \sum\limits_{j=1}^N  \sum_{i=1}^d \log a_{r_i}-\nu\ \sum\limits_{j=1}^N \sum_{i=1}^d  a_{r_i}^{-1}\left<R^{-1}_{j\{i,i\}}\right>+\text{ct}. 
\end{aligned}
\end{split}
\end{equation}
The resulting shape and scale parameters for $q^*(a_{r_{i}})^{(l+1)}$ therefore become
\begin{equation}
\begin{aligned}
\begin{split}
a_{qr}&=a_{pr}+\frac{r_{p}N}{2}, \\
b_{qr}&=b_{pr}+\nu \sum\limits_{j=1}^N \left<R^{-1}_{j\{i,i\}}\right>, \quad \text{where} \quad \left<R^{-1}_{j\{i,i\}}\right> =r\left< B_j^{-1} \right>.
\end{split}
\end{aligned}
\end{equation}
Looking at the update equations for the $R$ matrix in Appendix \ref{Appendix R}, we see that for multiple sequences, equation \eqref{UPDATE_Ra} becomes 
\begin{equation}
\begin{split}
\begin{aligned}
\log  q^*(\mathbf{R}|\{a_{r_{i}}\}_{i=1}^d)^{(l+1)} = &\sum\limits_{j=1}^N \log  q^*(R_j|\{a_{r_{i}}\}_{i=1}^d)^{(l+1)} \\
= &\sum\limits_{j=1}^N\underbrace{\left<\log p(R_j|\{a_{r_{i}}\}_{i=1}^d)\right>_{q(a_r)}}_{\text{term 1}} \\
&+ \underbrace{\sum\limits_{j=1}^N\sum\limits_{t=1}^T \left<\log p(Z_t(j)|C_t(j)\varphi_{t})\right>_{q(\varphi_1^T)^{(l)}}}_{\text{term 2}}+\text{ct},
\end{aligned}
\end{split}
\end{equation}
where each block-diagonal entry $R_j$ is conditional on a unique dataset $\{{Z_1^T}_{(j)}\}$, and will therefore present the same properties as the unique $R$ matrix in Appendix \ref{Appendix R}. The resulting parameters for the variational posterior inverse-Wishart distribution \\
$q^*(R_j|\{a_{r_{i}}\}_{i=1}^d)^{(l+1)}$ can therefore be rewritten as 
\begin{equation}
\begin{aligned}
\begin{split}
B_j&=2\,\nu \,\text{diag}[\frac{1}{\left<a_{r_1}\right>} ,\dots, \frac{1}{\left<a_{r_d}\right>}]+\sum\limits_{t=1}^T [(Z_t-C_t\mu_t){(Z_t-C_t\mu_t)+C_t\Sigma_tC_t^{'}}^{'}R],\\
r&=r_p+T.
\end{split}
\end{aligned}
\end{equation}

\begin{figure}[tb]
\centering
\includegraphics[scale=0.5]{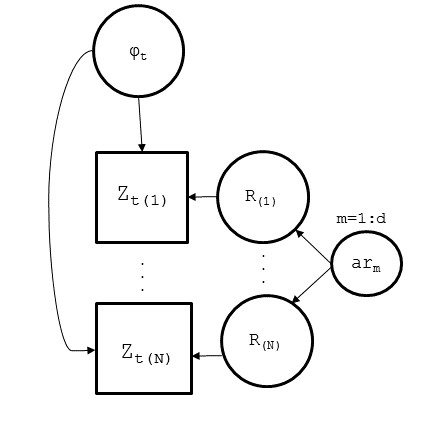}
\caption{Part of the DAG representation of the full model given in Figure~\ref{DAG_Figure2} that is modified to account for  multiple trials.}
\label{DAG_MULT}
\end{figure}

\section{Granger-Causality Detection Results}
\label{Appendix Granger causality}

\subsection{Data generated with slowly-varying parameters and normal errors}
\label{Data generated with slowly varying parameters and normal errors}

\begin{figure}[!htb]
\centering
\hspace*{-1in}
\includegraphics[width=6.4in]{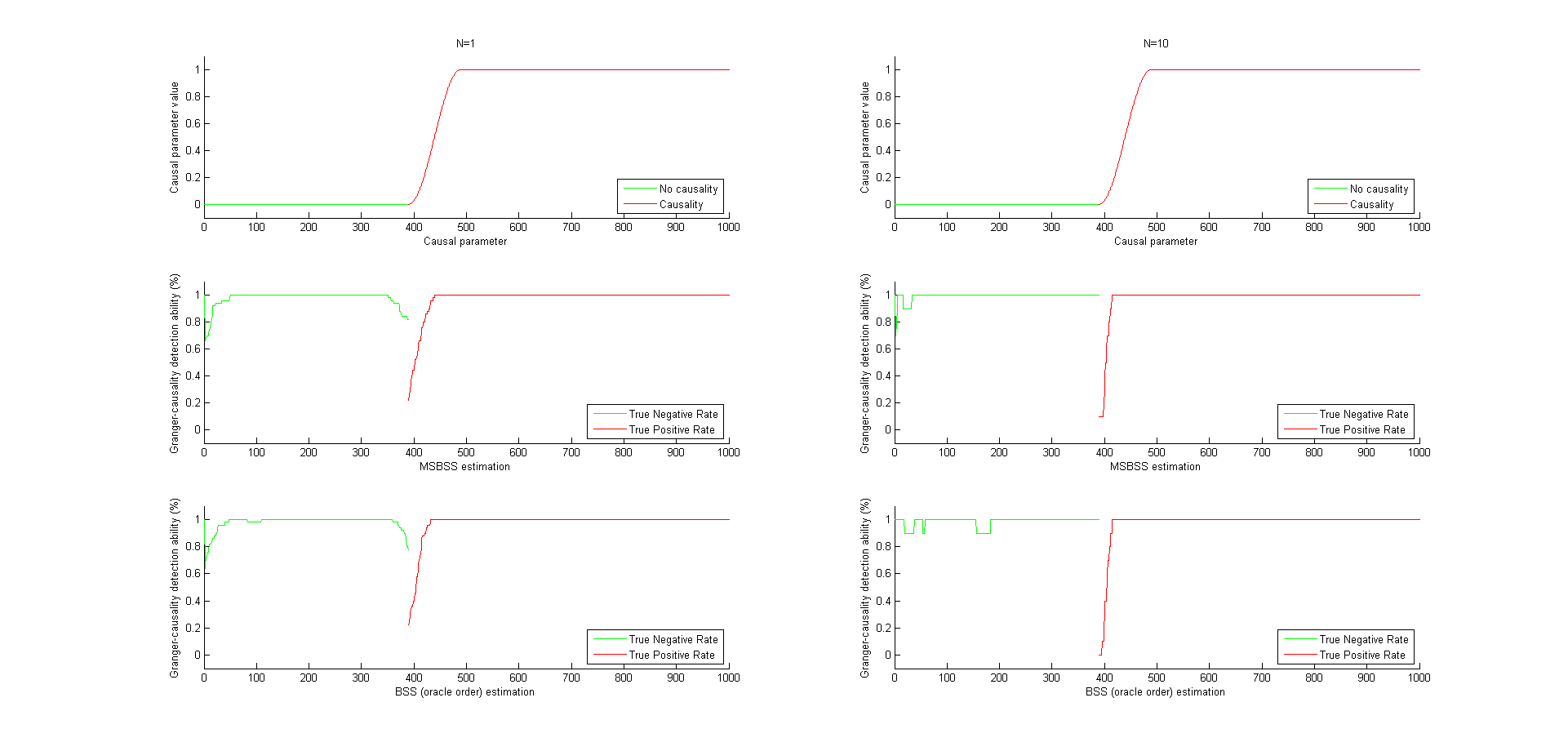}
\caption{Granger-causality detection ability for order $1$, series length $1000$ and causal parameter $1$.}
\label{SINUS_GC12}
\end{figure}

\begin{figure}[!htb]
\centering
\hspace*{-1in}
\includegraphics[width=6.4in]{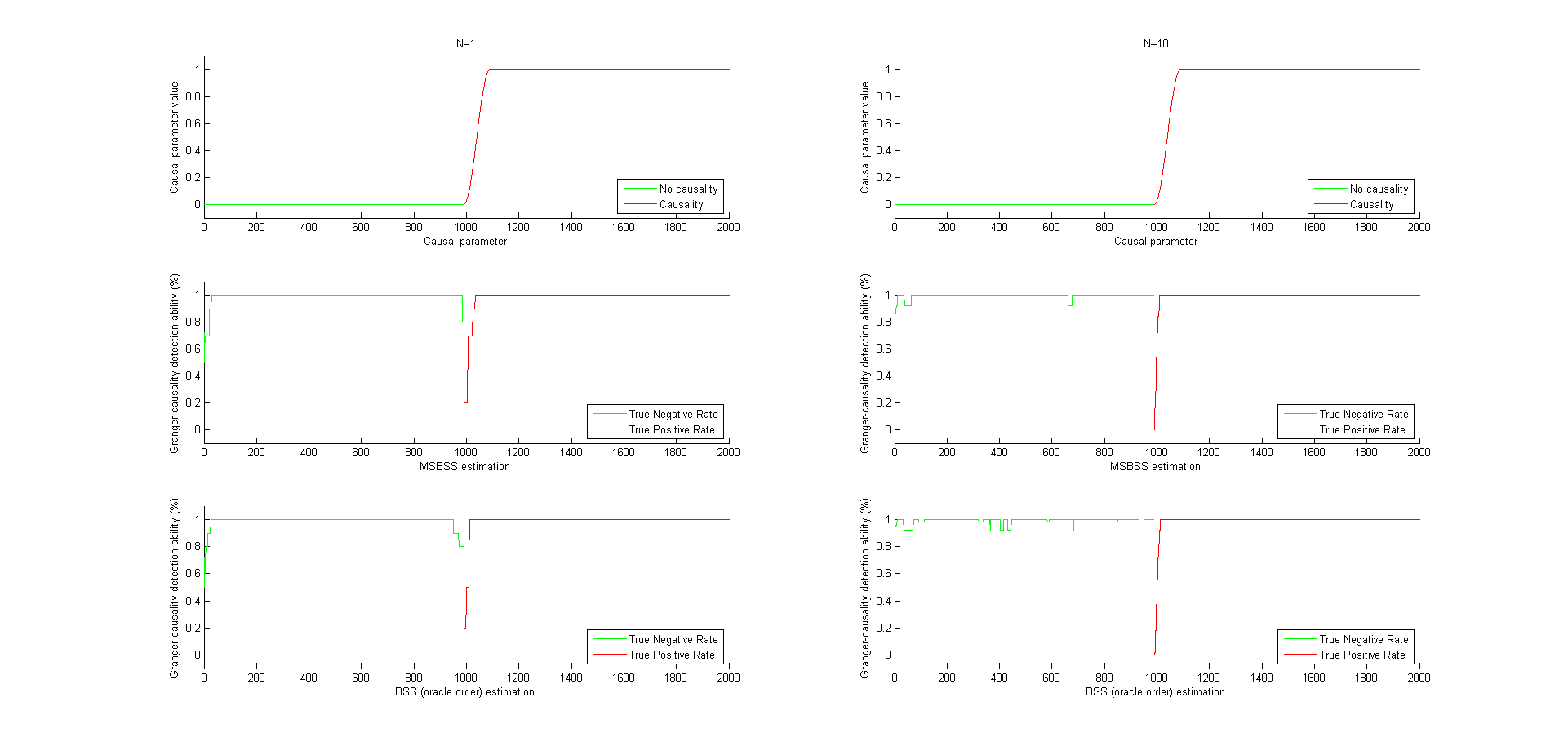}
\caption{Granger-causality detection ability for order $1$, series length $2000$ and causal parameter $1$.}
\label{SINUS_GC13}
\end{figure}

\begin{figure}[!htb]
\centering
\hspace*{-1in}
\includegraphics[width=6.4in]{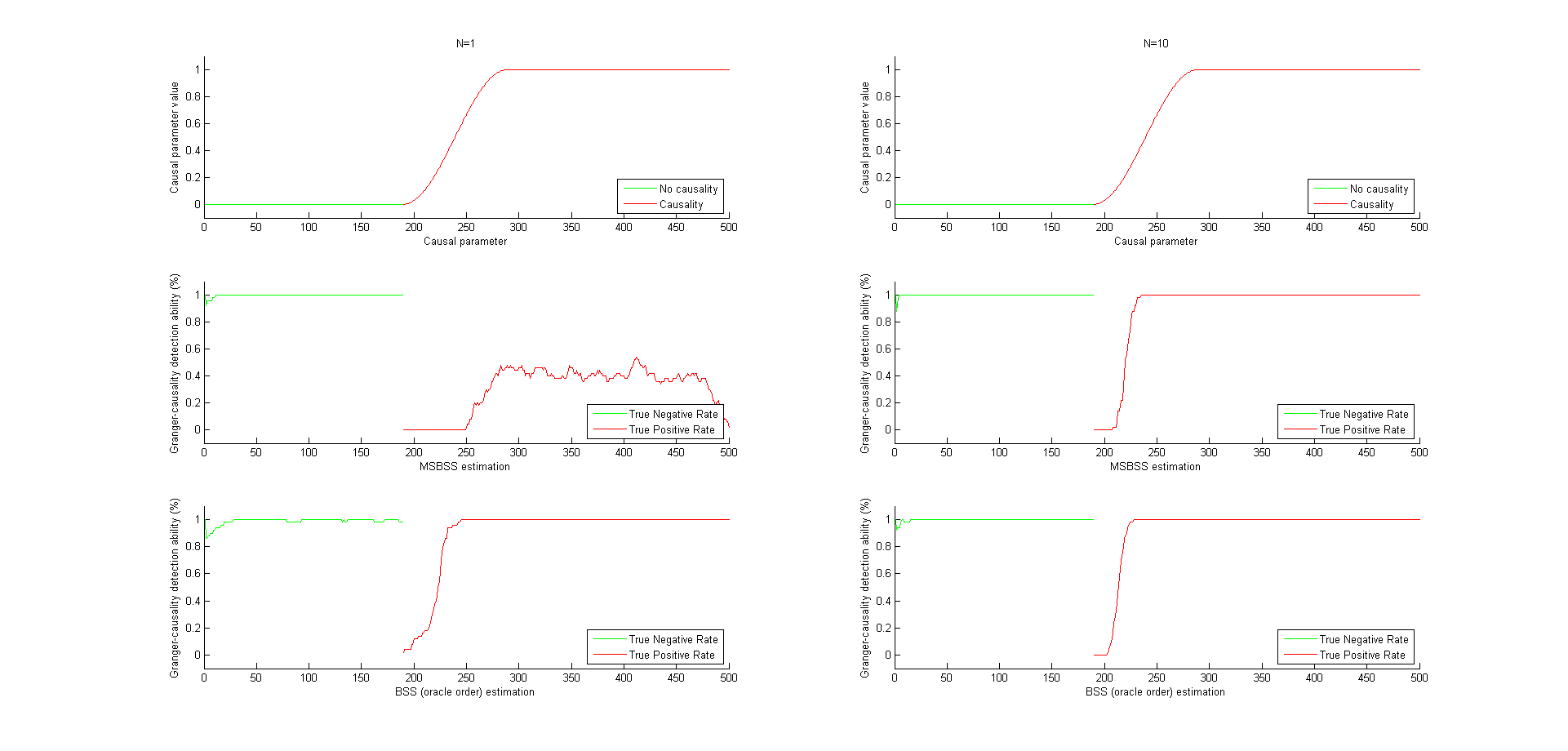}
\caption{Granger-causality detection ability for order $2$, series length $500$ and causal parameter $1$.}
\label{SINUS_GC21}
\end{figure}

\begin{figure}[!htb]
\centering
\hspace*{-1in}
\includegraphics[width=6.4in]{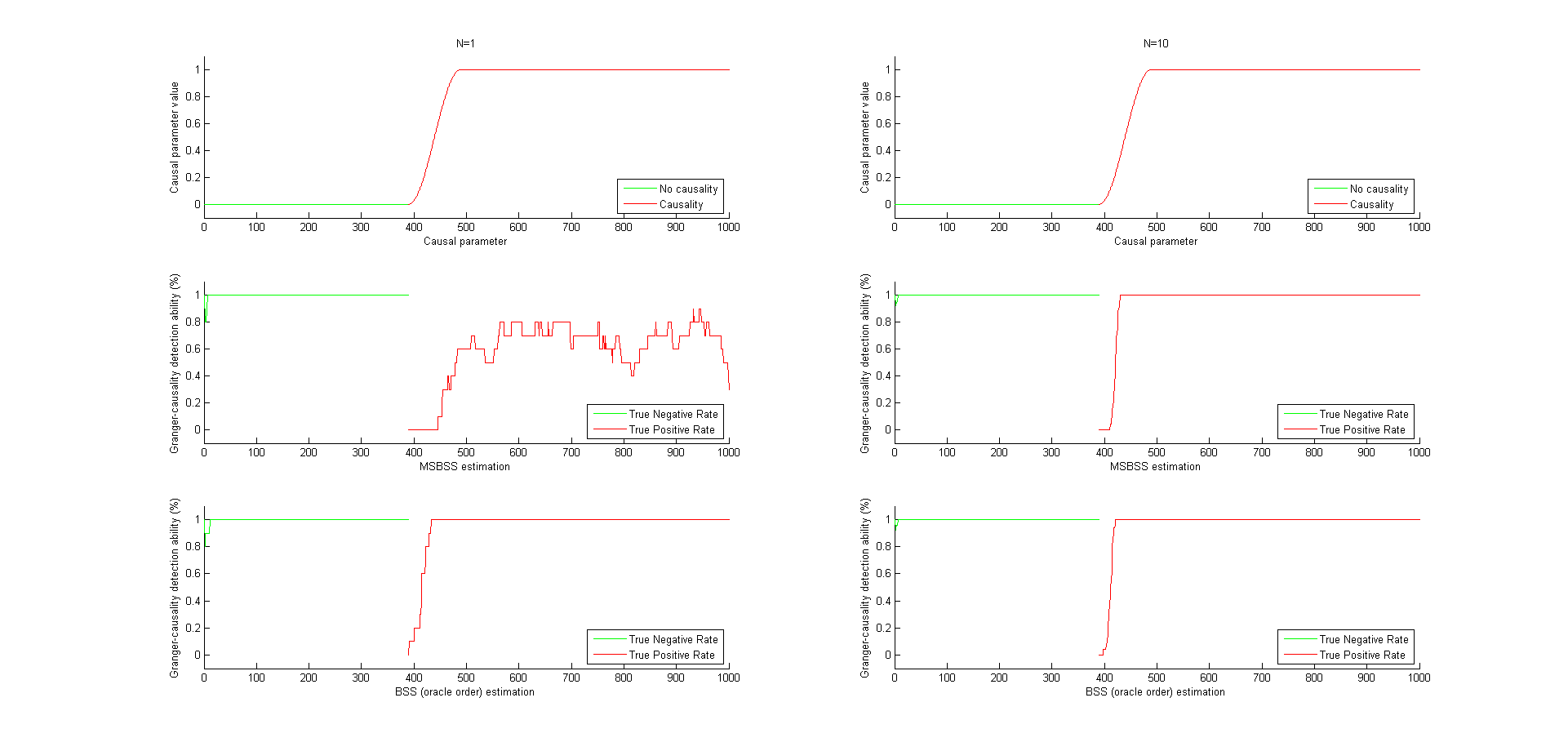}
\caption{Granger-causality detection ability for order $2$, series length $1000$ and causal parameter $1$.}
\label{SINUS_GC22}
\end{figure}

\begin{figure}[!htb]
\centering
\hspace*{-1in}
\includegraphics[width=6.4in]{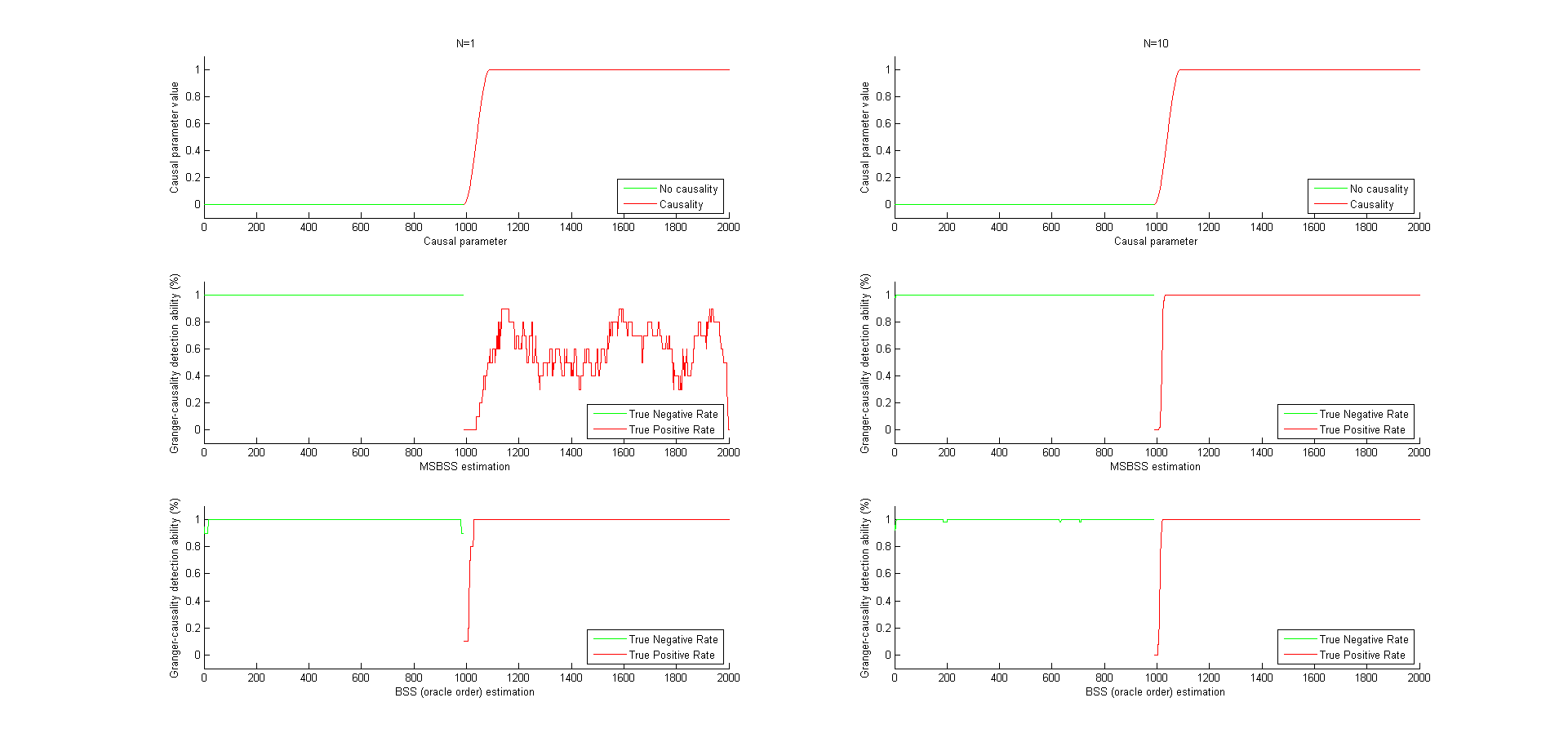}
\caption{Granger-causality detection ability for order $2$, series length $2000$ and causal parameter $1$.}
\label{SINUS_GC23}
\end{figure}

\begin{figure}[!htb]
\centering
\hspace*{-1in}
\includegraphics[width=6.4in]{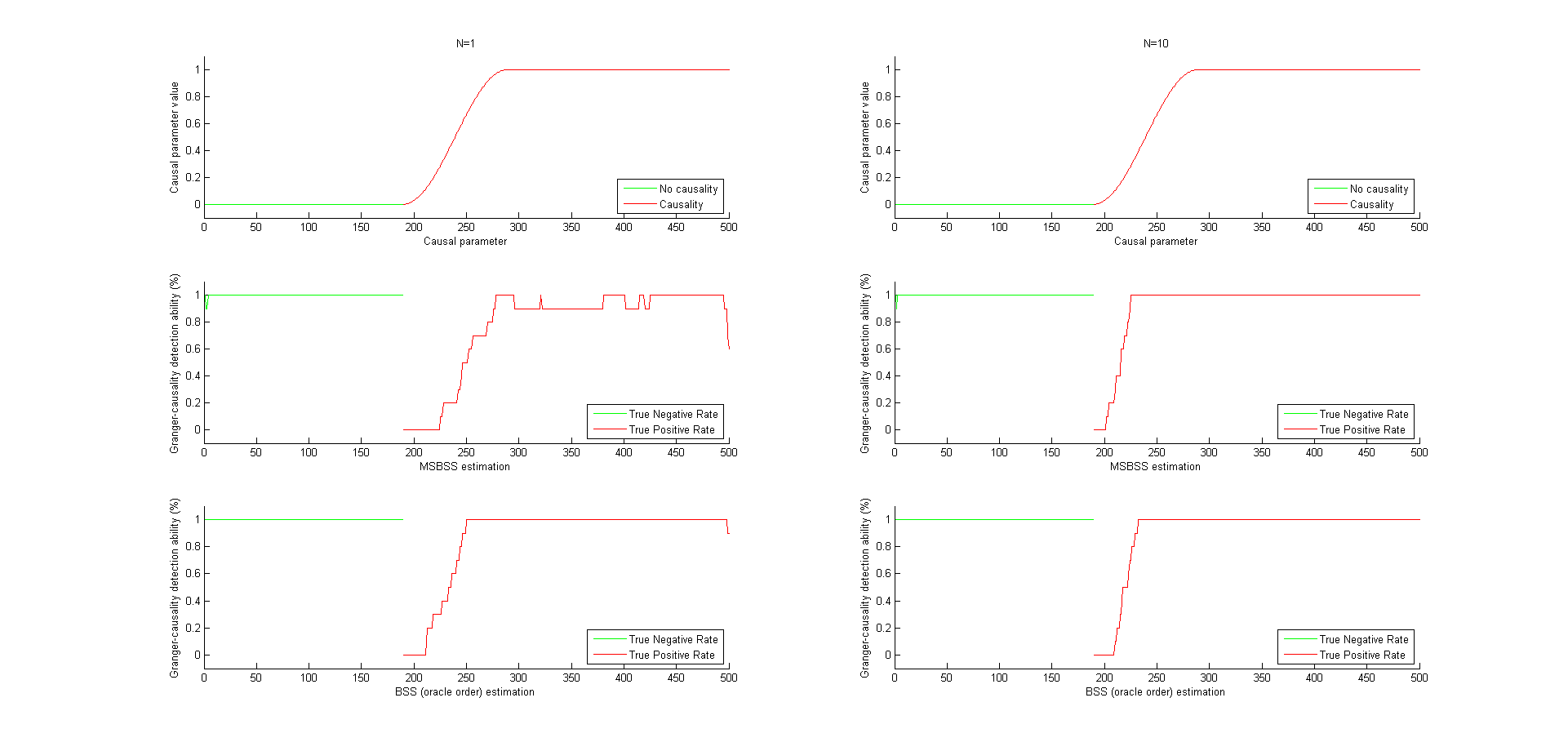}
\caption{Granger-causality detection ability for order $4$, series length $500$ and causal parameter $1$.}
\label{SINUS_GC31}
\end{figure}

\begin{figure}[!htb]
\centering
\hspace*{-1in}
\includegraphics[width=6.4in]{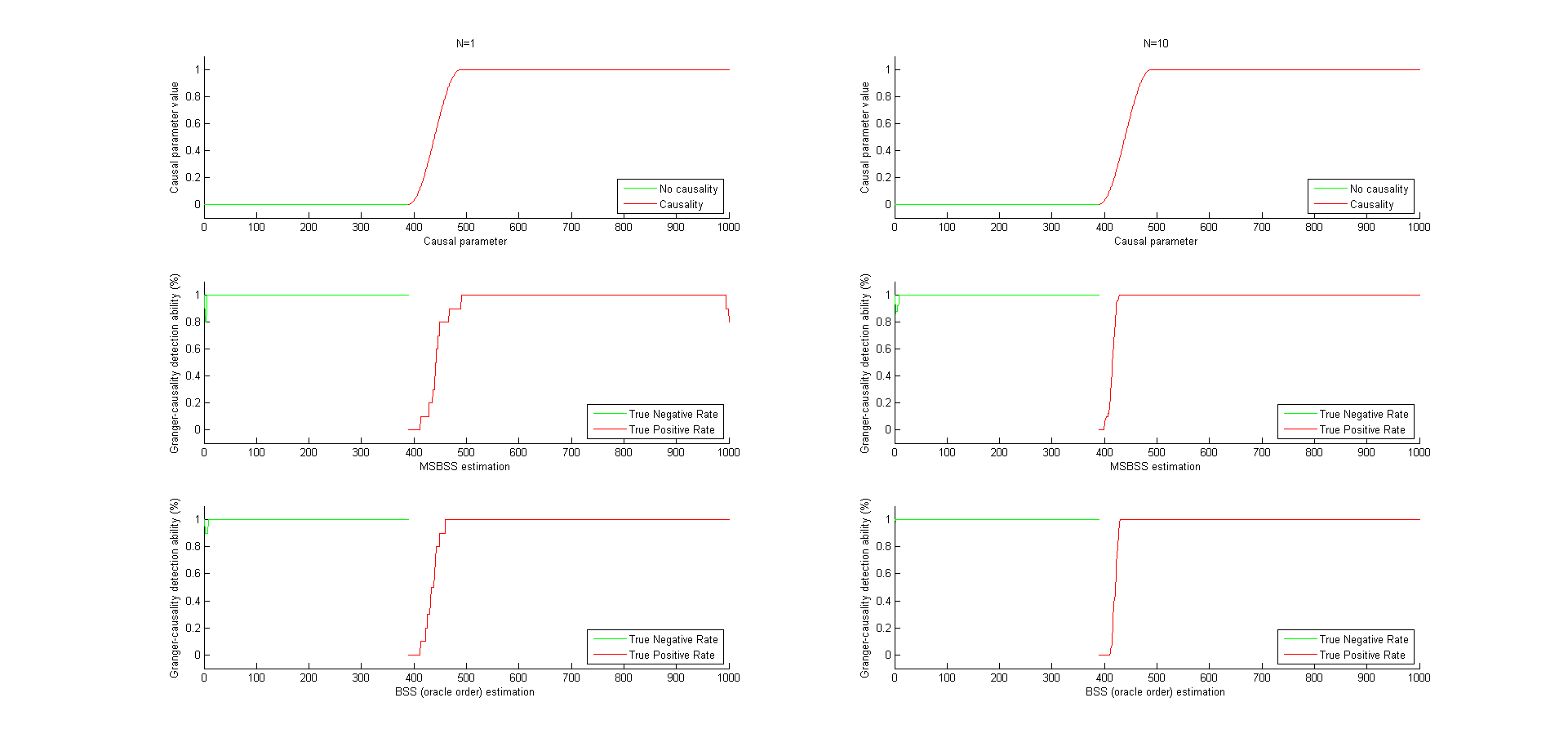}
\caption{Granger-causality detection ability for order $4$, series length $1000$ and causal parameter $1$.}
\label{SINUS_GC32}
\end{figure}

\begin{figure}[!htb]
\centering
\hspace*{-1in}
\includegraphics[width=6.4in]{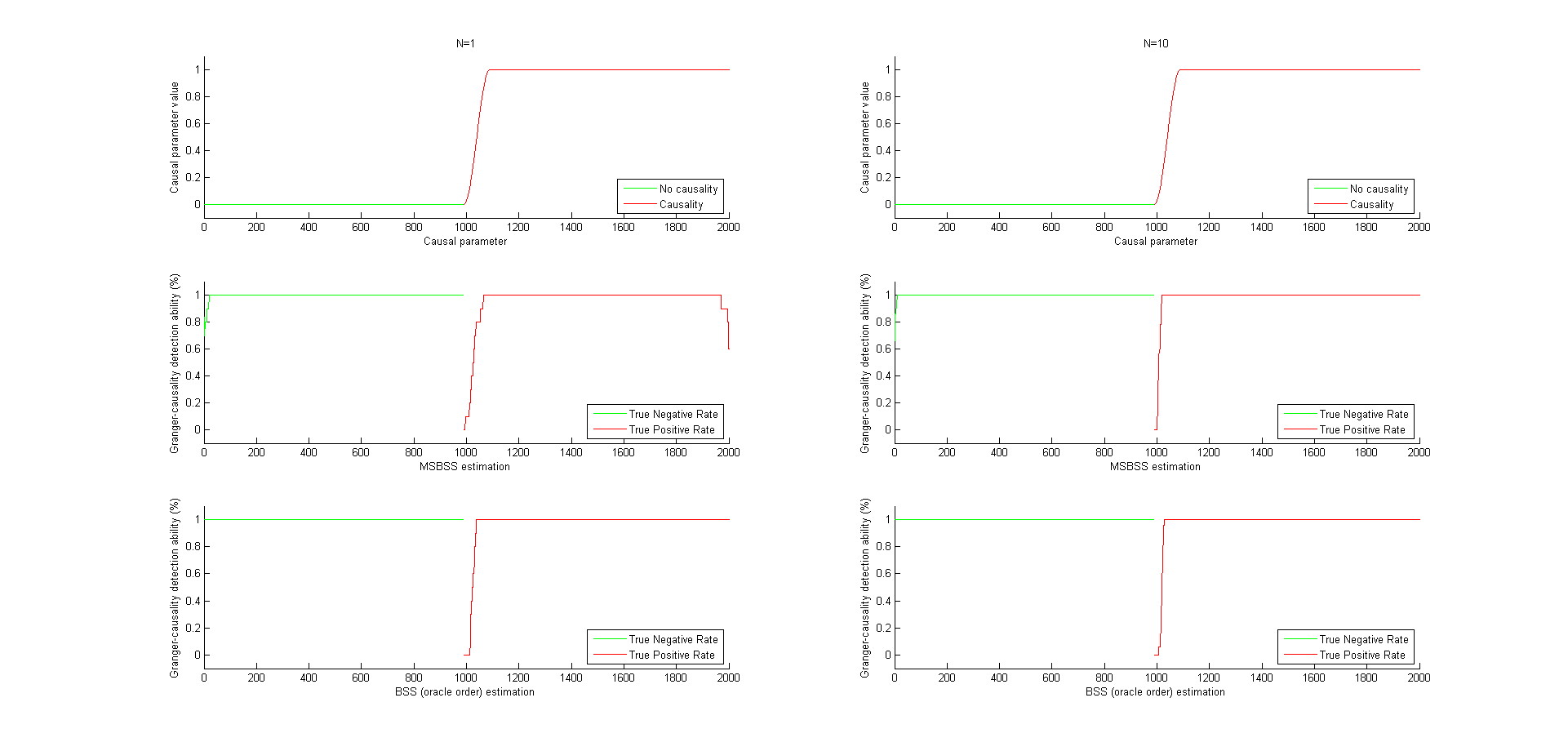}
\caption{Granger-causality detection ability for order $4$, series length $2000$ and causal parameter $1$.}
\label{SINUS_GC33}
\end{figure}

\begin{figure}[!htb]
\centering
\hspace*{-1in}
\includegraphics[width=6.4in]{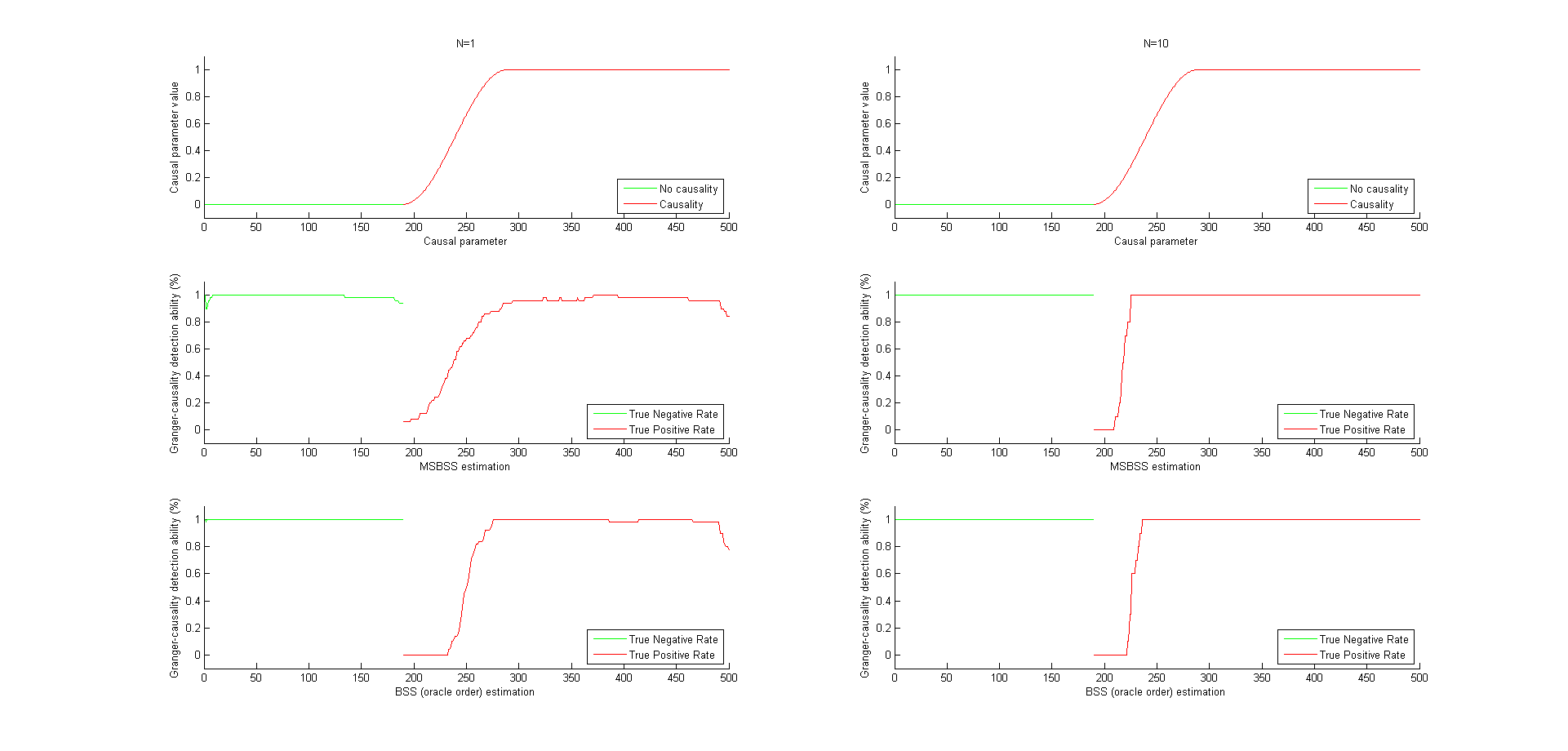}
\caption{Granger-causality detection ability for order $8$, series length $500$ and causal parameter $1$.}
\label{SINUS_GC41}
\end{figure}

\begin{figure}[!htb]
\centering
\hspace*{-1in}
\includegraphics[width=6.4in]{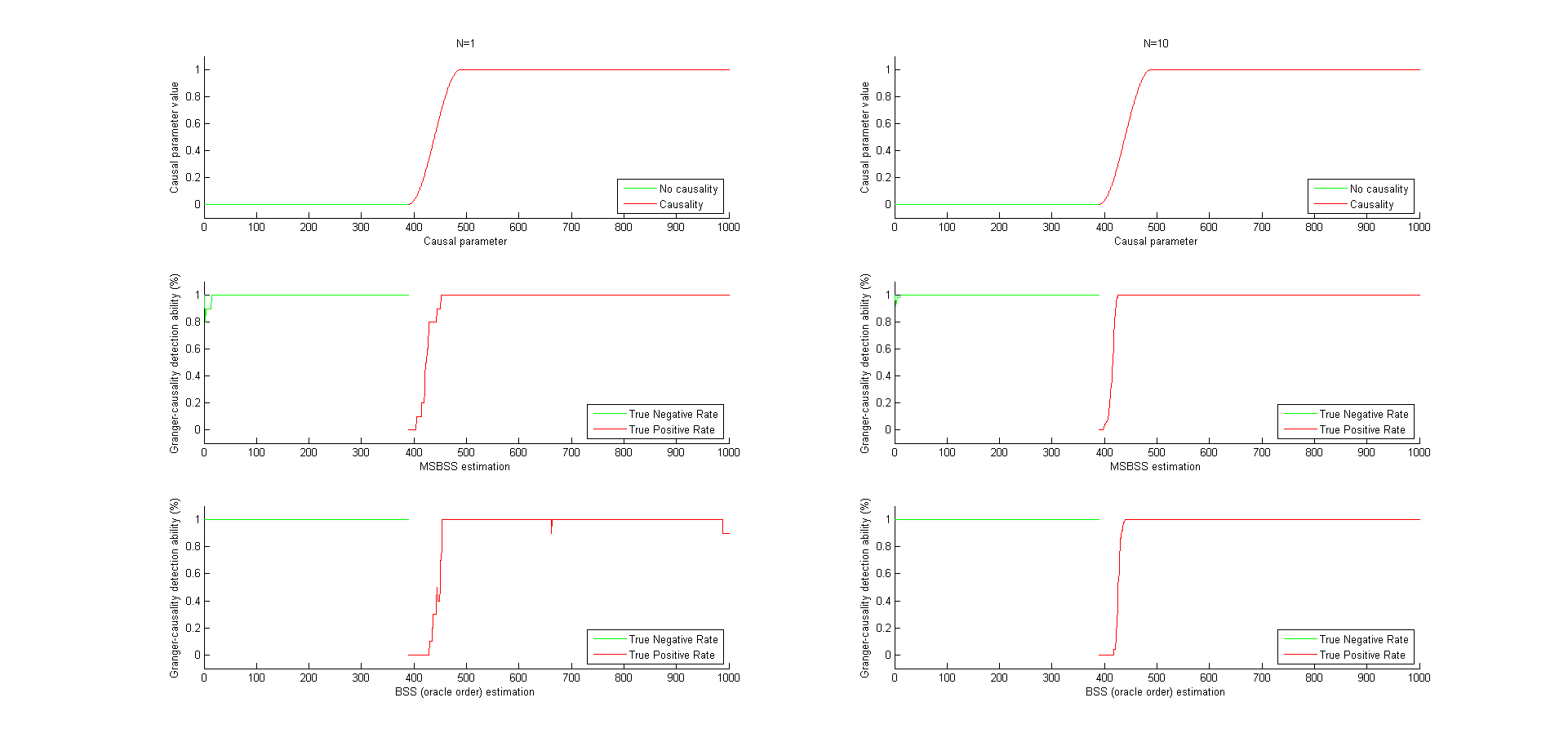}
\caption{Granger-causality detection ability for order $8$, series length $1000$ and causal parameter $1$.}
\label{SINUS_GC42}
\end{figure}

\begin{figure}[!htb]
\centering
\hspace*{-1in}
\includegraphics[width=6.4in]{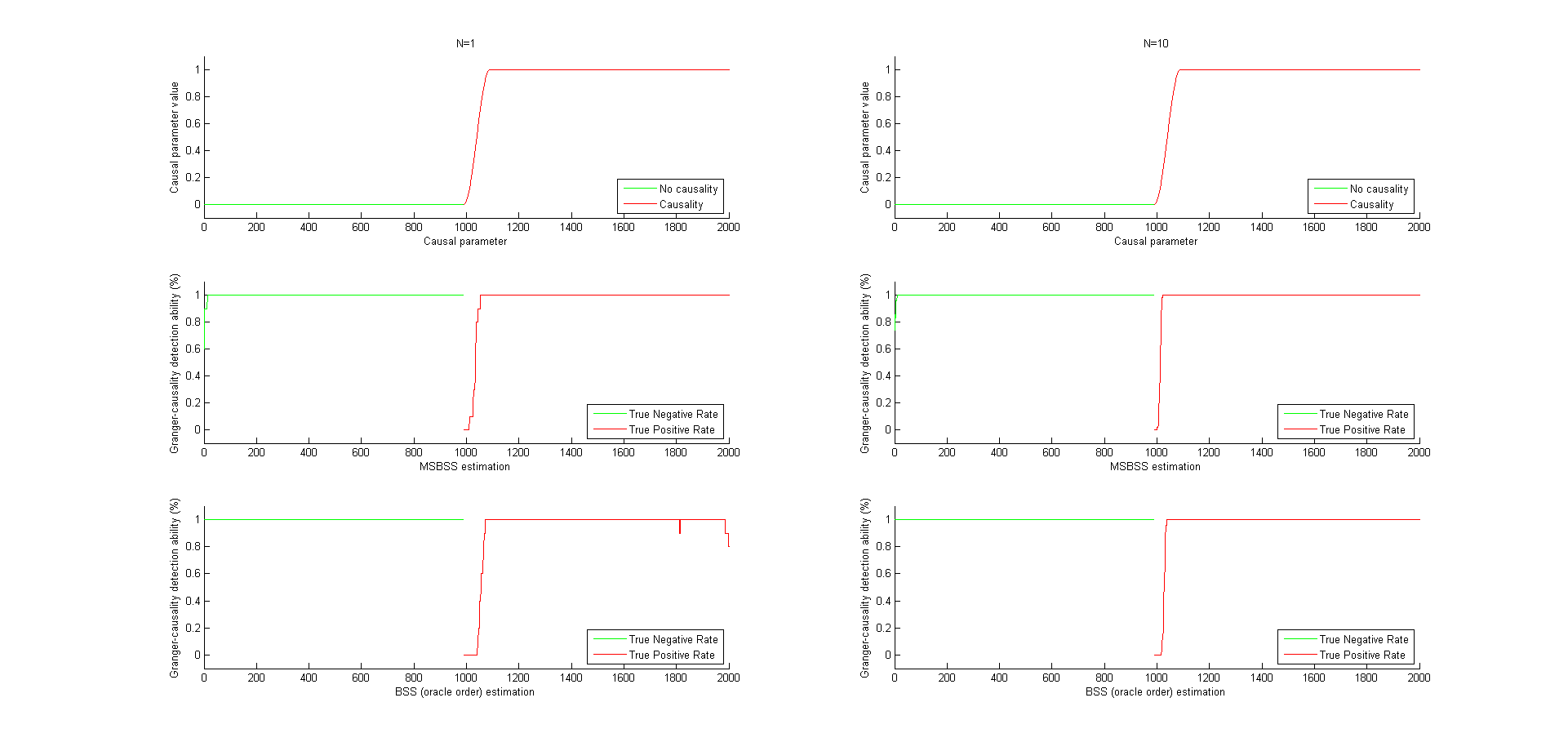}
\caption{Granger-causality detection ability for order $8$, series length $2000$ and causal parameter $1$.}
\label{SINUS_GC43}
\end{figure}

\begin{figure}[!htb]
\centering
\hspace*{-1in}
\includegraphics[width=6.4in]{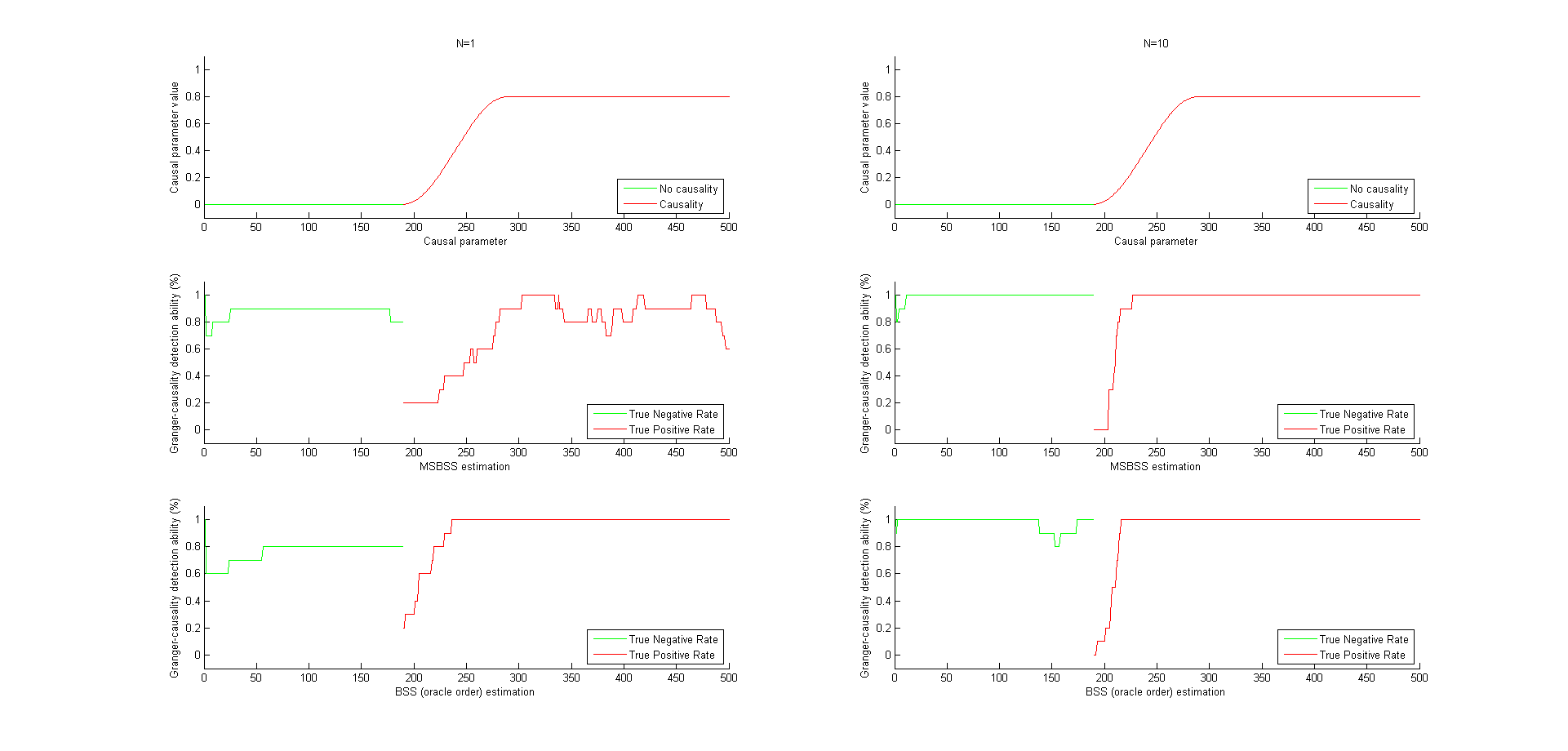}
\caption{Granger-causality detection ability for order $1$, series length $500$ and causal parameter $0.8$.}
\label{GC_1_1_1}
\end{figure}

\begin{figure}[!htb]
\centering
\hspace*{-1in}
\includegraphics[width=6.4in]{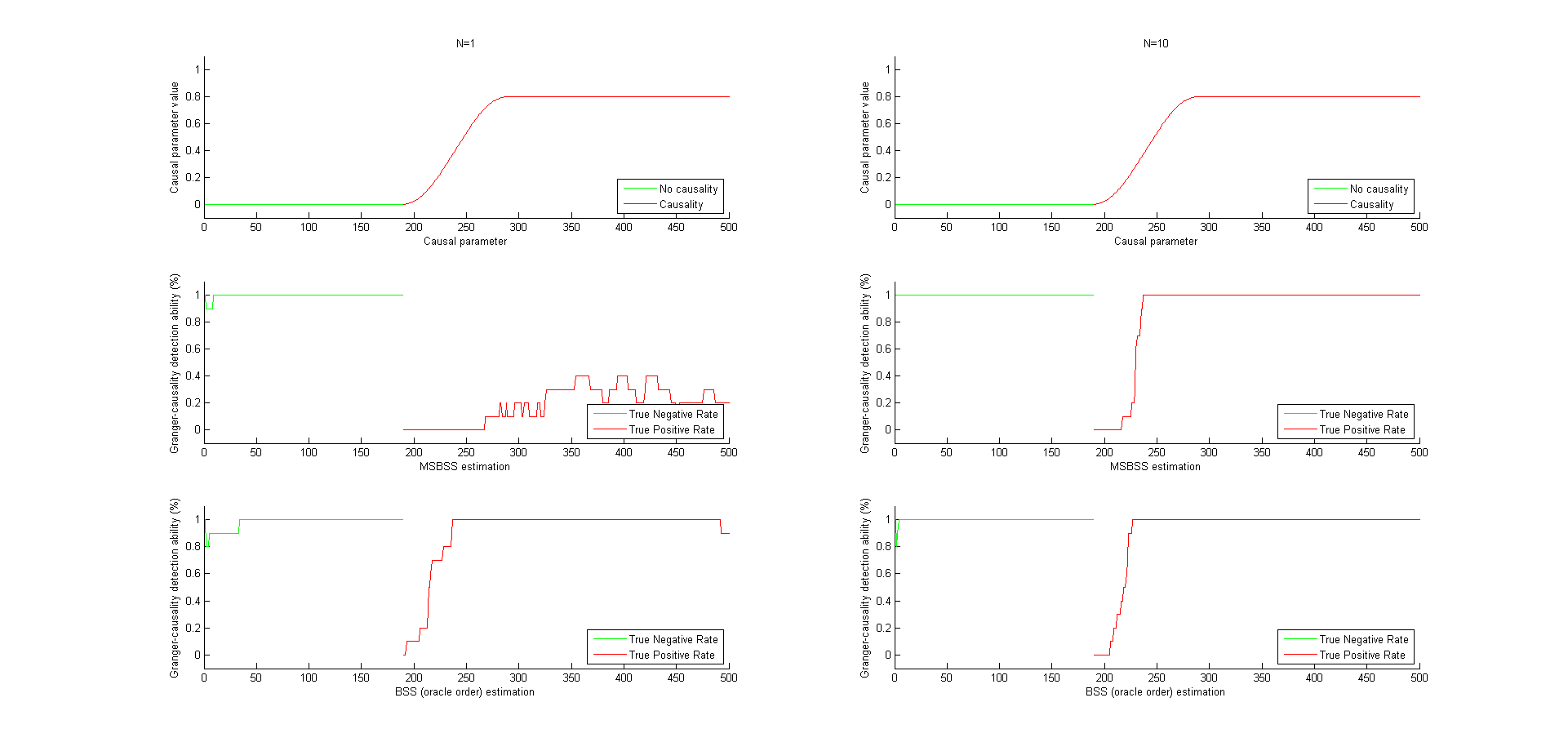}
\caption{Granger-causality detection ability for order $2$, series length $500$ and causal parameter $0.8$.}
\label{GC_2_1_1}
\end{figure}

\begin{figure}[!htb]
\centering
\hspace*{-1in}
\includegraphics[width=6.4in]{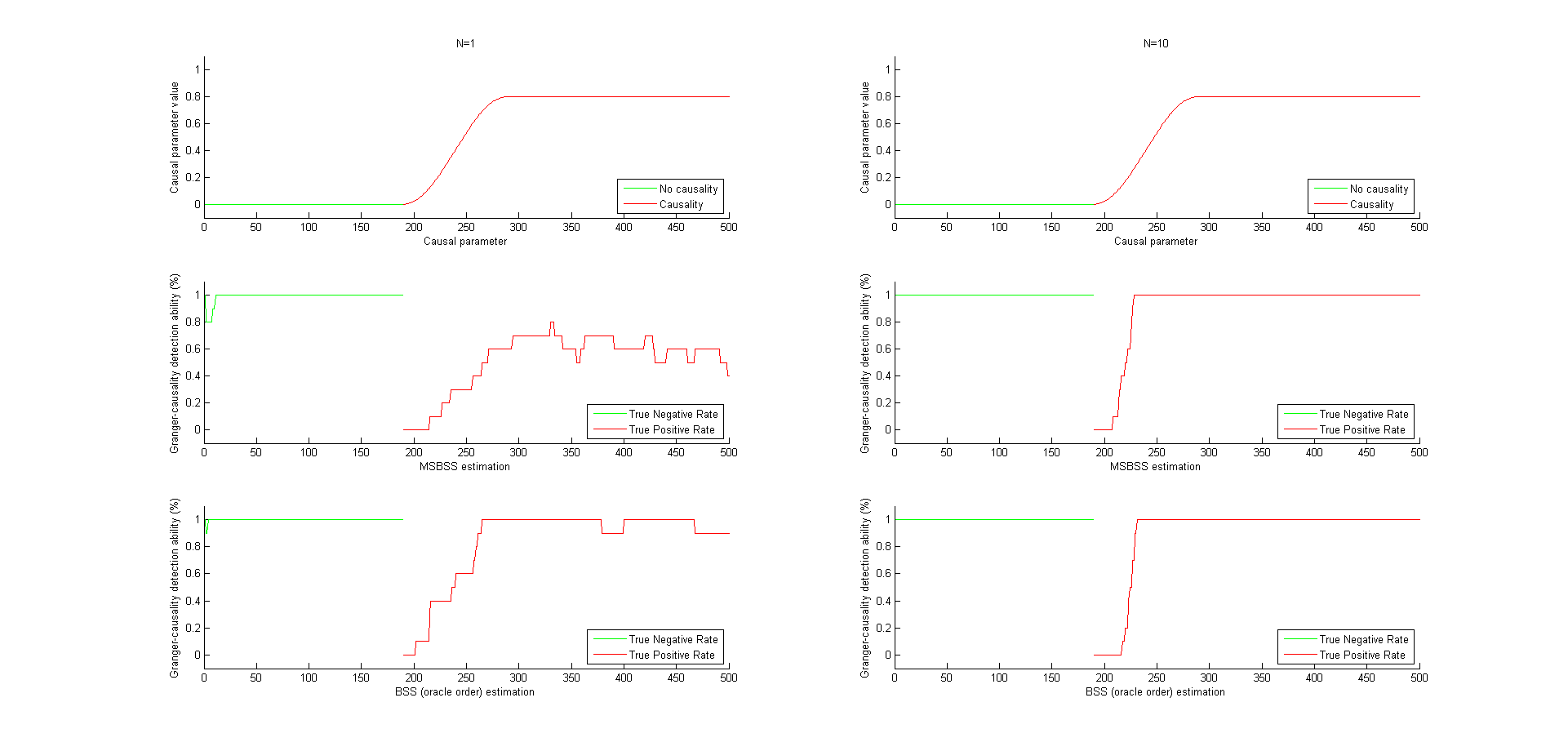}
\caption{Granger-causality detection ability for order $4$, series length $500$ and causal parameter $0.8$.}
\label{GC_3_1_1}
\end{figure}

\begin{figure}[!htb]
\centering
\hspace*{-1in}
\includegraphics[width=6.4in]{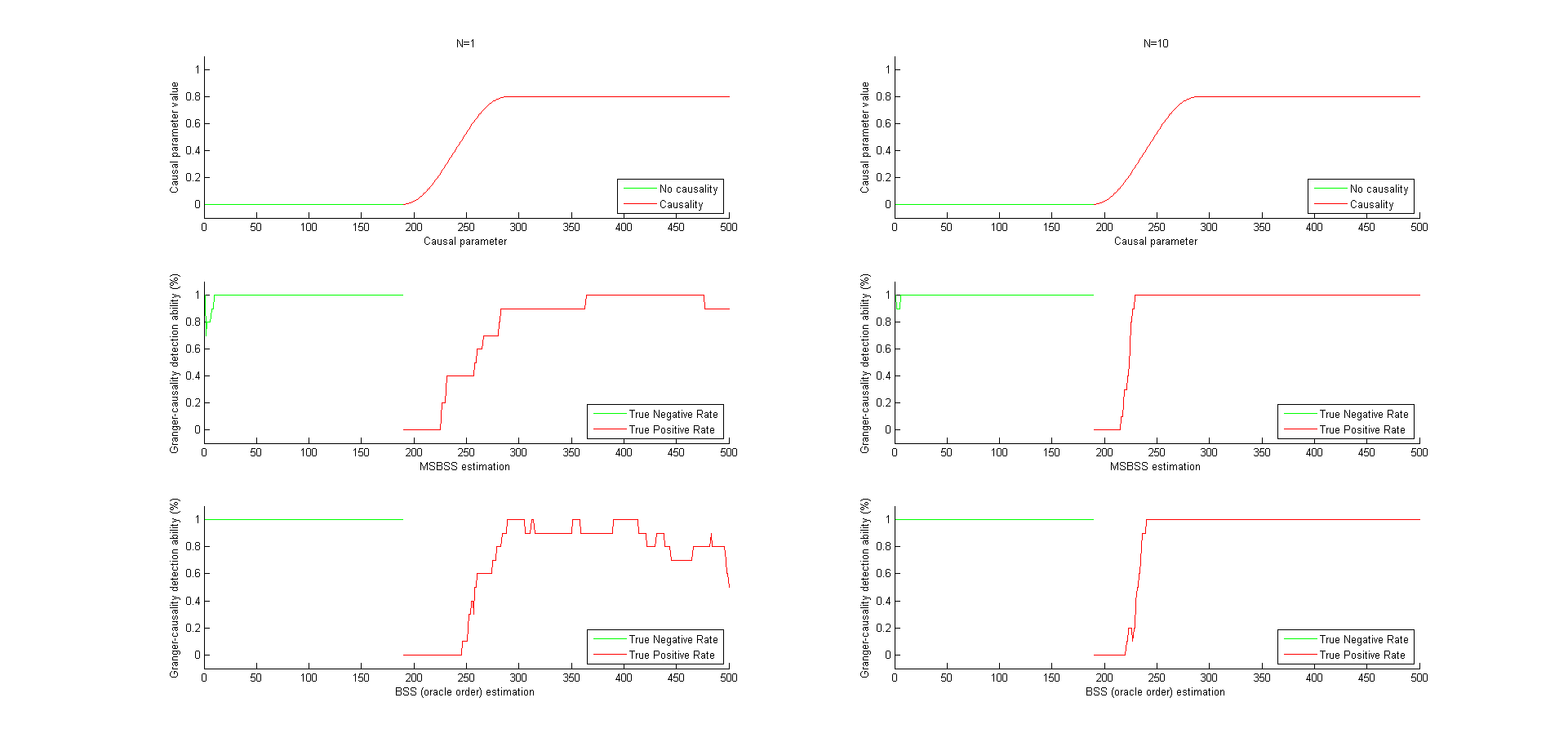}
\caption{Granger-causality detection ability for order $8$, series length $500$ and causal parameter $0.8$.}
\label{GC_4_1_1}
\end{figure}

\begin{figure}[!htb]
\centering
\hspace*{-1in}
\includegraphics[width=6.4in]{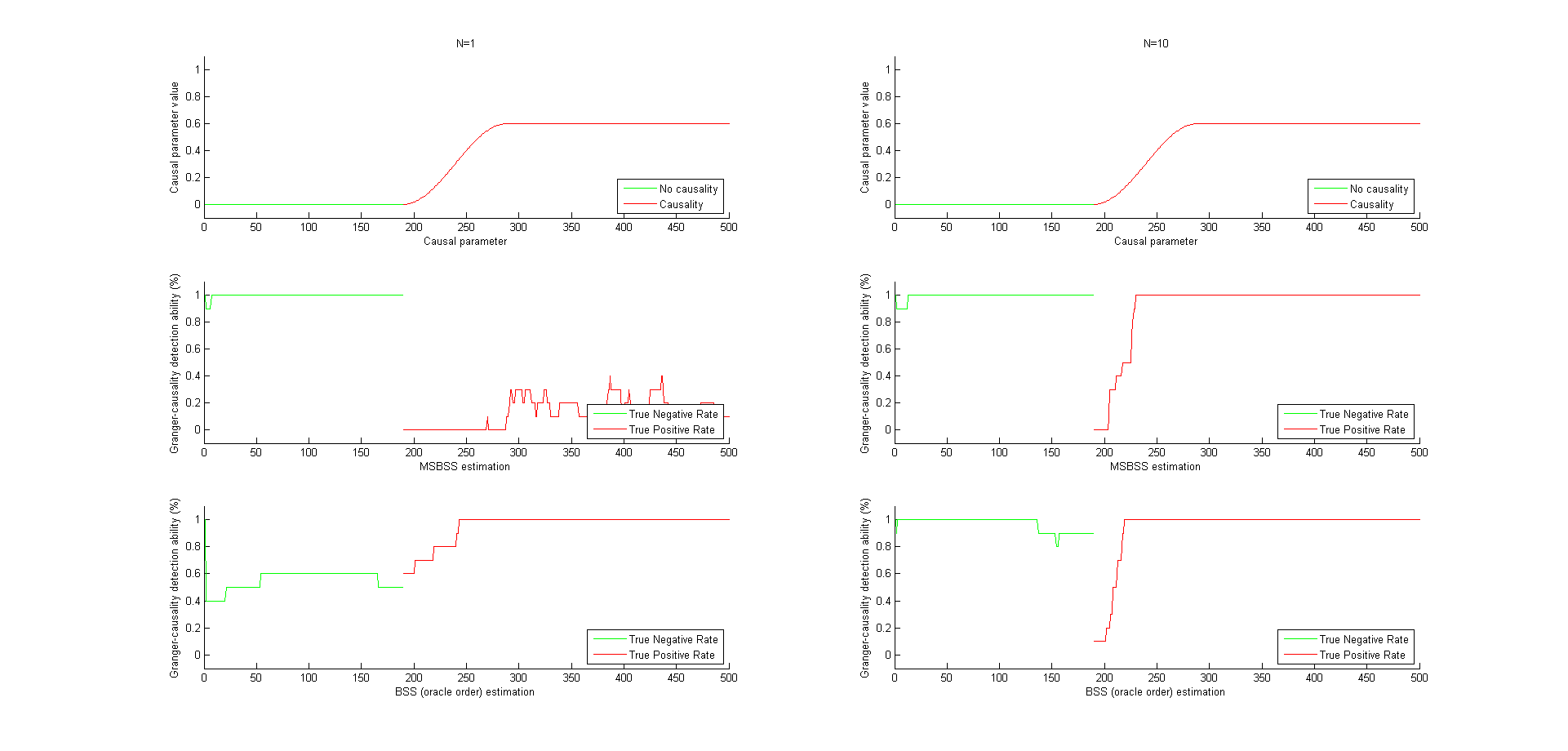}
\caption{Granger-causality detection ability for order $1$, series length $500$ and causal parameter $0.6$.}
\label{GC_1_1_2}
\end{figure}

\begin{figure}[!htb]
\centering
\hspace*{-1in}
\includegraphics[width=6.4in]{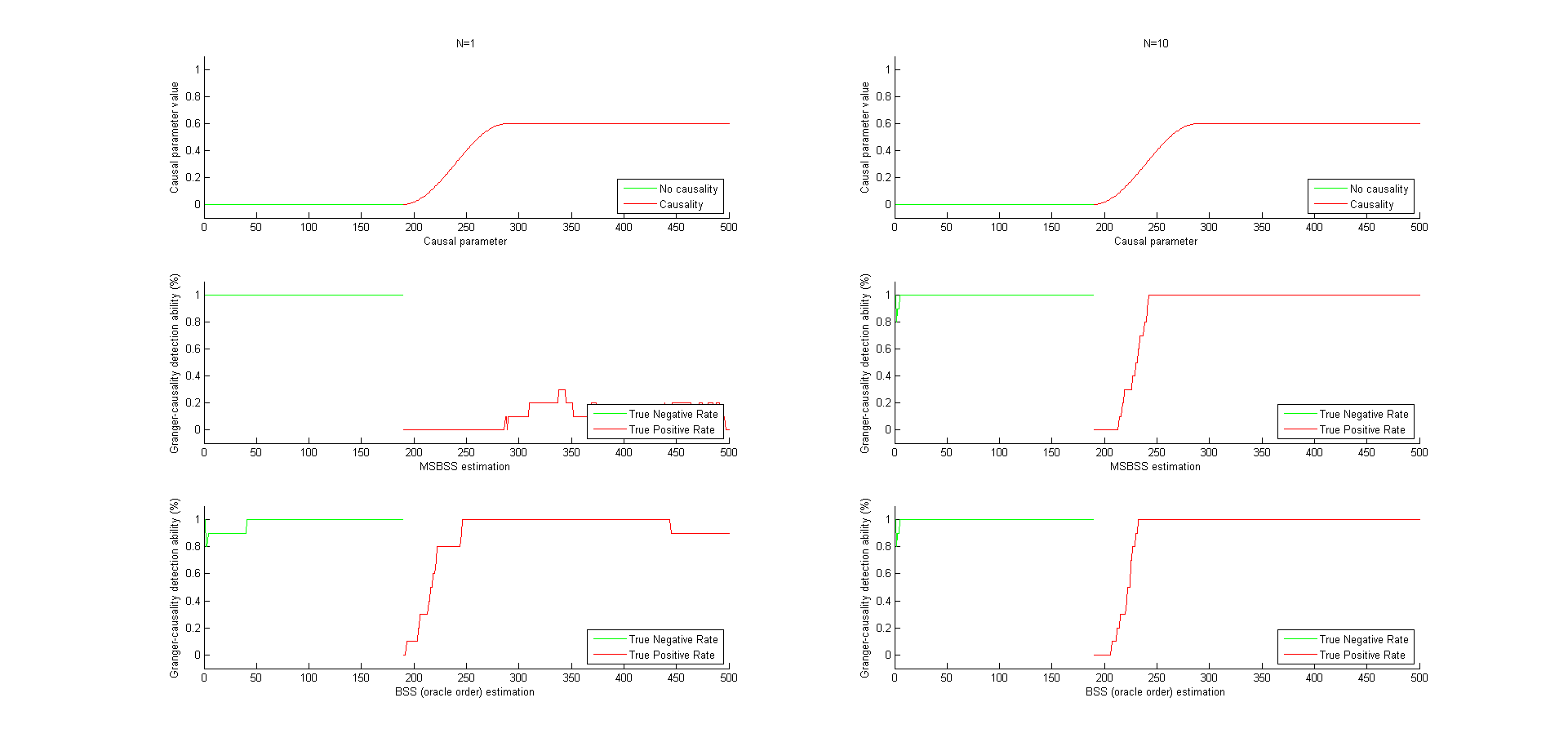}
\caption{Granger-causality detection ability for order $2$, series length $500$ and causal parameter $0.6$.}
\label{GC_2_1_2}
\end{figure}

\begin{figure}[!htb]
\centering
\hspace*{-1in}
\includegraphics[width=6.4in]{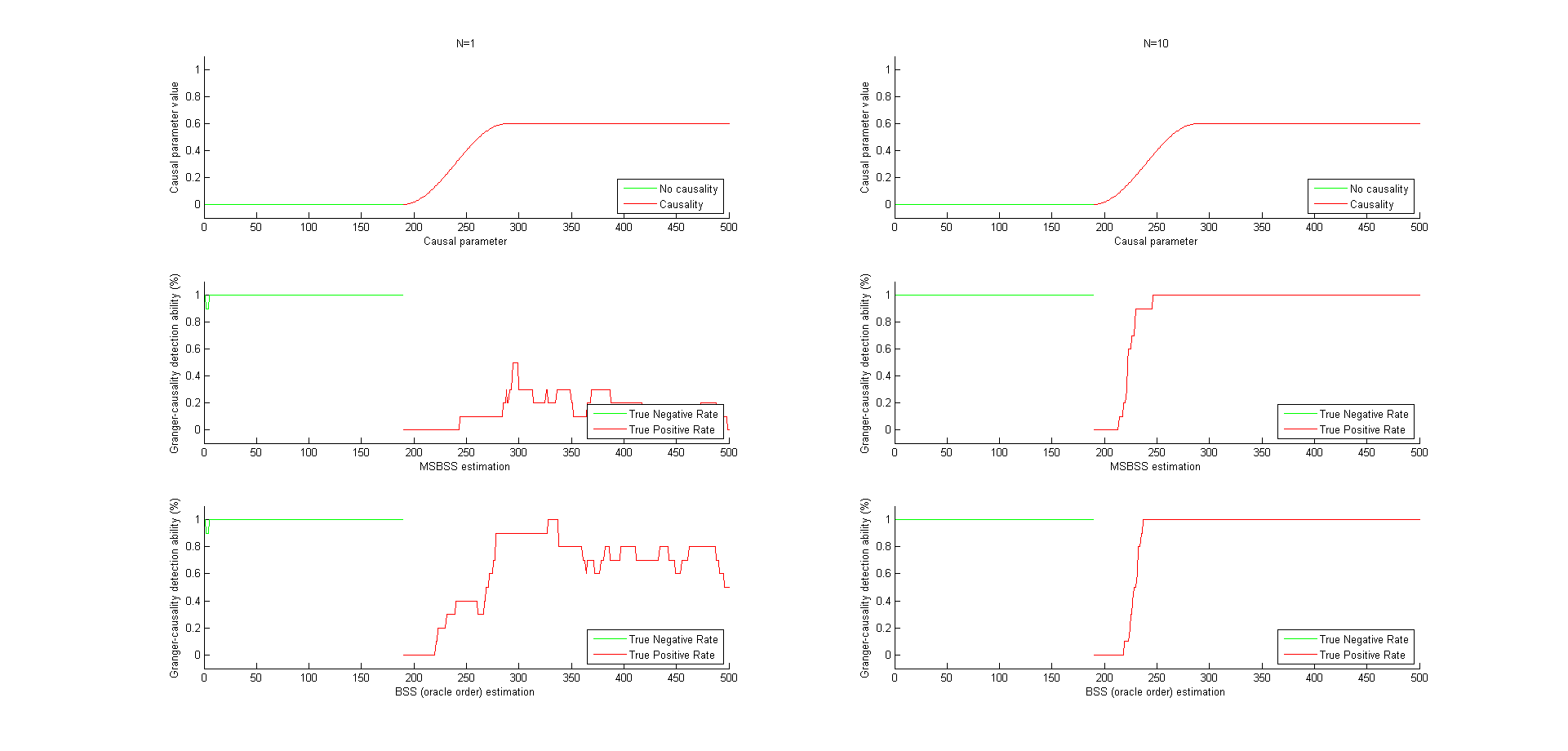}
\caption{Granger-causality detection ability for order $4$, series length $500$ and causal parameter $0.6$.}
\label{GC_3_1_2}
\end{figure}

\begin{figure}[!htb]
\centering
\hspace*{-1in}
\includegraphics[width=6.4in]{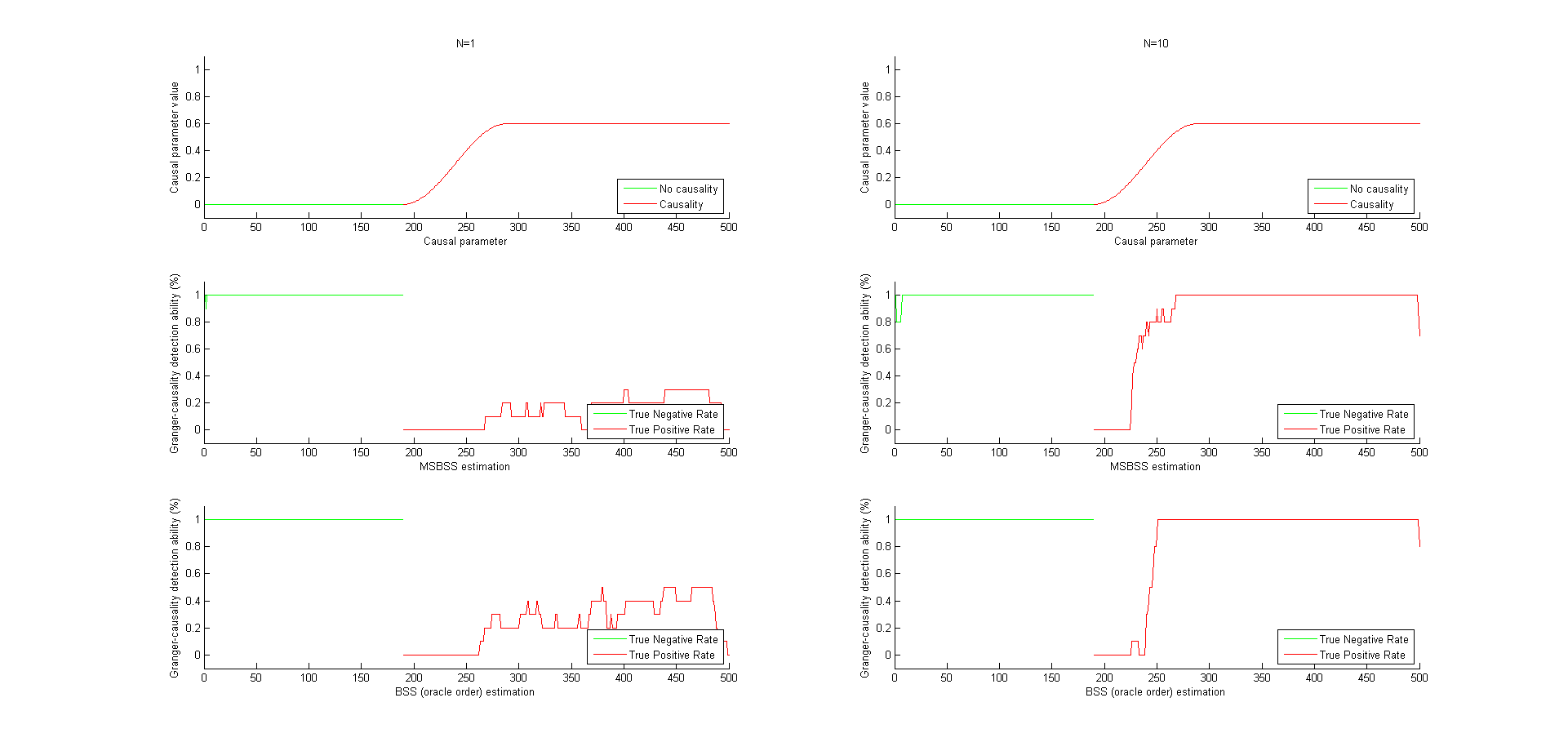}
\caption{Granger-causality detection ability for order $8$, series length $500$ and causal parameter $0.6$.}
\label{GC_4_1_2}
\end{figure}

\begin{figure}[!htb]
\centering
\hspace*{-1in}
\includegraphics[width=6.4in]{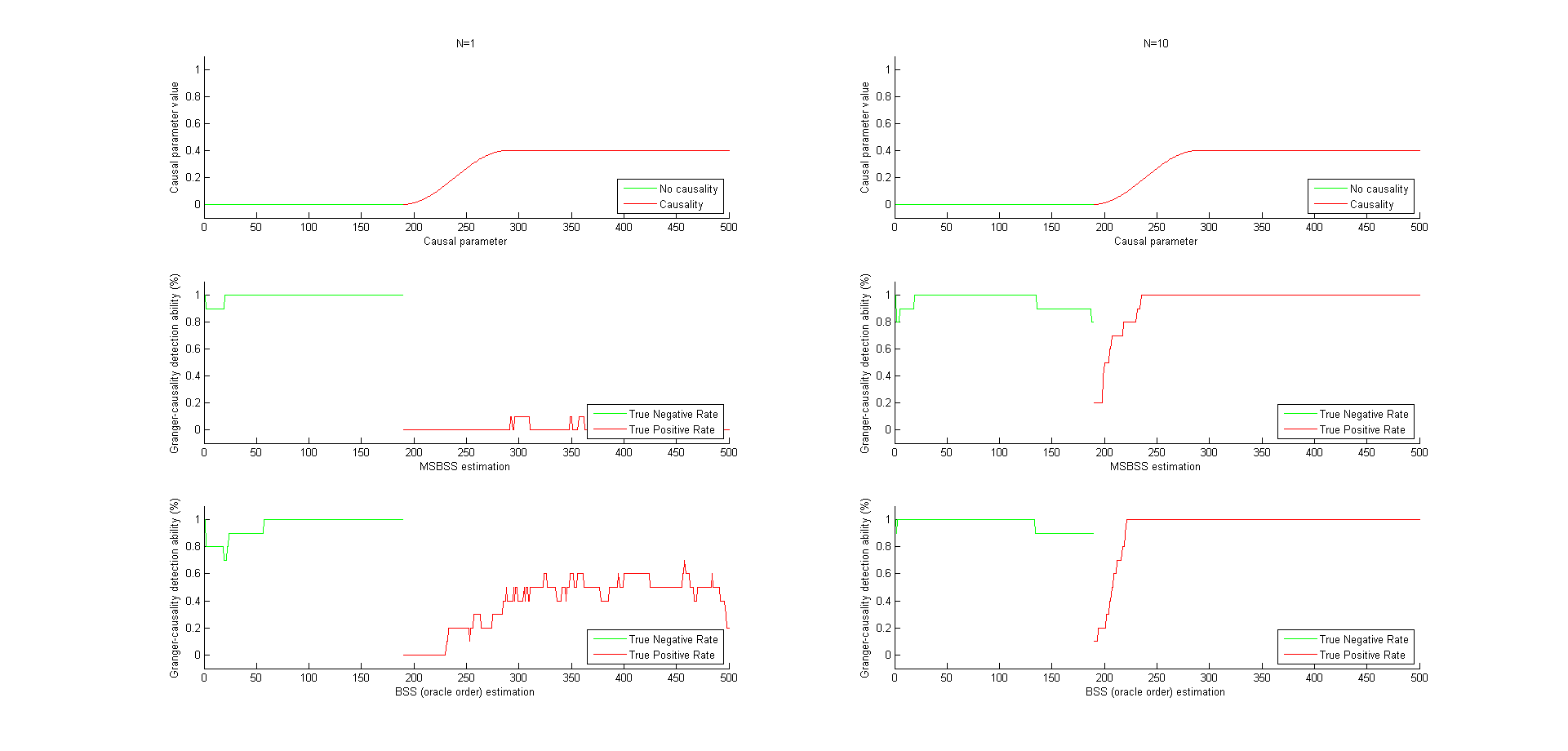}
\caption{Granger-causality detection ability for order $1$, series length $500$ and causal parameter $0.4$.}
\label{GC_1_1_3}
\end{figure}

\begin{figure}[!htb]
\centering
\hspace*{-1in}
\includegraphics[width=6.4in]{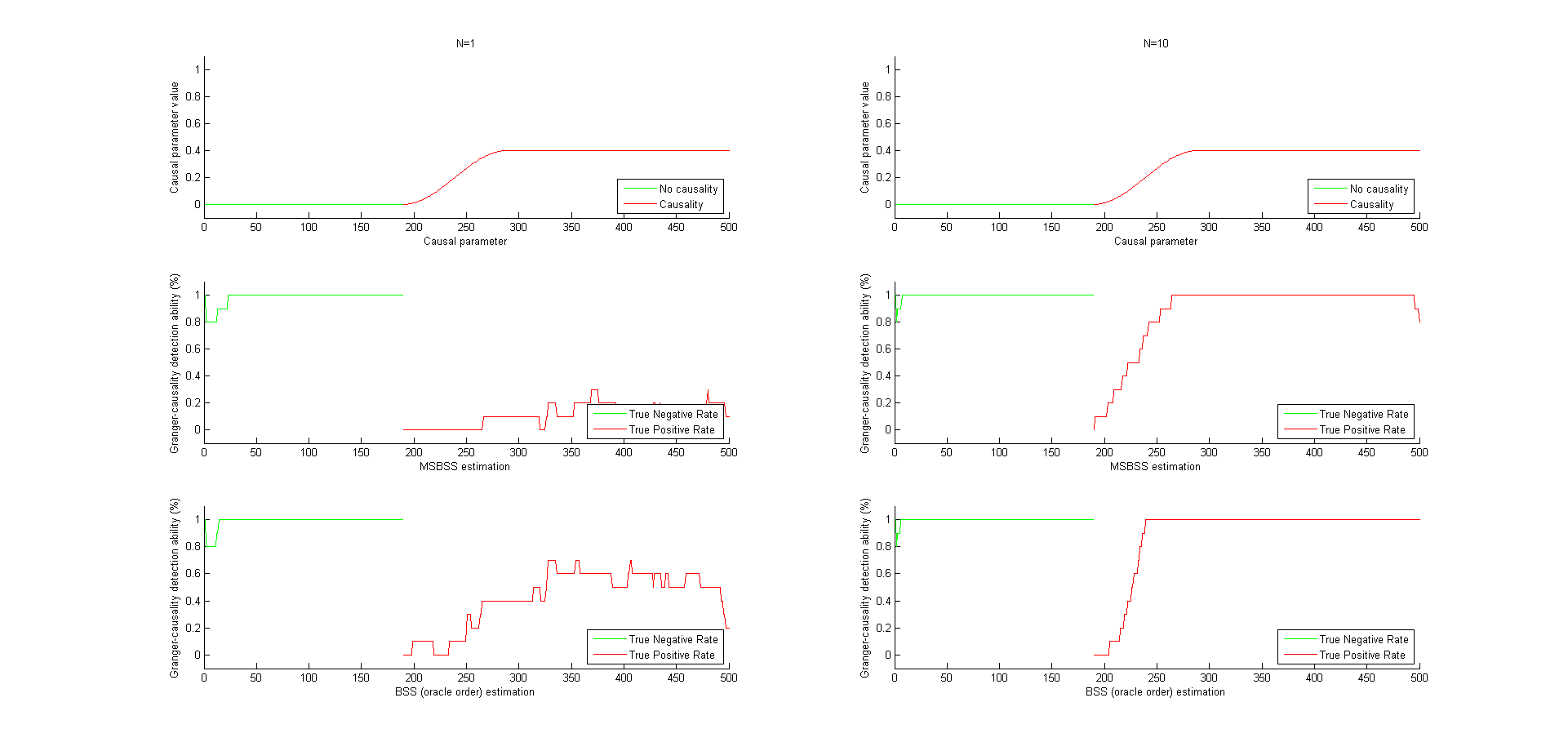}
\caption{Granger-causality detection ability for order $2$, series length $500$ and causal parameter $0.4$.}
\label{GC_2_1_3}
\end{figure}

\begin{figure}[!htb]
\centering
\hspace*{-1in}
\includegraphics[width=6.4in]{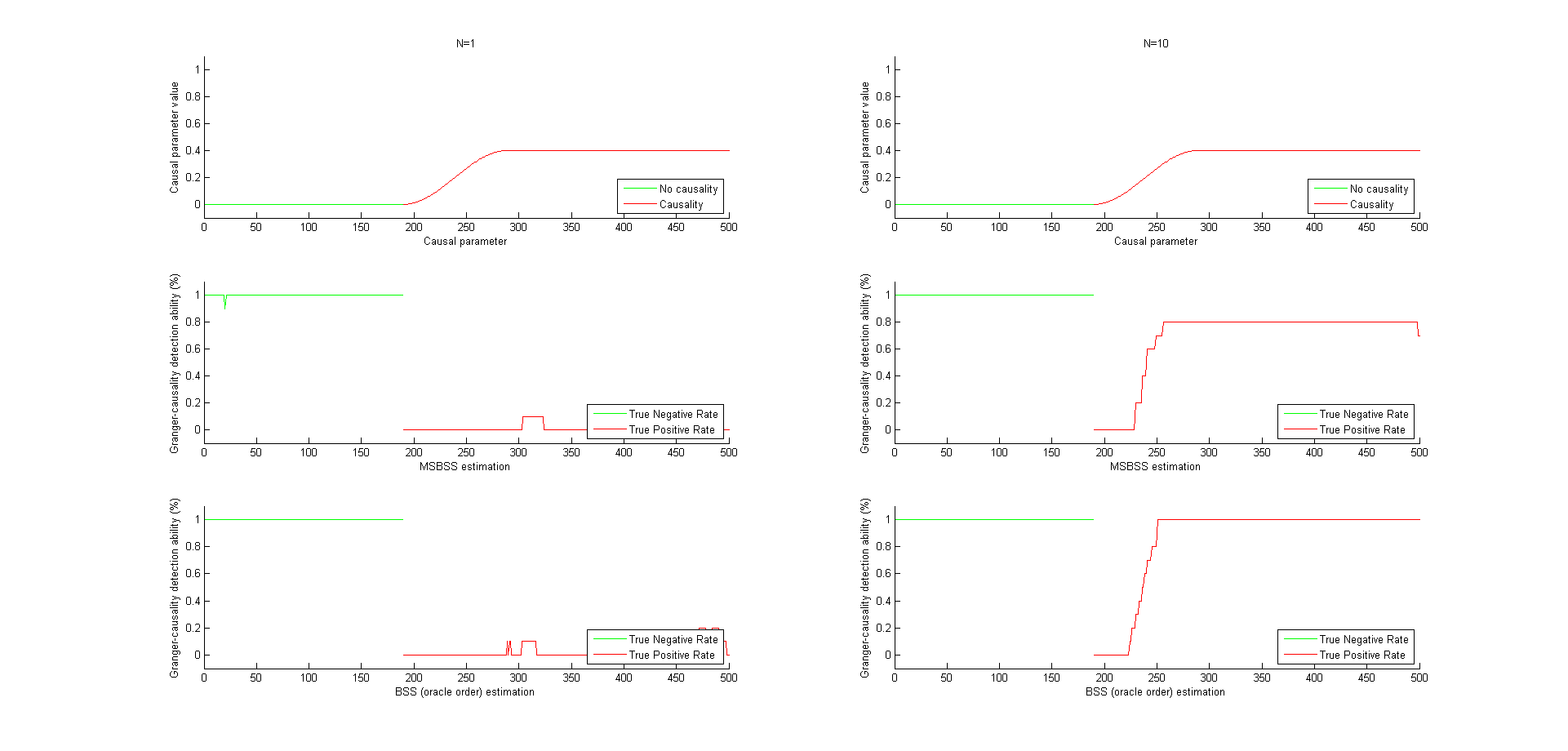}
\caption{Granger-causality detection ability for order $4$, series length $500$ and causal parameter $0.4$.}
\label{GC_3_1_3}
\end{figure}

\begin{figure}[!htb]
\centering
\hspace*{-1in}
\includegraphics[width=6.4in]{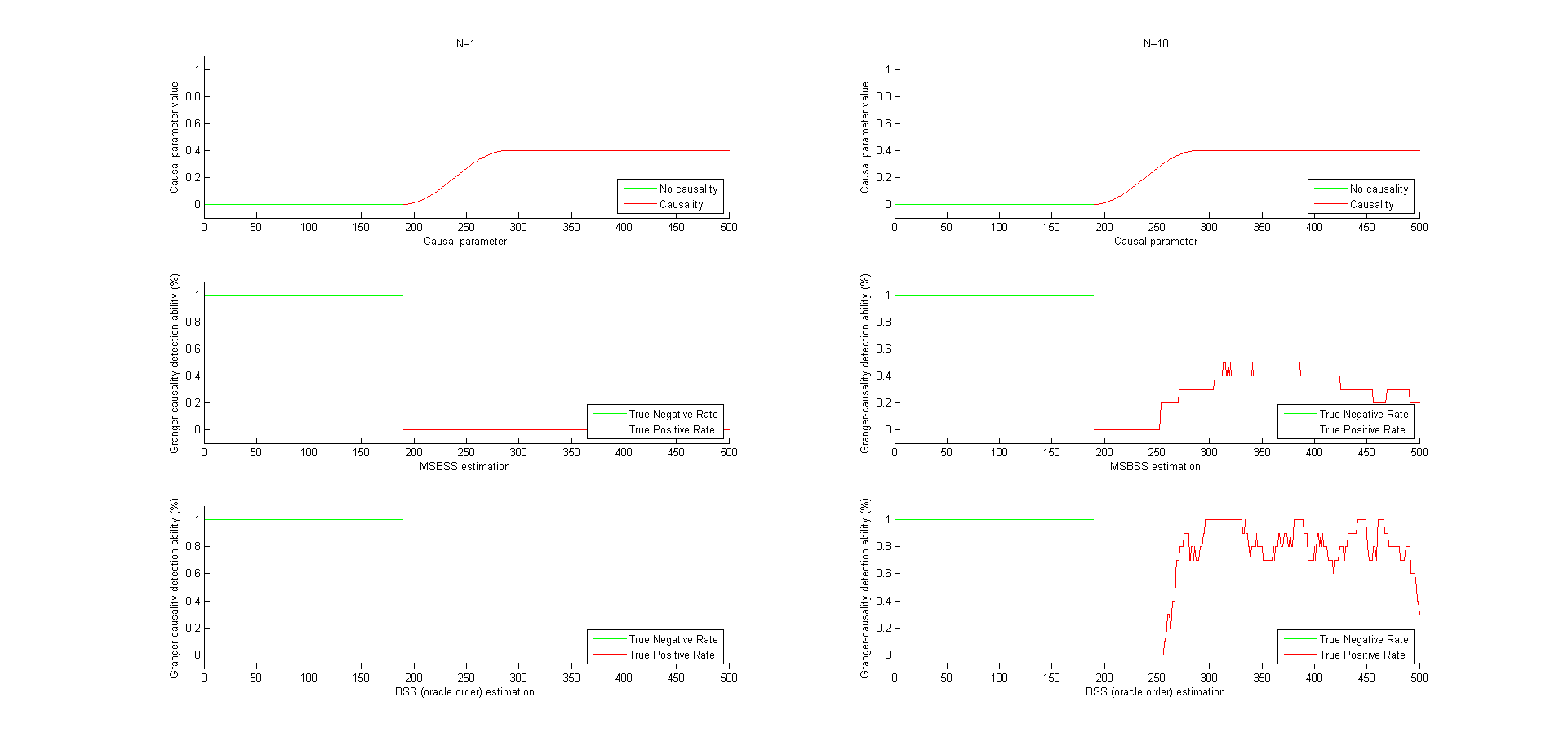}
\caption{Granger-causality detection ability for order $8$, series length $500$ and causal parameter $0.4$.}
\label{GC_4_1_3}
\end{figure}

\begin{figure}[!htb]
\centering
\hspace*{-1in}
\includegraphics[width=6.4in]{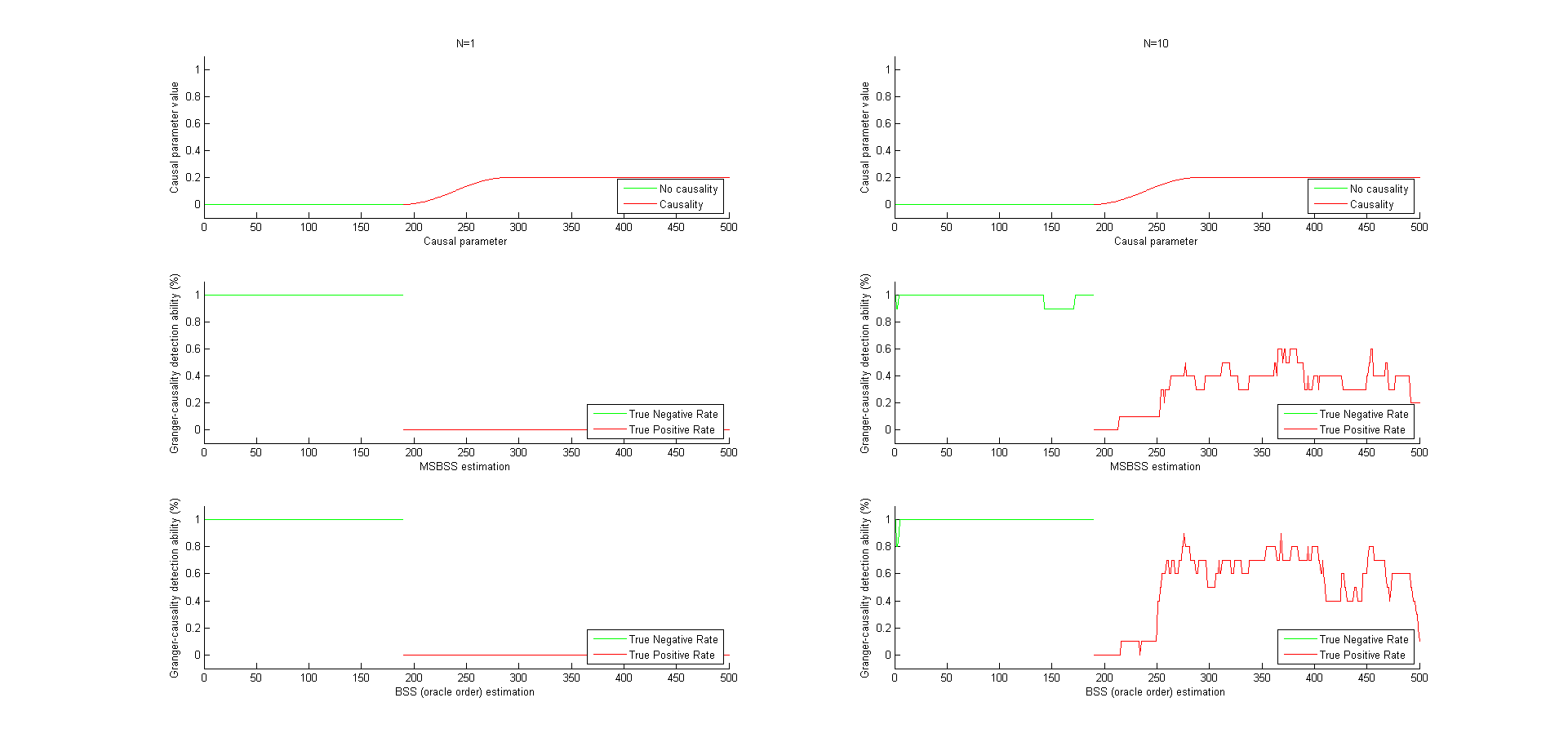}
\caption{Granger-causality detection ability for order $2$, series length $500$ and causal parameter $0.2$.}
\label{GC_2_1_4}
\end{figure}

\begin{figure}[!htb]
\centering
\hspace*{-1in}
\includegraphics[width=6.4in]{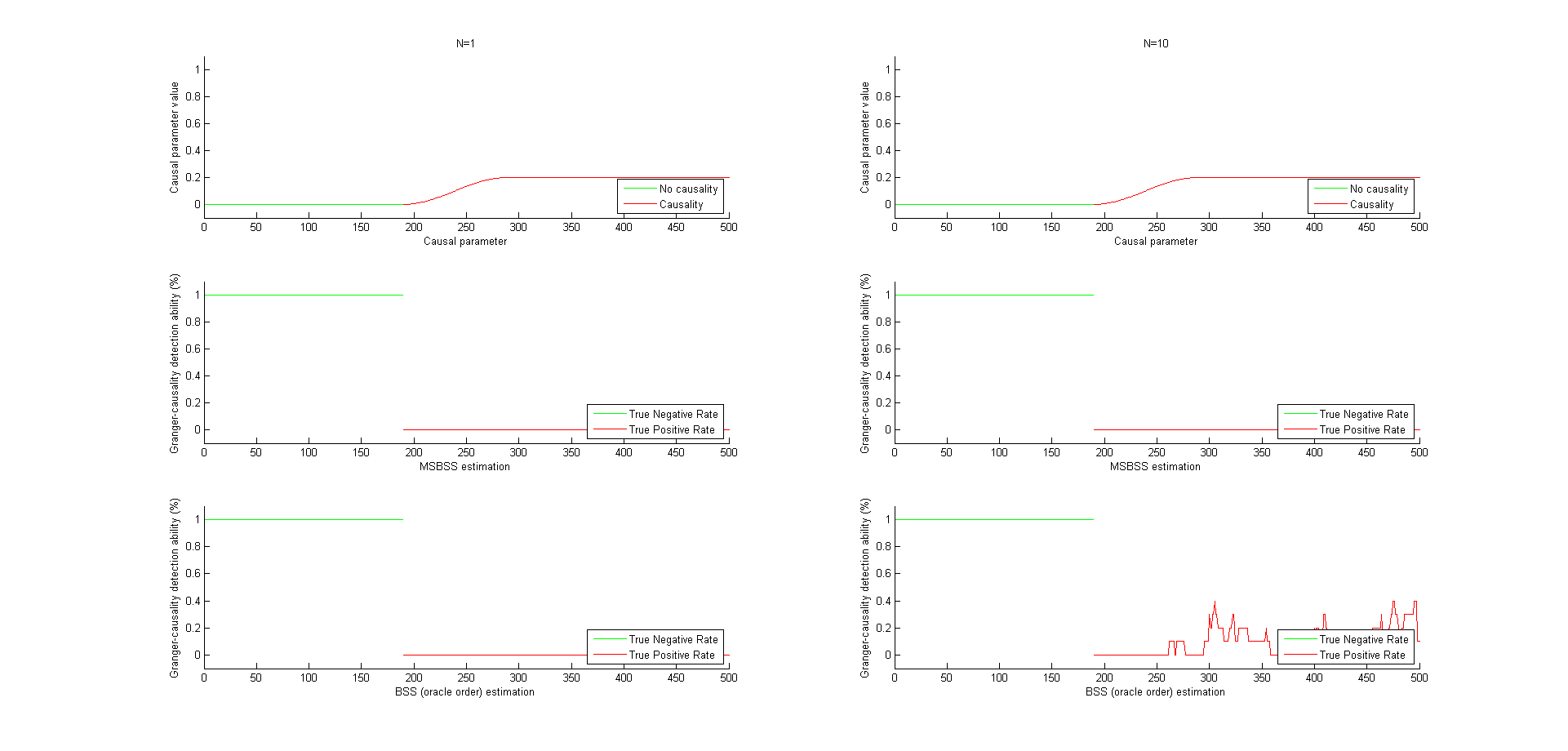}
\caption{Granger-causality detection ability for order $4$, series length $500$ and causal parameter $0.2$.}
\label{GC_3_1_4}
\end{figure}

\begin{figure}[!htb]
\centering
\hspace*{-1in}
\includegraphics[width=6.4in]{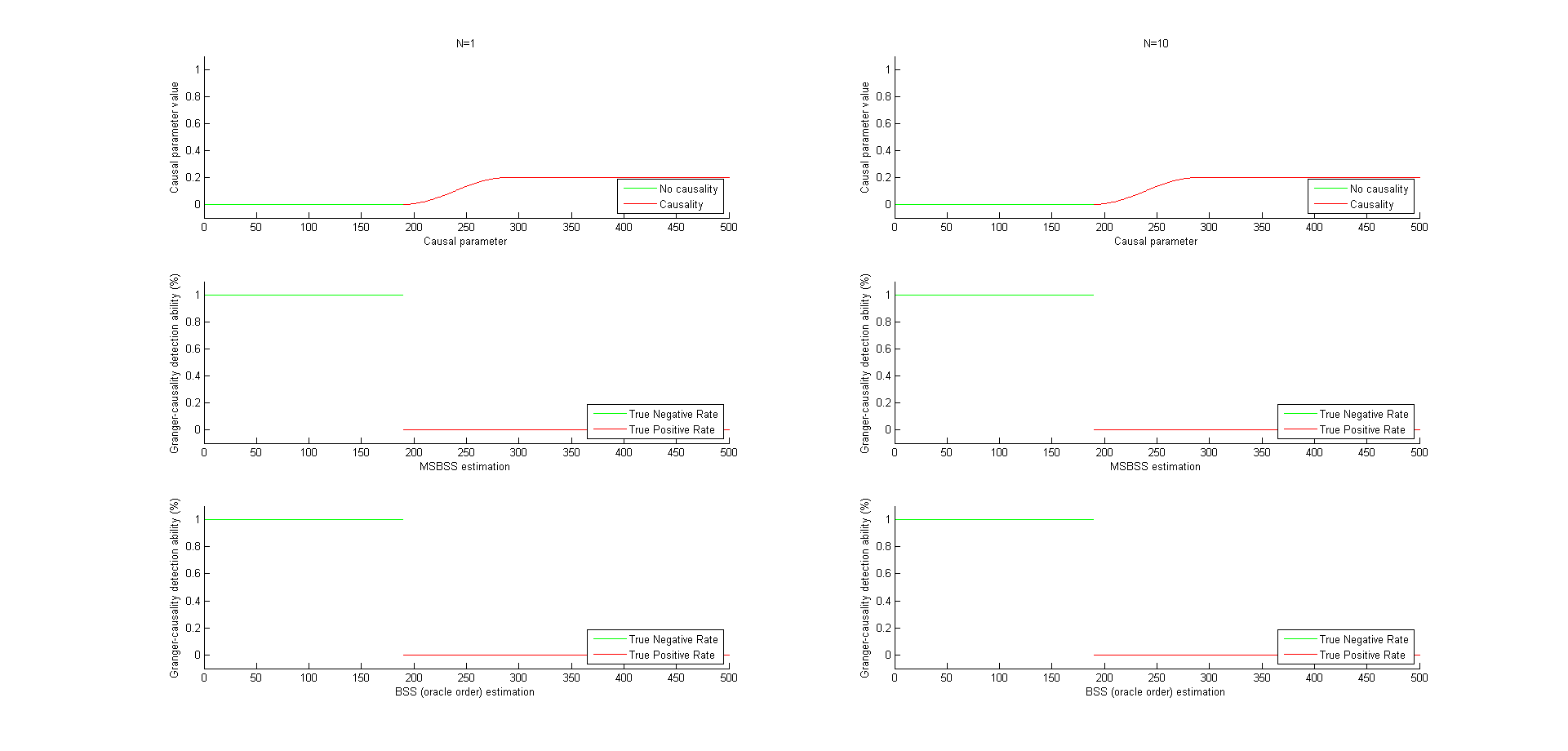}
\caption{Granger-causality detection ability for order $8$, series length $500$ and causal parameter $0.2$.}
\label{GC_4_1_4}
\end{figure}

\clearpage

\subsection{Data generated with slowly-varying parameters and non-normal errors}
\label{Data generated with slowly varying parameters and non-normal errors}

\begin{figure}[!htb]
\centering
\hspace*{-1in}
\includegraphics[width=6.4in]{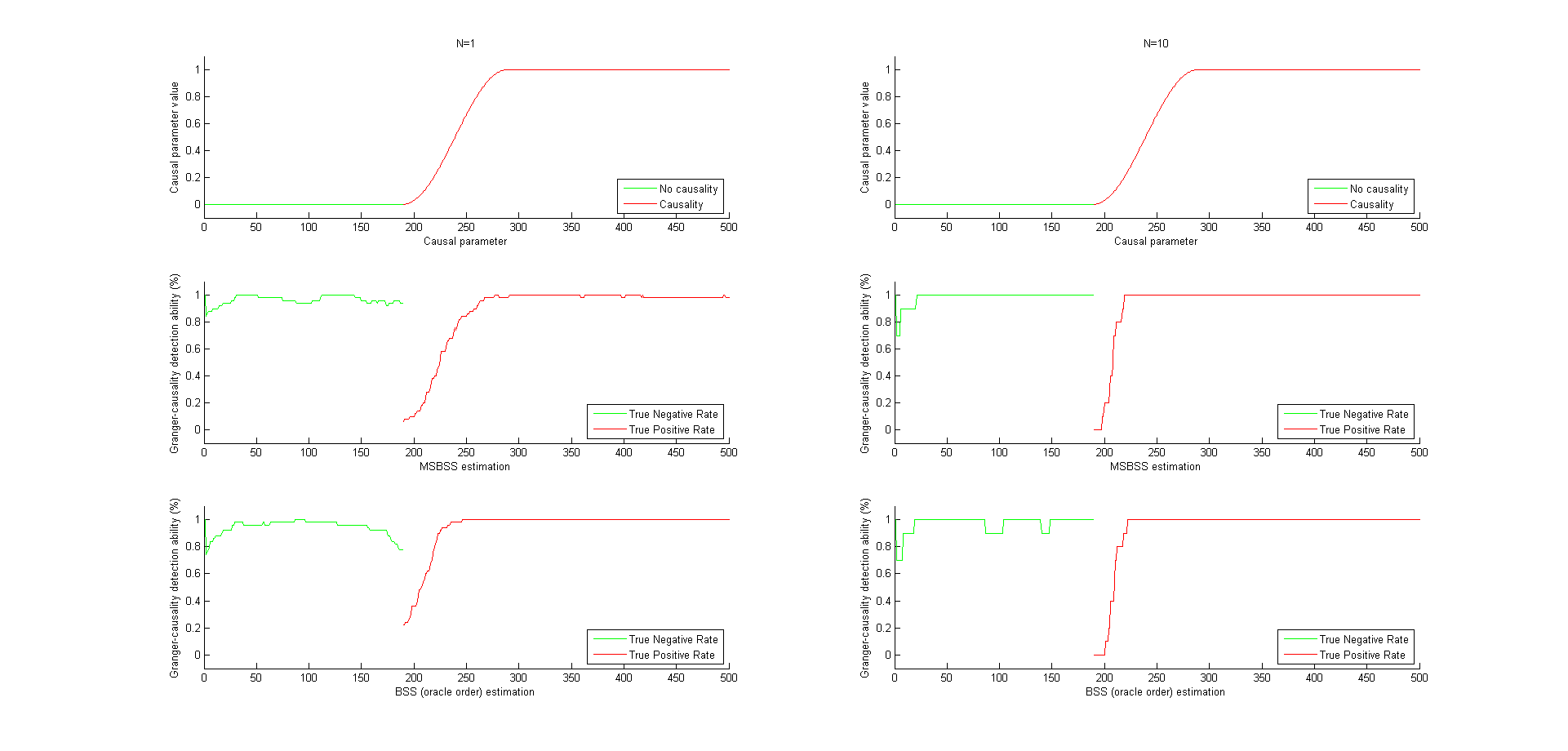}
\caption{Granger-causality detection ability for a non-normal error simulation of order $1$, series length $500$ and causal parameter $1$.}
\label{T_1_1_1}
\end{figure}

\begin{figure}[!htb]
\centering
\hspace*{-1in}
\includegraphics[width=6.4in]{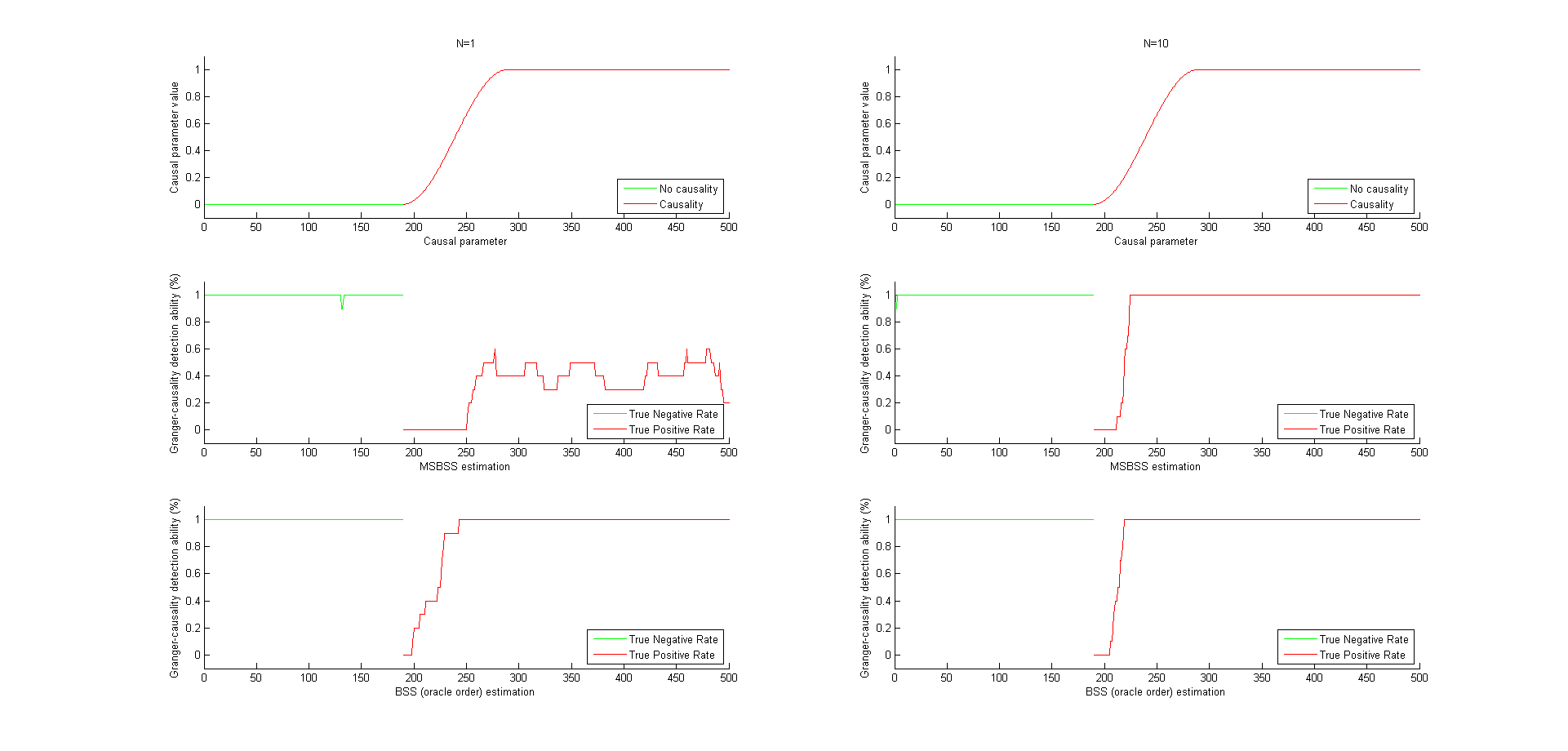}
\caption{Granger-causality detection ability for a non-normal error simulation of order $2$ and series length $500$ and causal parameter $1$.}
\label{T_2_1_1}
\end{figure}

\begin{figure}[!htb]
\centering
\hspace*{-1in}
\includegraphics[width=6.4in]{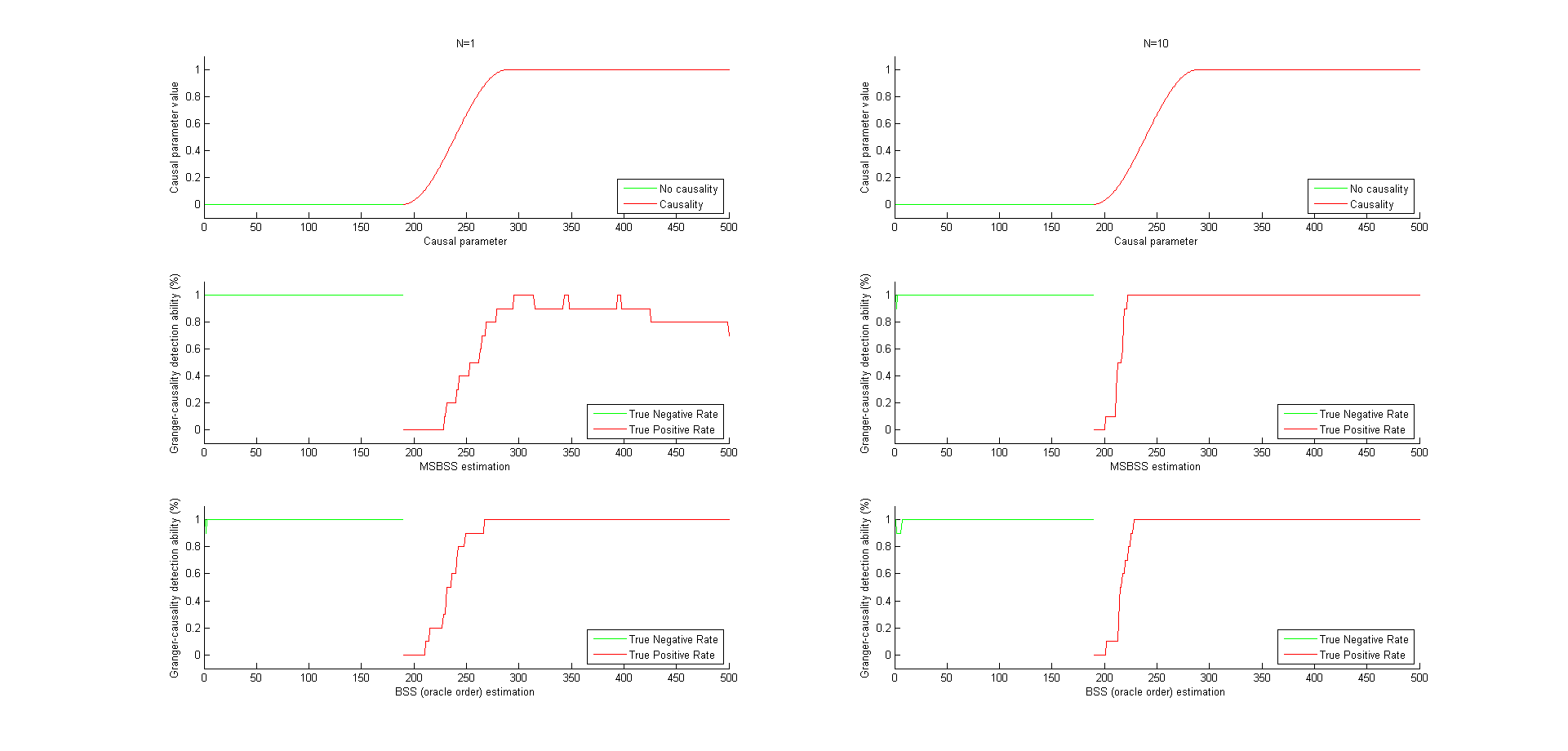}
\caption{Granger-causality detection ability for a non-normal error simulation of order $4$ and series length $500$ and causal parameter $1$.}
\label{T_3_1_1}
\end{figure}

\begin{figure}[!htb]
\centering
\hspace*{-1in}
\includegraphics[width=6.4in]{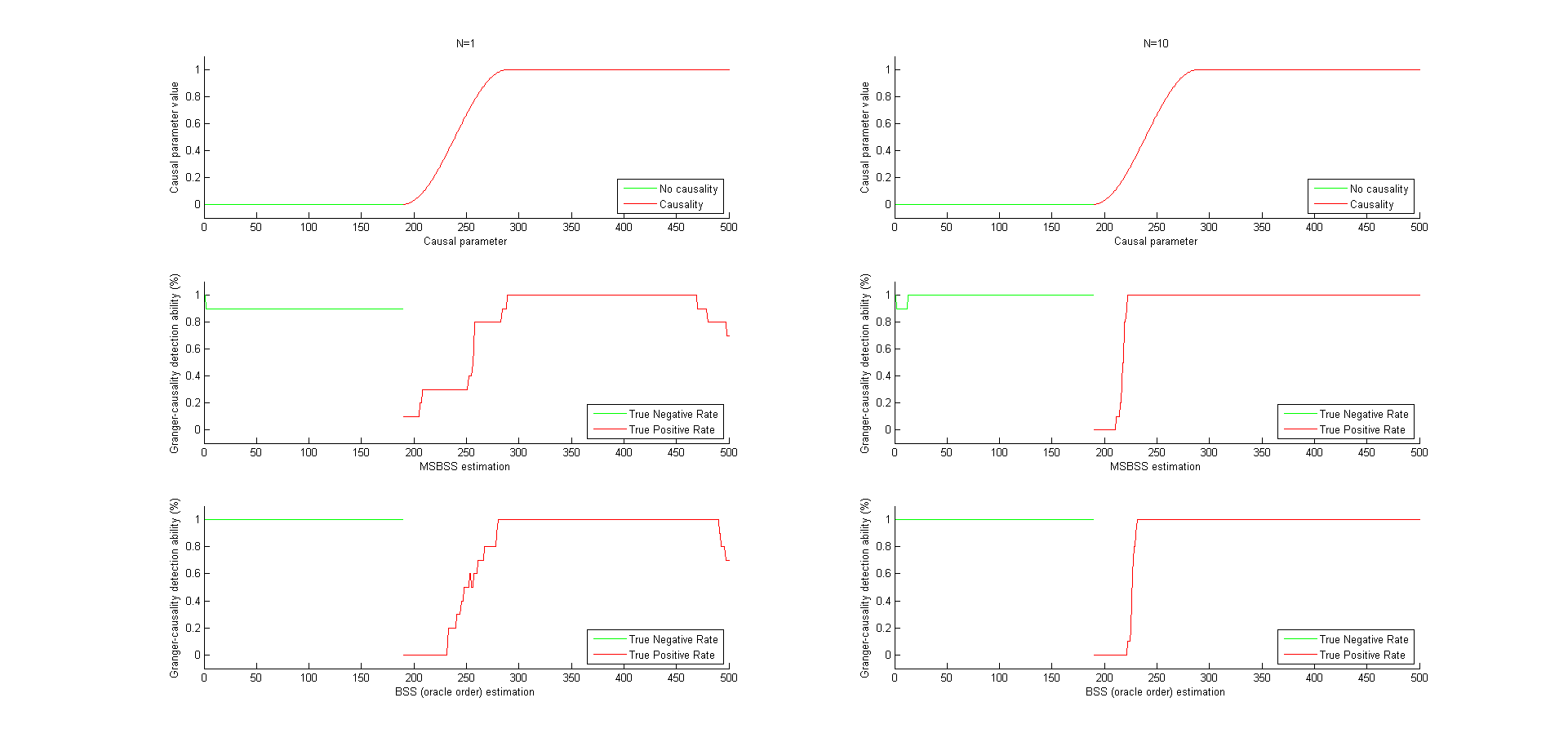}
\caption{Granger-causality detection ability for a non-normal error simulation of order $8$ and series length $500$ and causal parameter $1$.}
\label{T_4_1_1}
\end{figure}

\begin{figure}[!htb]
\centering
\hspace*{-1in}
\includegraphics[width=6.4in]{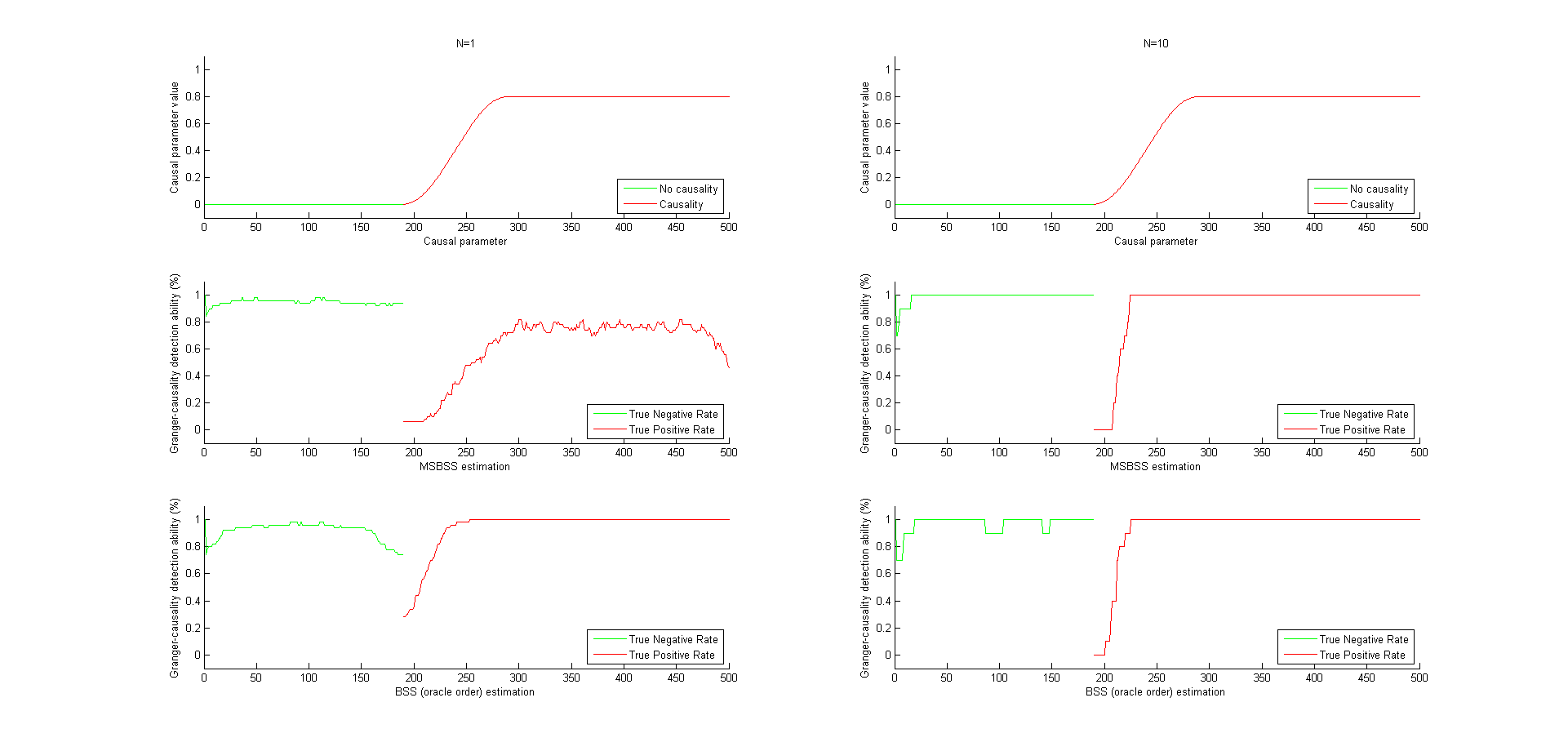}
\caption{Granger-causality detection ability for a non-normal error simulation of order $1$ and series length $500$ and causal parameter $0.8$.}
\label{T_1_1_2}
\end{figure}

\begin{figure}[!htb]
\centering
\hspace*{-1in}
\includegraphics[width=6.4in]{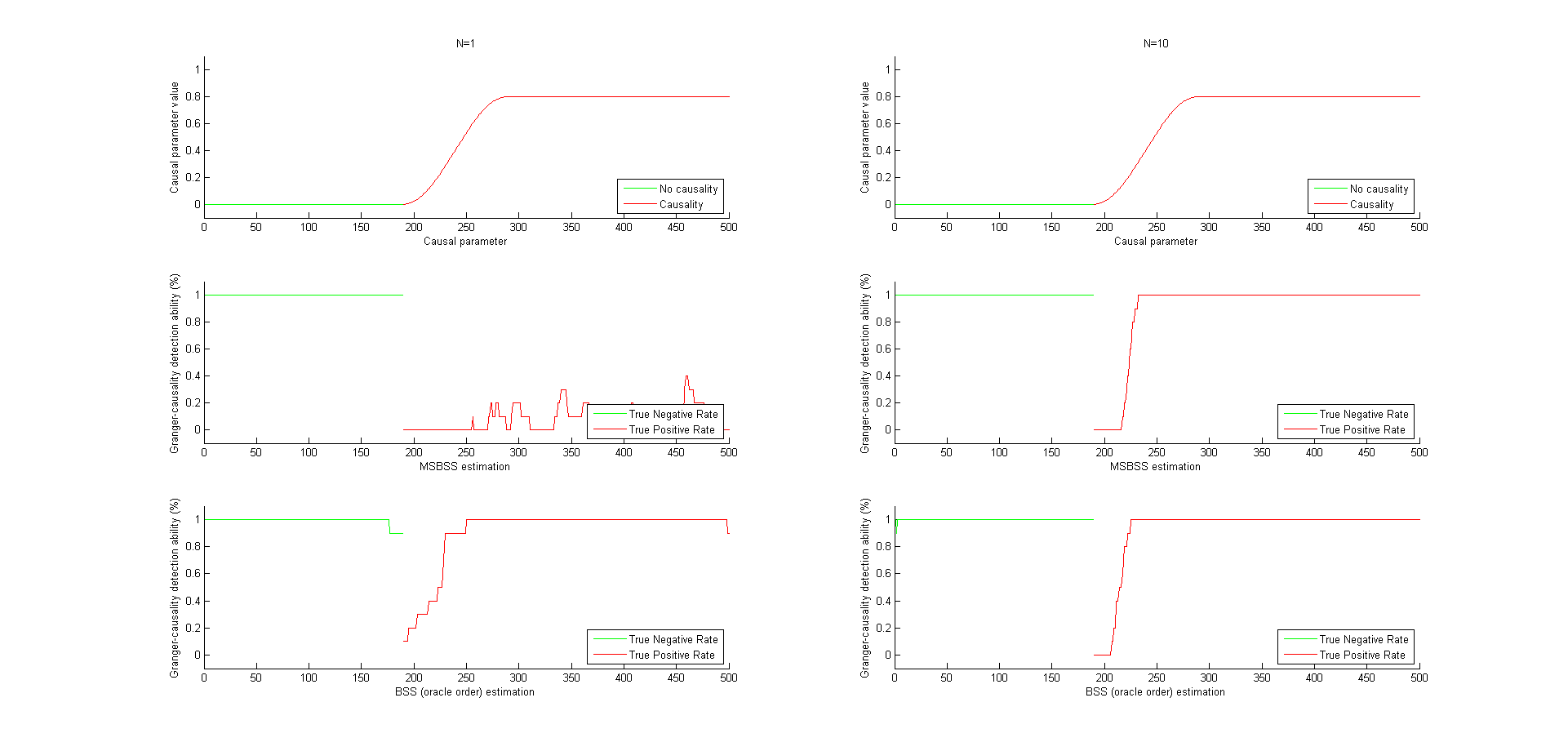}
\caption{Granger-causality detection ability for a non-normal error simulation of order $2$ and series length $500$ and causal parameter $0.8$.}
\label{T_2_1_2}
\end{figure}

\begin{figure}[!htb]
\centering
\hspace*{-1in}
\includegraphics[width=6.4in]{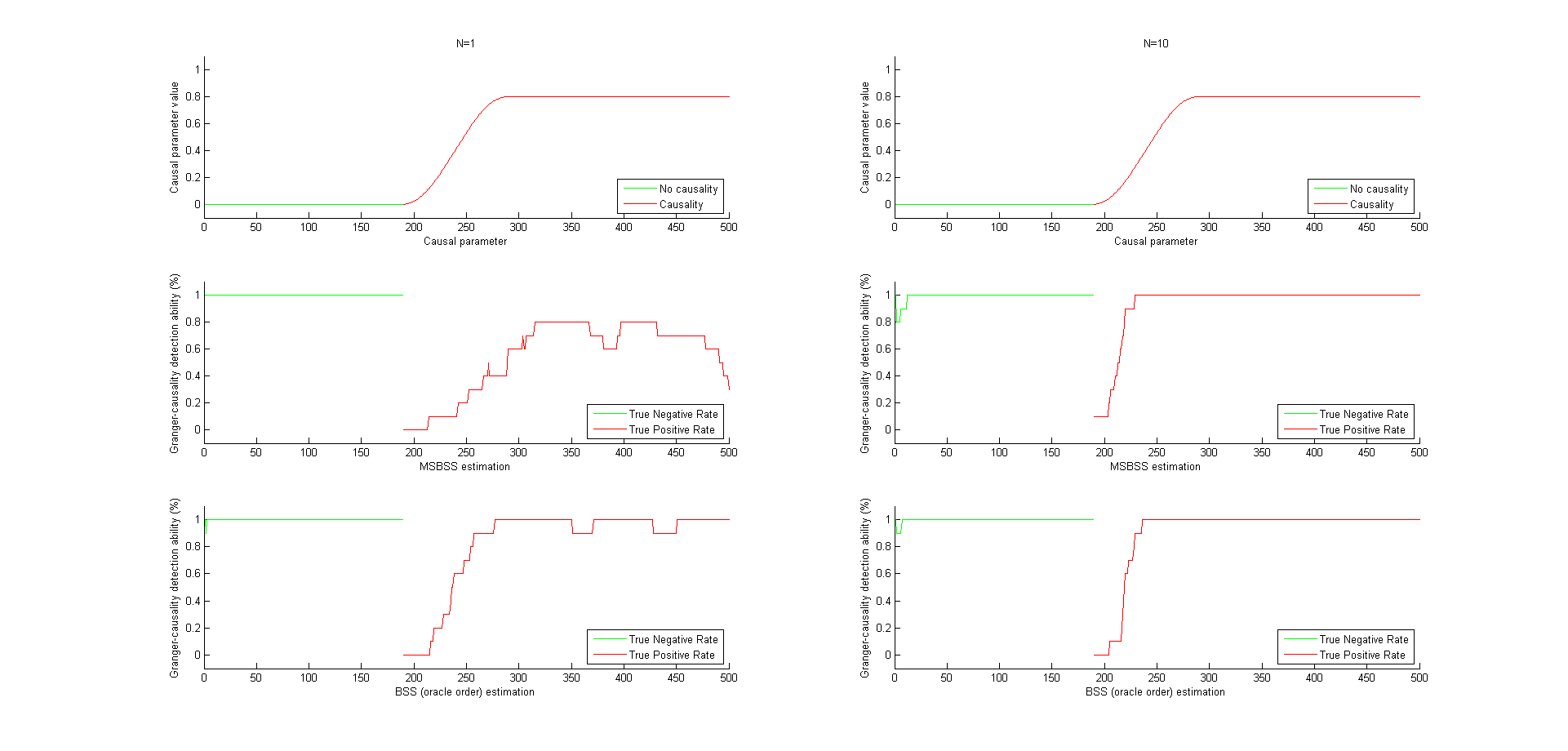}
\caption{Granger-causality detection ability for a non-normal error simulation of order $4$ and series length $500$ and causal parameter $0.8$.}
\label{T_3_1_2}
\end{figure}

\begin{figure}[!htb]
\centering
\hspace*{-1in}
\includegraphics[width=6.4in]{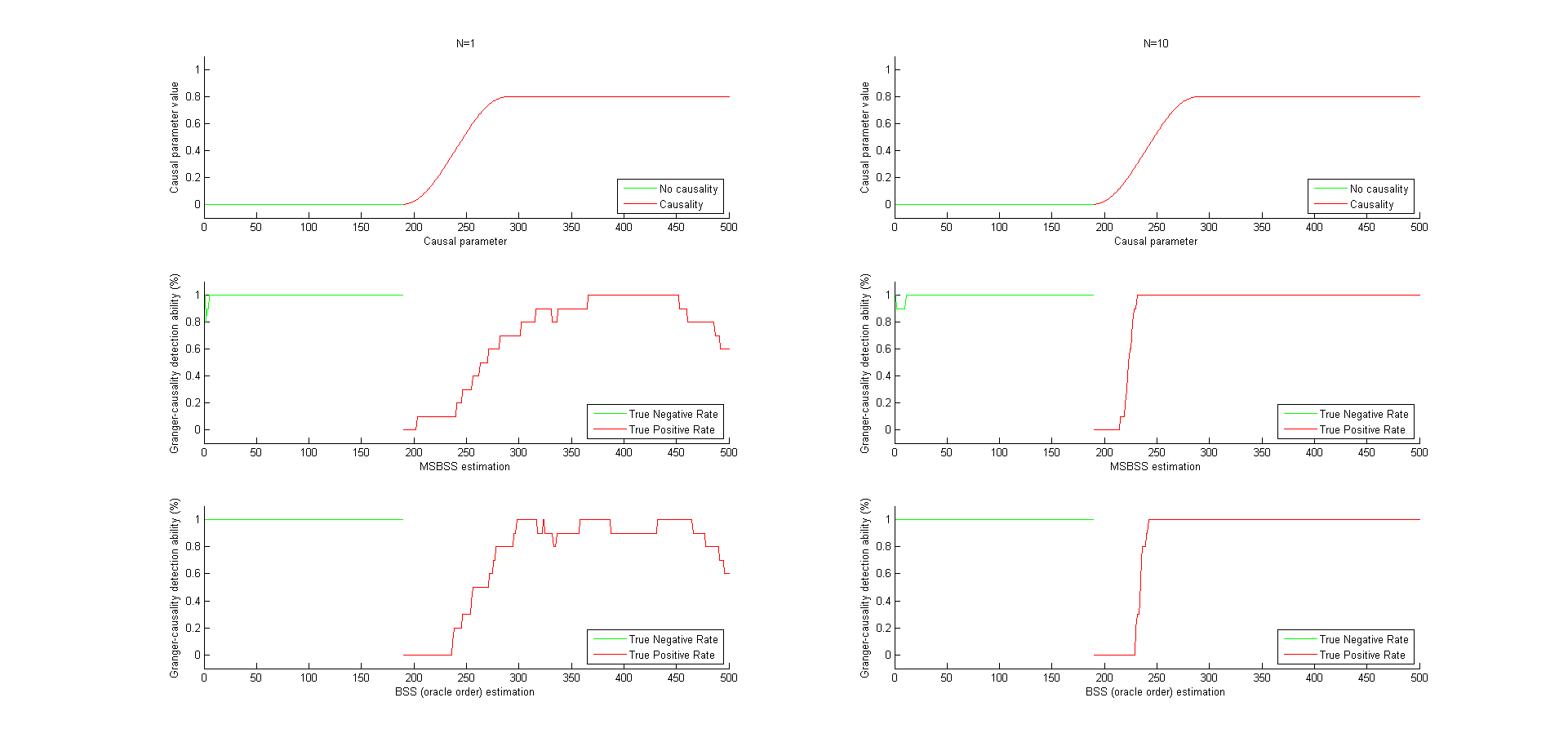}
\caption{Granger-causality detection ability for a non-normal error simulation of order $8$ and series length $500$ and causal parameter $0.8$.}
\label{T_4_1_2}
\end{figure}

\begin{figure}[!htb]
\centering
\hspace*{-1in}
\includegraphics[width=6.4in]{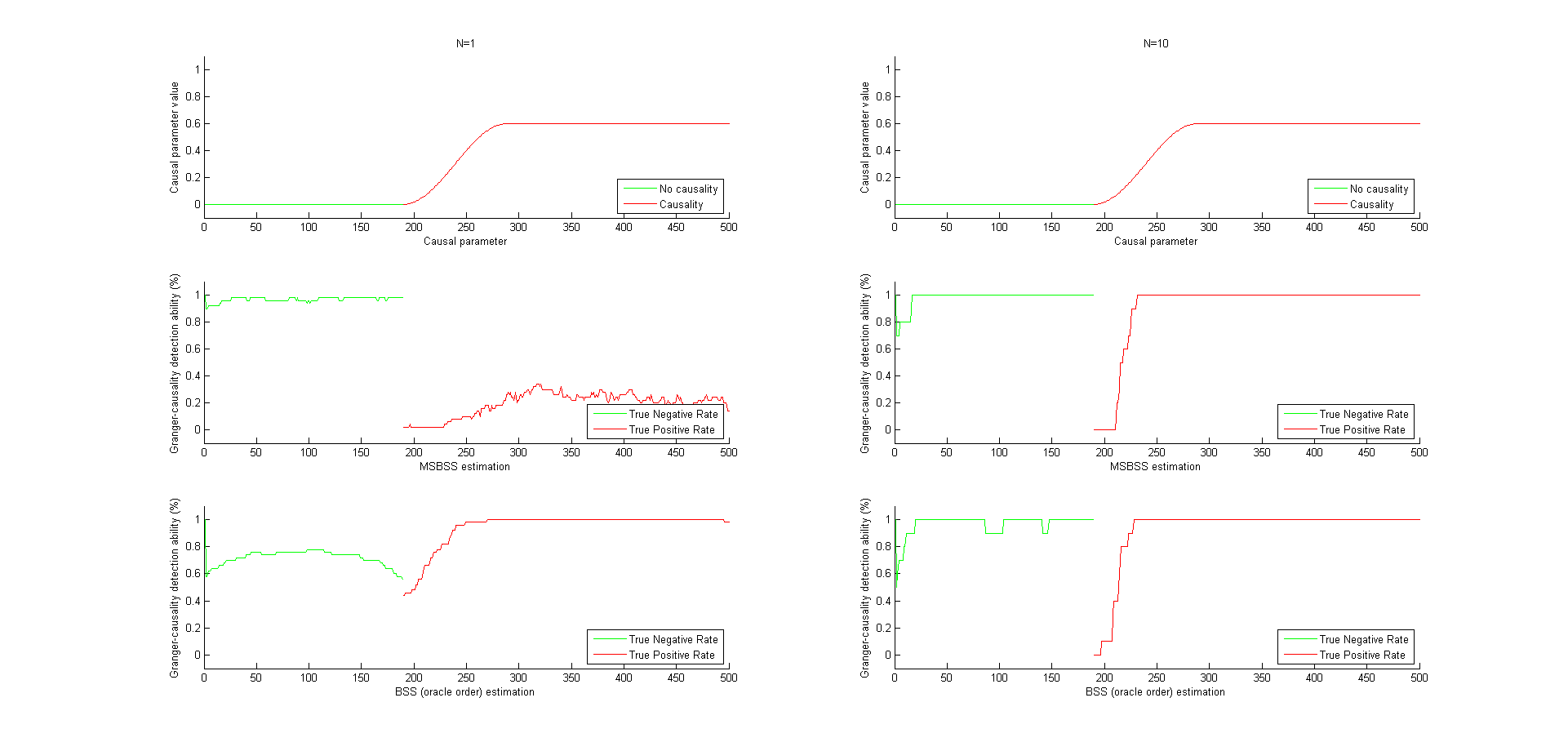}
\caption{Granger-causality detection ability for a non-normal error simulation of order $1$ and series length $500$ and causal parameter $0.6$.}
\label{T_1_1_3}
\end{figure}

\begin{figure}[!htb]
\centering
\hspace*{-1in}
\includegraphics[width=6.4in]{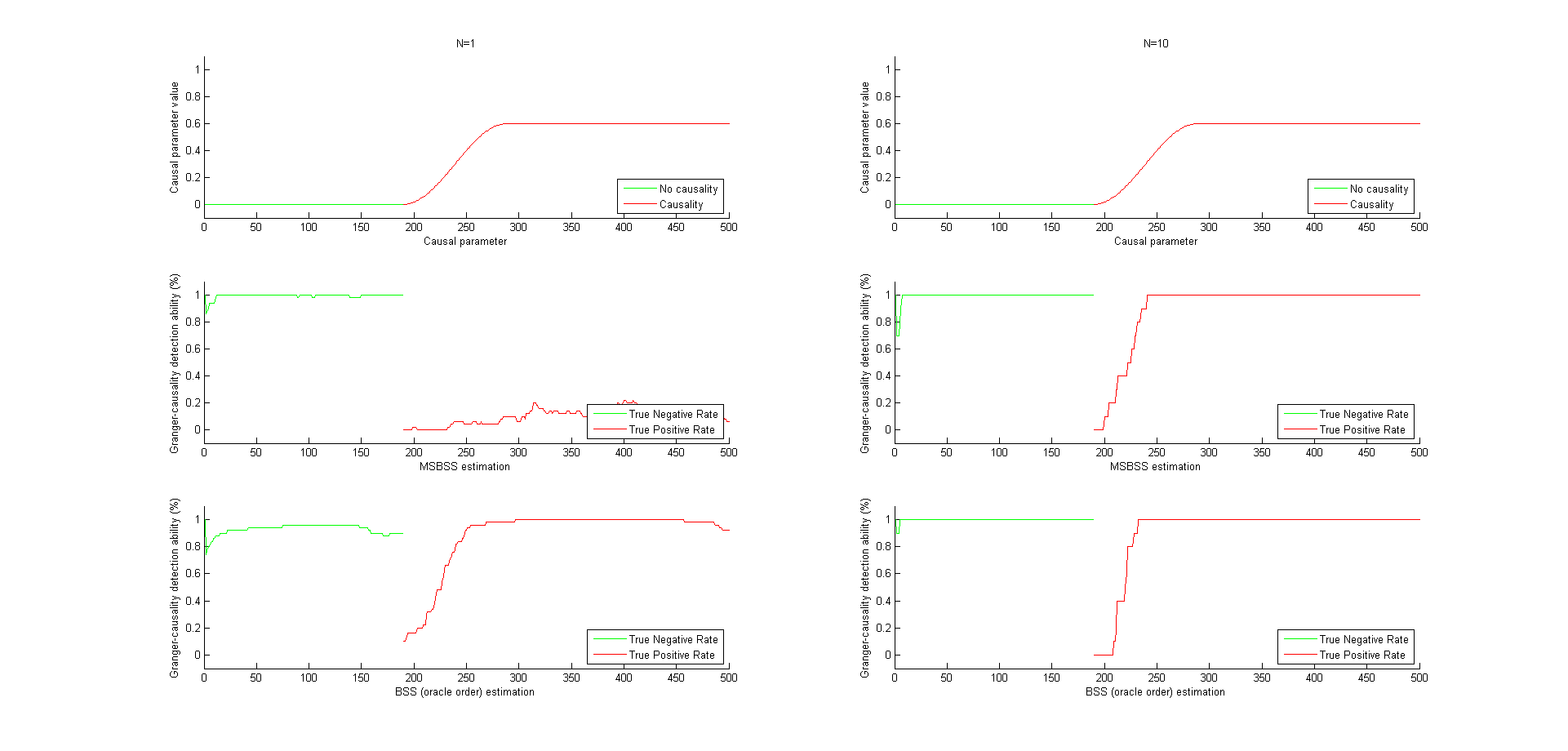}
\caption{Granger-causality detection ability for a non-normal error simulation of order $2$ and series length $500$ and causal parameter $0.6$.}
\label{T_2_1_3}
\end{figure}

\begin{figure}[!htb]
\centering
\hspace*{-1in}
\includegraphics[width=6.4in]{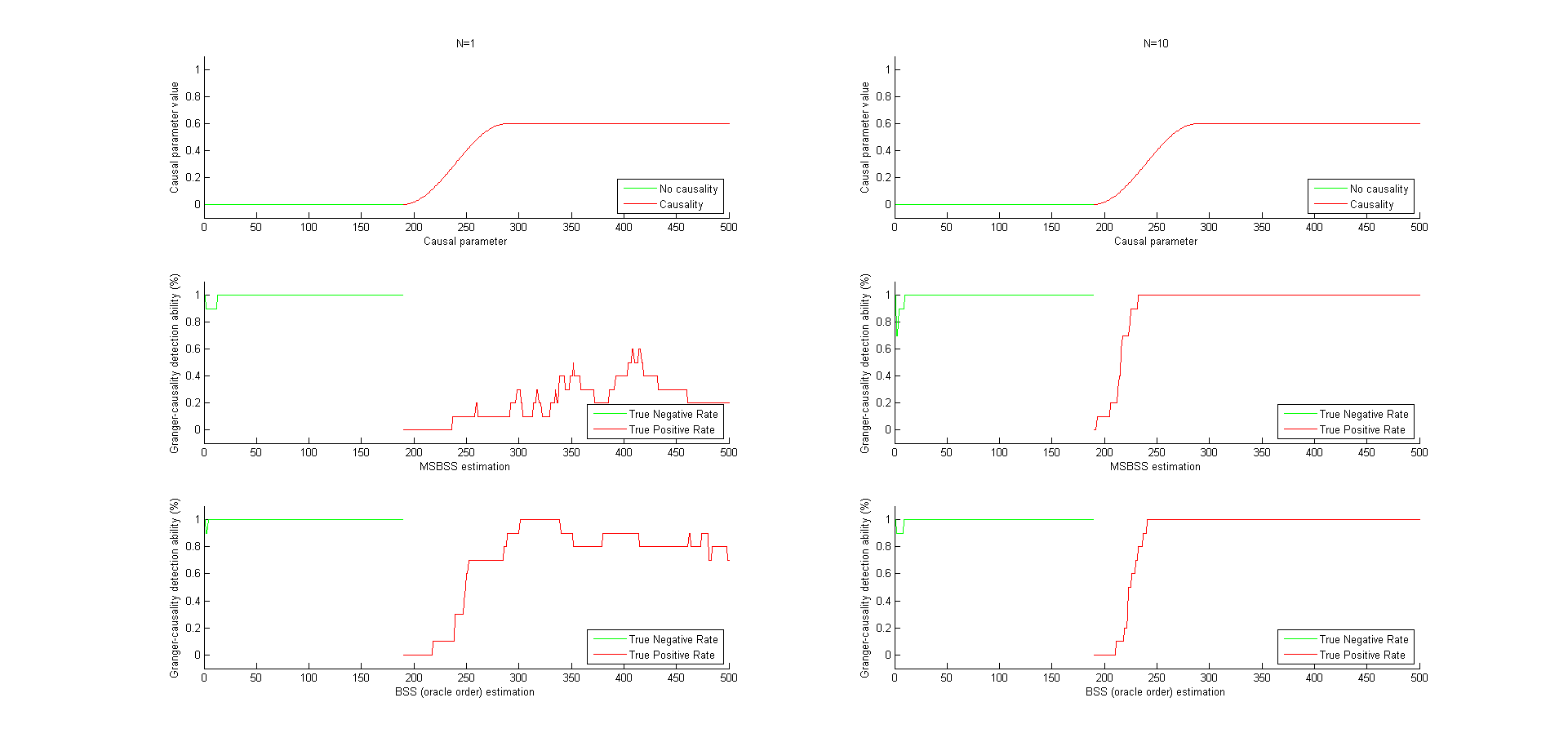}
\caption{Granger-causality detection ability for a non-normal error simulation of order $4$ and series length $500$ and causal parameter $0.6$.}
\label{T_3_1_3}
\end{figure}

\begin{figure}[!htb]
\centering
\hspace*{-1in}
\includegraphics[width=6.4in]{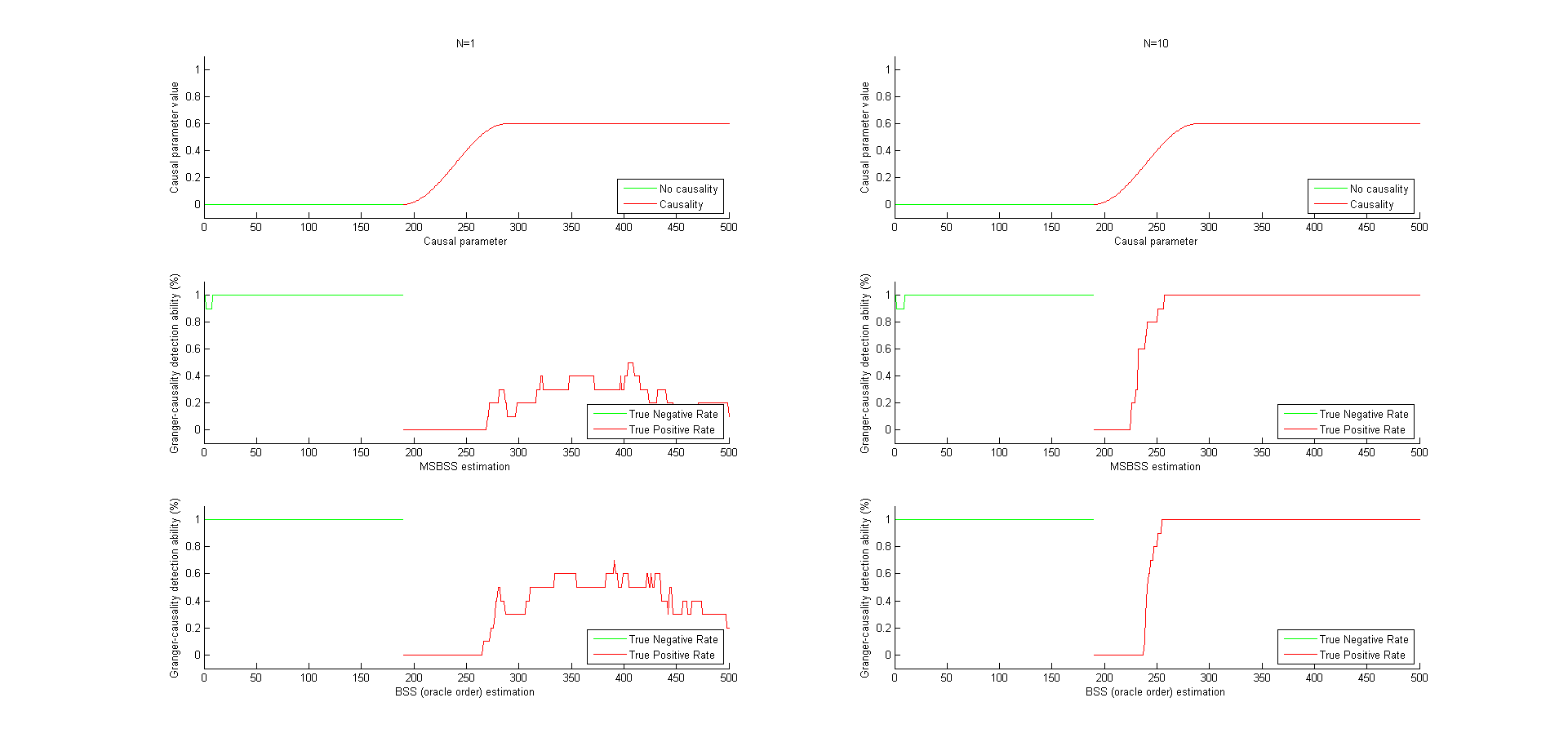}
\caption{Granger-causality detection ability for a non-normal error simulation of order $8$ and series length $500$ and causal parameter $0.6$.}
\label{T_4_1_3}
\end{figure}

\begin{figure}[!htb]
\centering
\hspace*{-1in}
\includegraphics[width=6.4in]{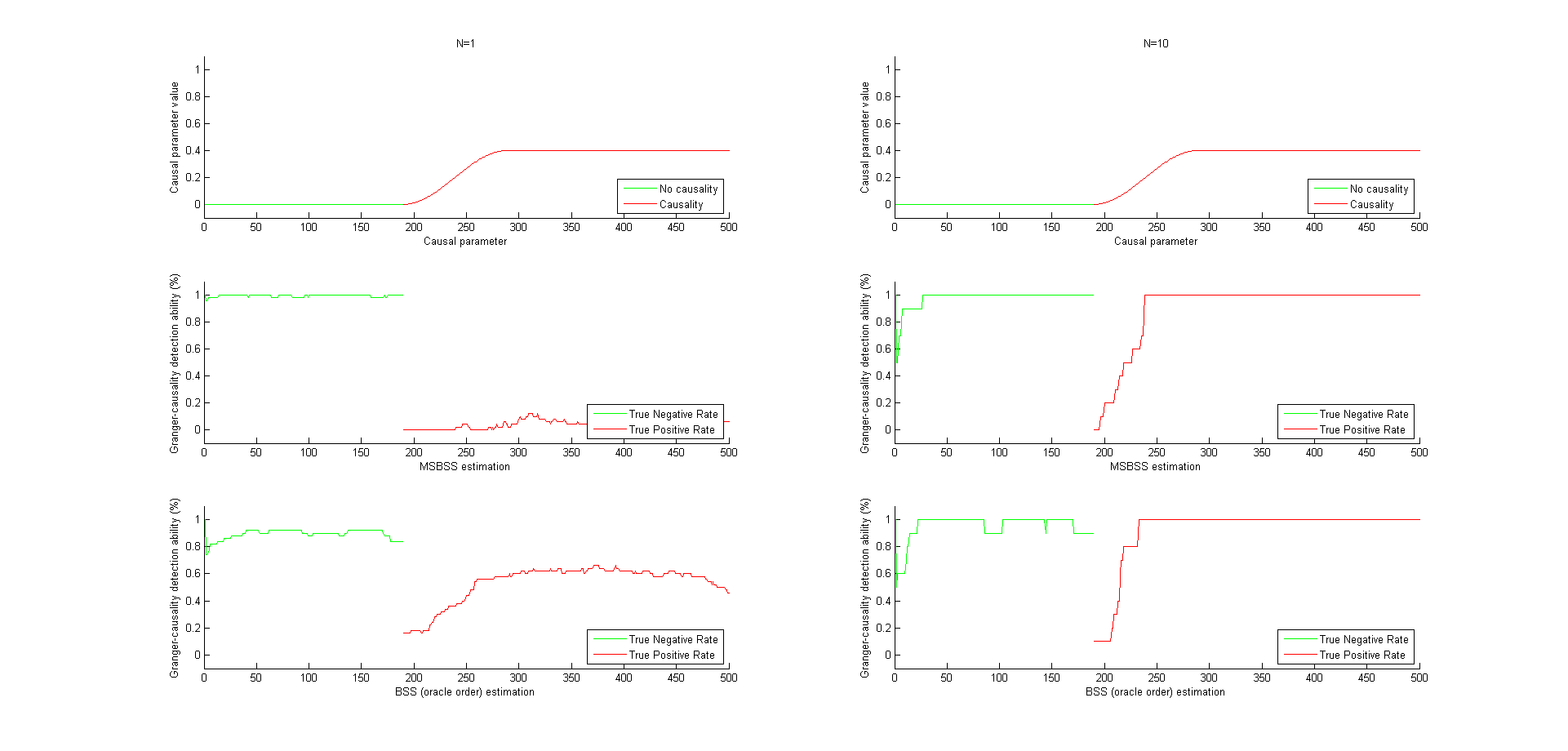}
\caption{Granger-causality detection ability for a non-normal error simulation of order $1$, series length $500$ and causal parameter $0.4$.}
\label{T_1_1_4}
\end{figure}

\begin{figure}[!htb]
\centering
\hspace*{-1in}
\includegraphics[width=6.4in]{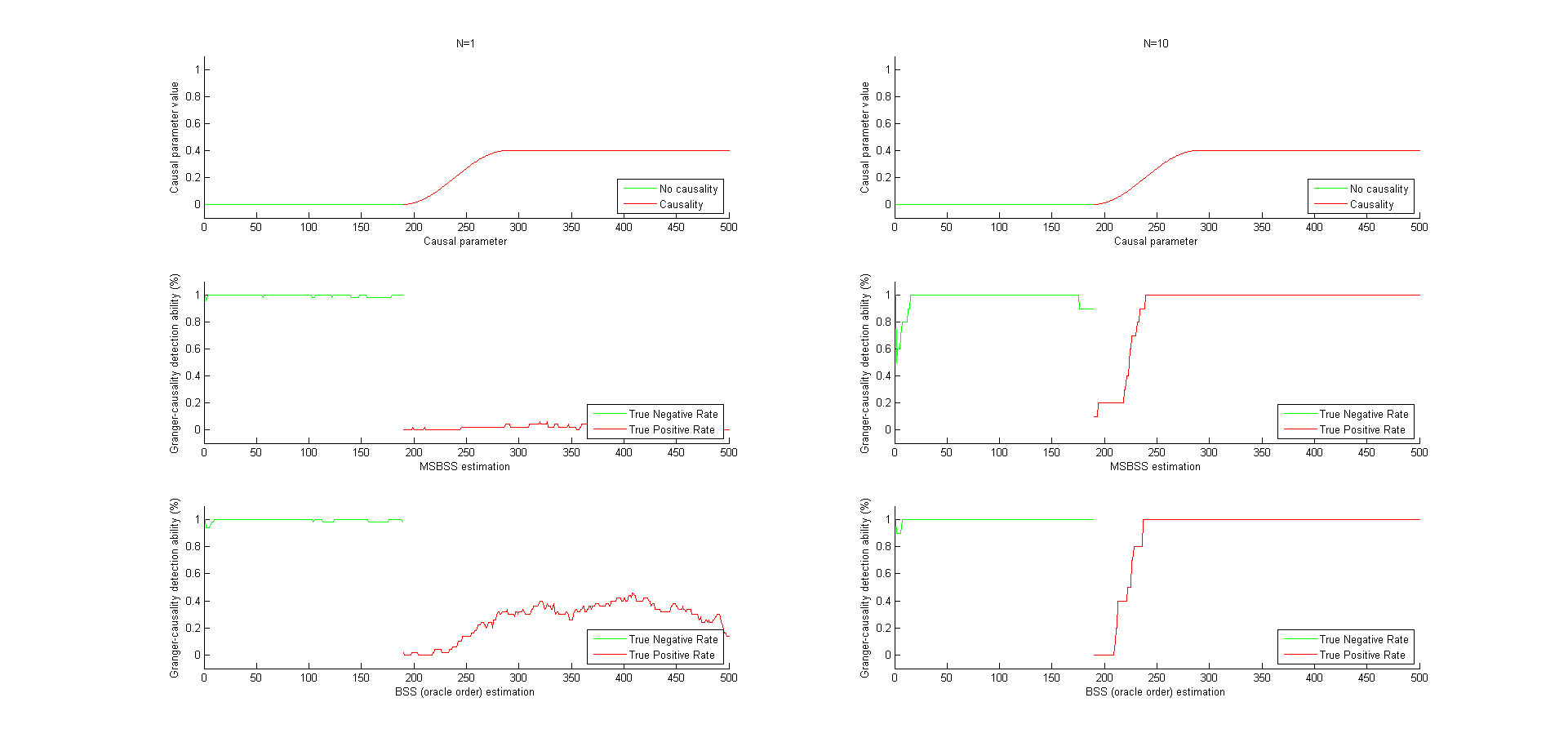}
\caption{Granger-causality detection ability for a non-normal error simulation of order $2$ and series length $500$ and causal parameter $0.4$.}
\label{T_2_1_4}
\end{figure}

\begin{figure}[!htb]
\centering
\hspace*{-1in}
\includegraphics[width=6.4in]{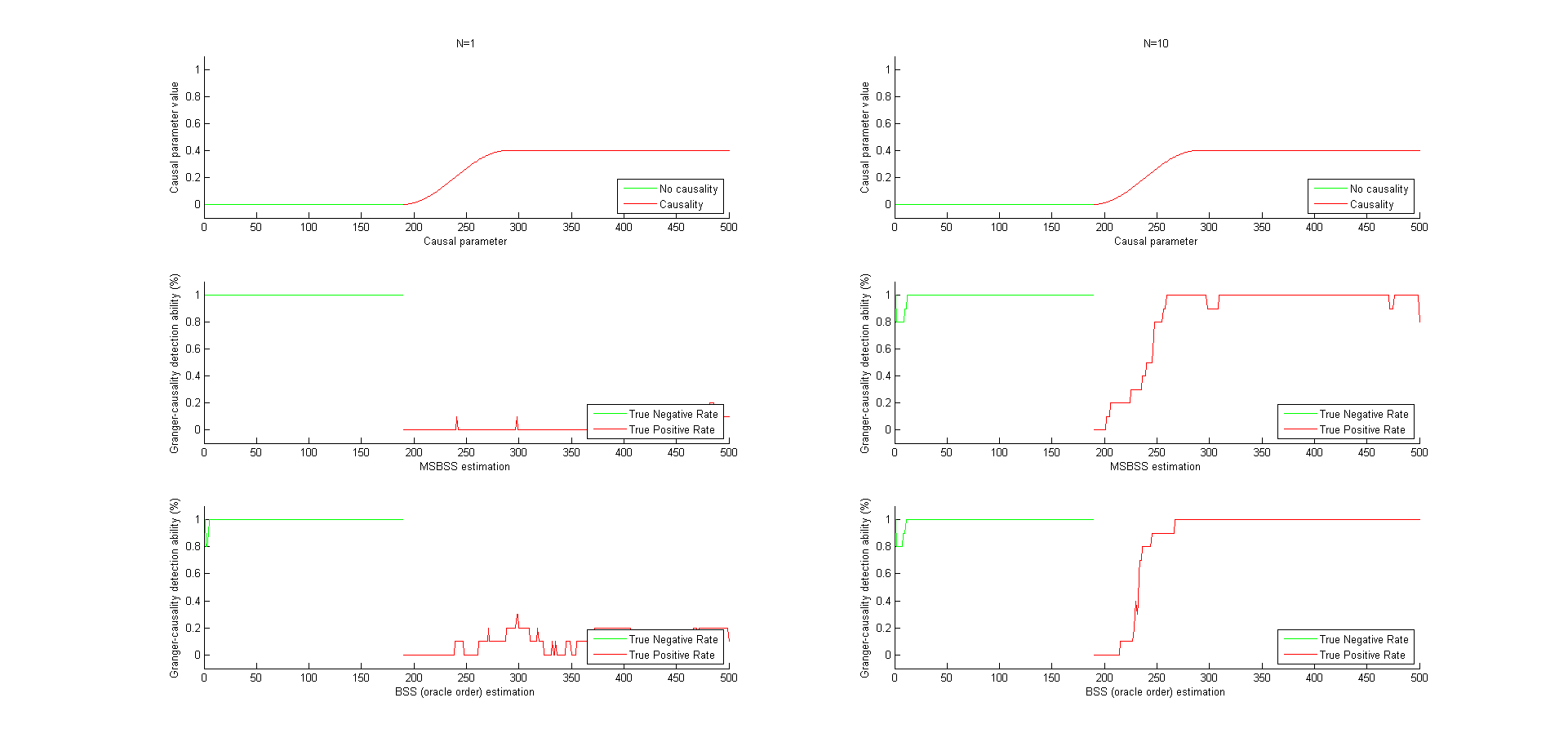}
\caption{Granger-causality detection ability for a non-normal error simulation of order $4$ and series length $500$ and causal parameter $0.4$.}
\label{T_3_1_4}
\end{figure}

\begin{figure}[!htb]
\centering
\hspace*{-1in}
\includegraphics[width=6.4in]{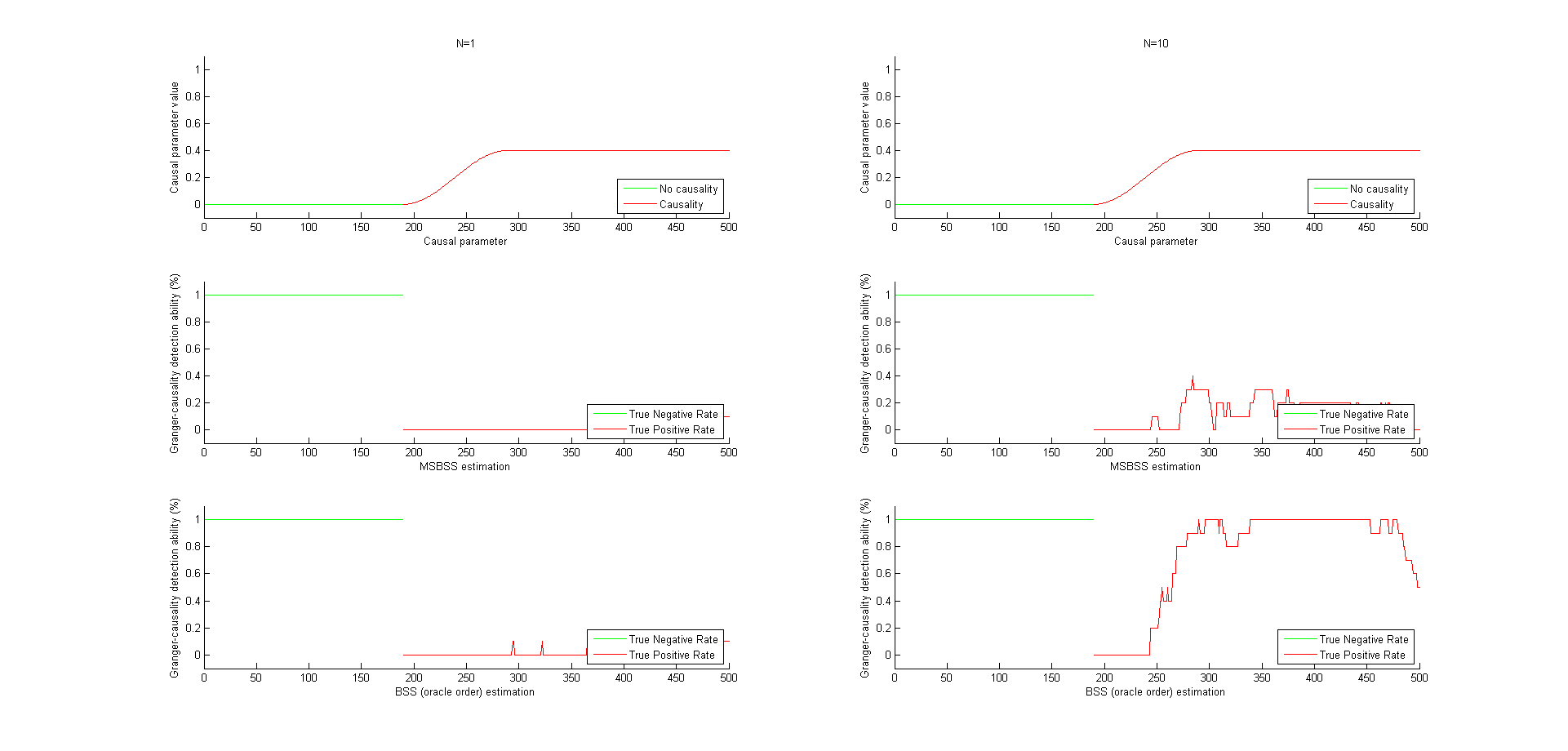}
\caption{Granger-causality detection ability for a non-normal error simulation of order $8$ and series length $500$ and causal parameter $0.4$.}
\label{T_4_1_4}
\end{figure}

\begin{figure}[!htb]
\centering
\hspace*{-1in}
\includegraphics[width=6.4in]{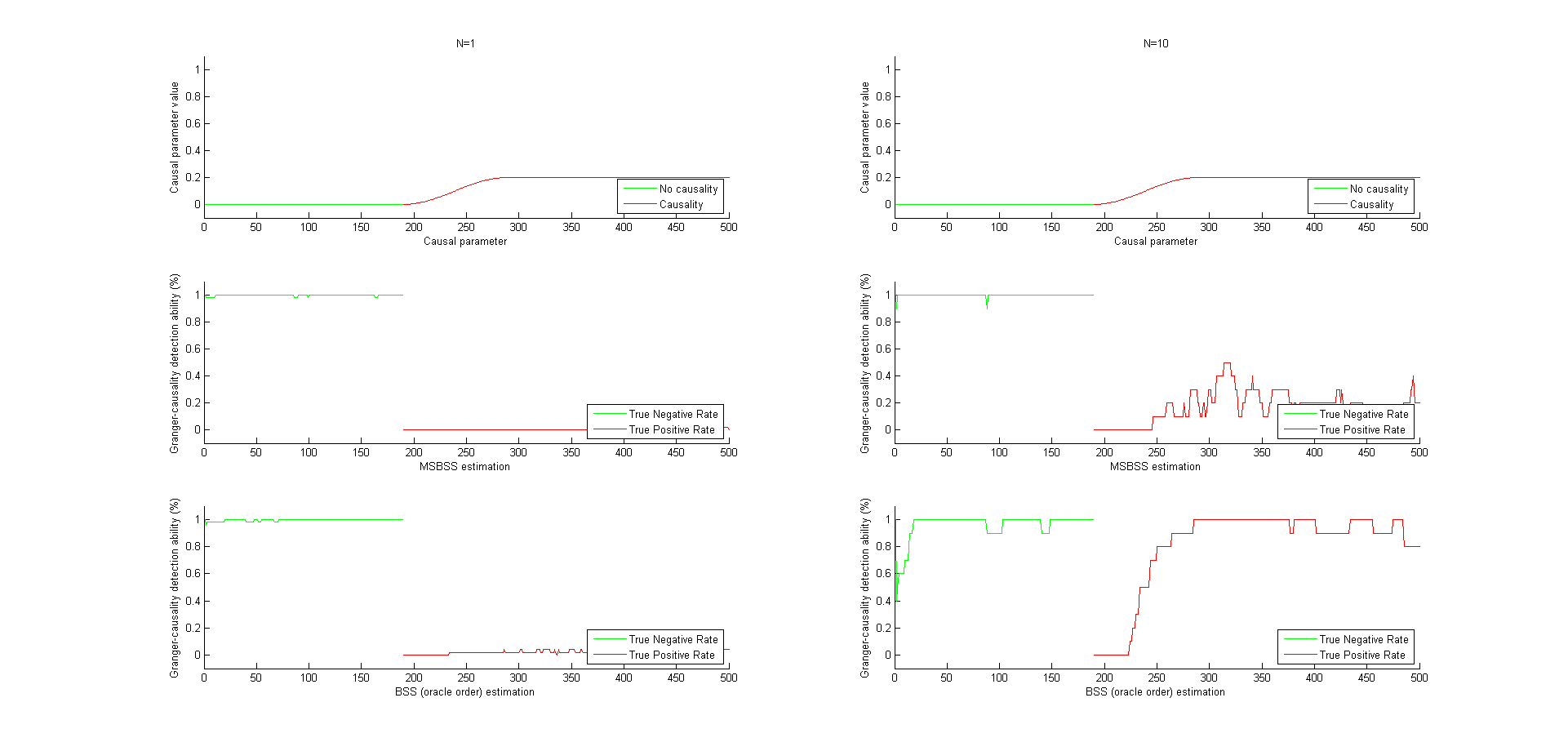}
\caption{Granger-causality detection ability for a non-normal error simulation of order $1$ and series length $500$ and causal parameter $0.2$.}
\label{T_1_1_5}
\end{figure}

\begin{figure}[!htb]
\centering
\hspace*{-1in}
\includegraphics[width=6.4in]{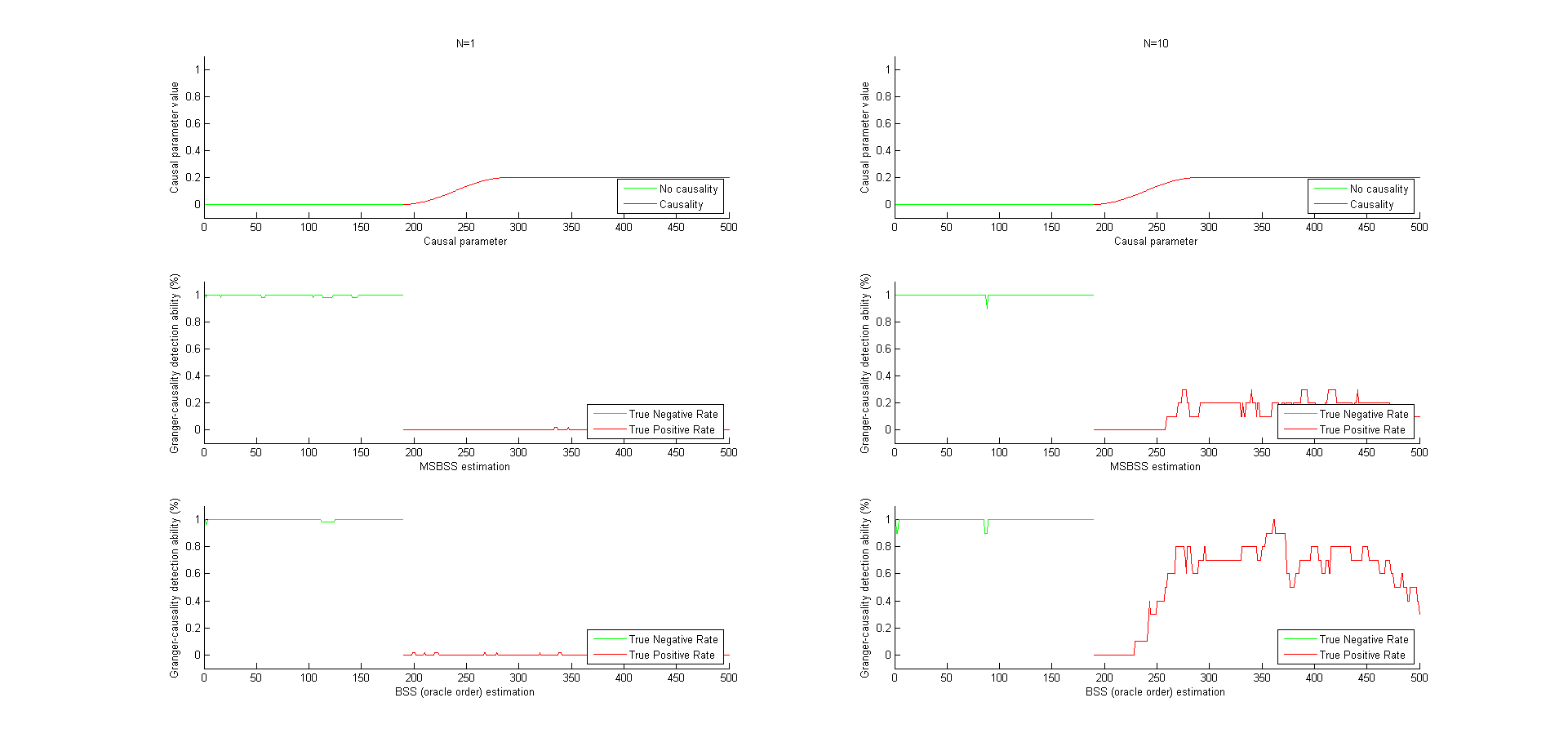}
\caption{Granger-causality detection ability for a non-normal error simulation of order $2$ and series length $500$ and causal parameter $0.2$.}
\label{T_2_1_5}
\end{figure}

\begin{figure}[!htb]
\centering
\hspace*{-1in}
\includegraphics[width=6.4in]{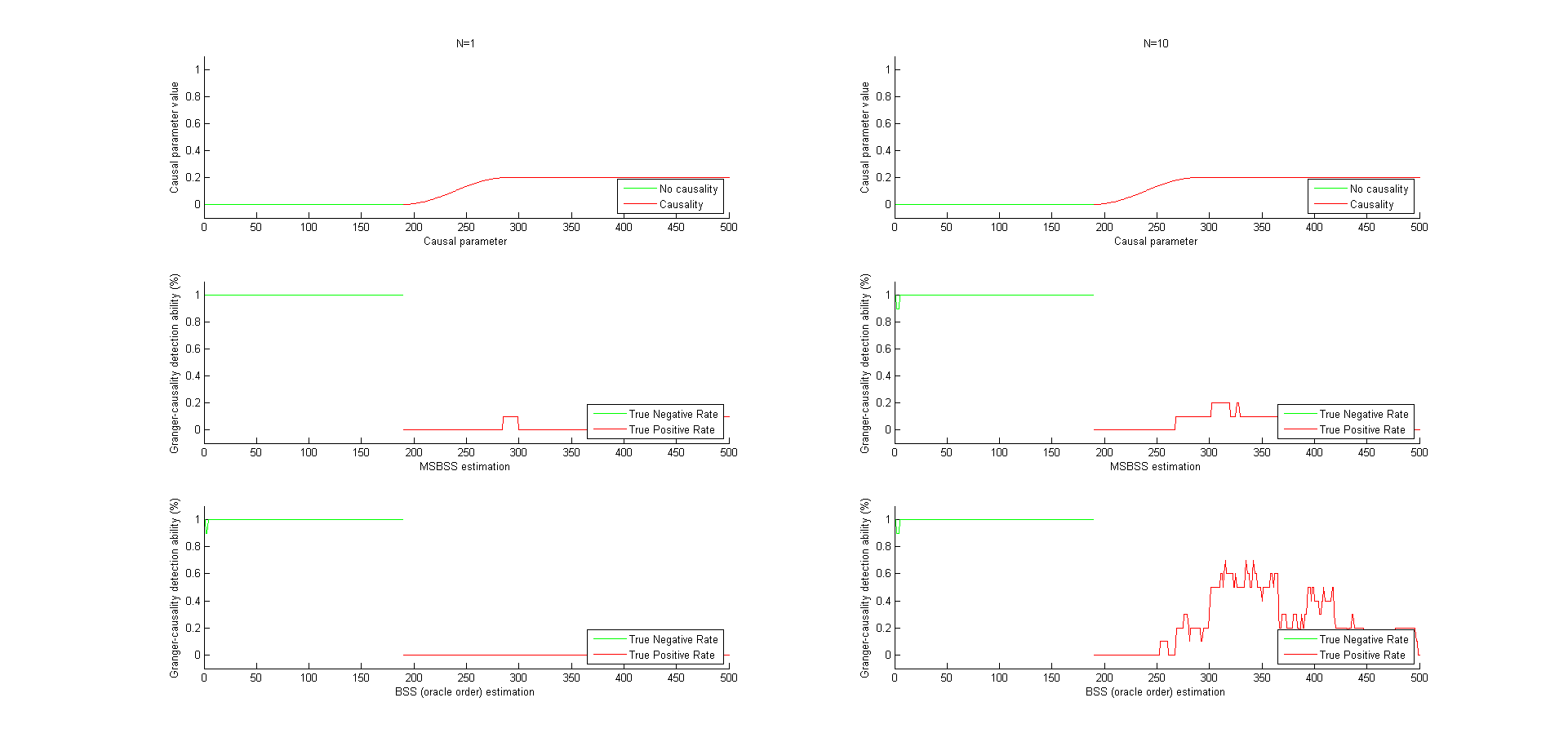}
\caption{Granger-causality detection ability for a non-normal error simulation of order $4$ and series length $500$ and causal parameter $0.2$.}
\label{T_3_1_5}
\end{figure}

\begin{figure}[!htb]
\centering
\hspace*{-1in}
\includegraphics[width=6.4in]{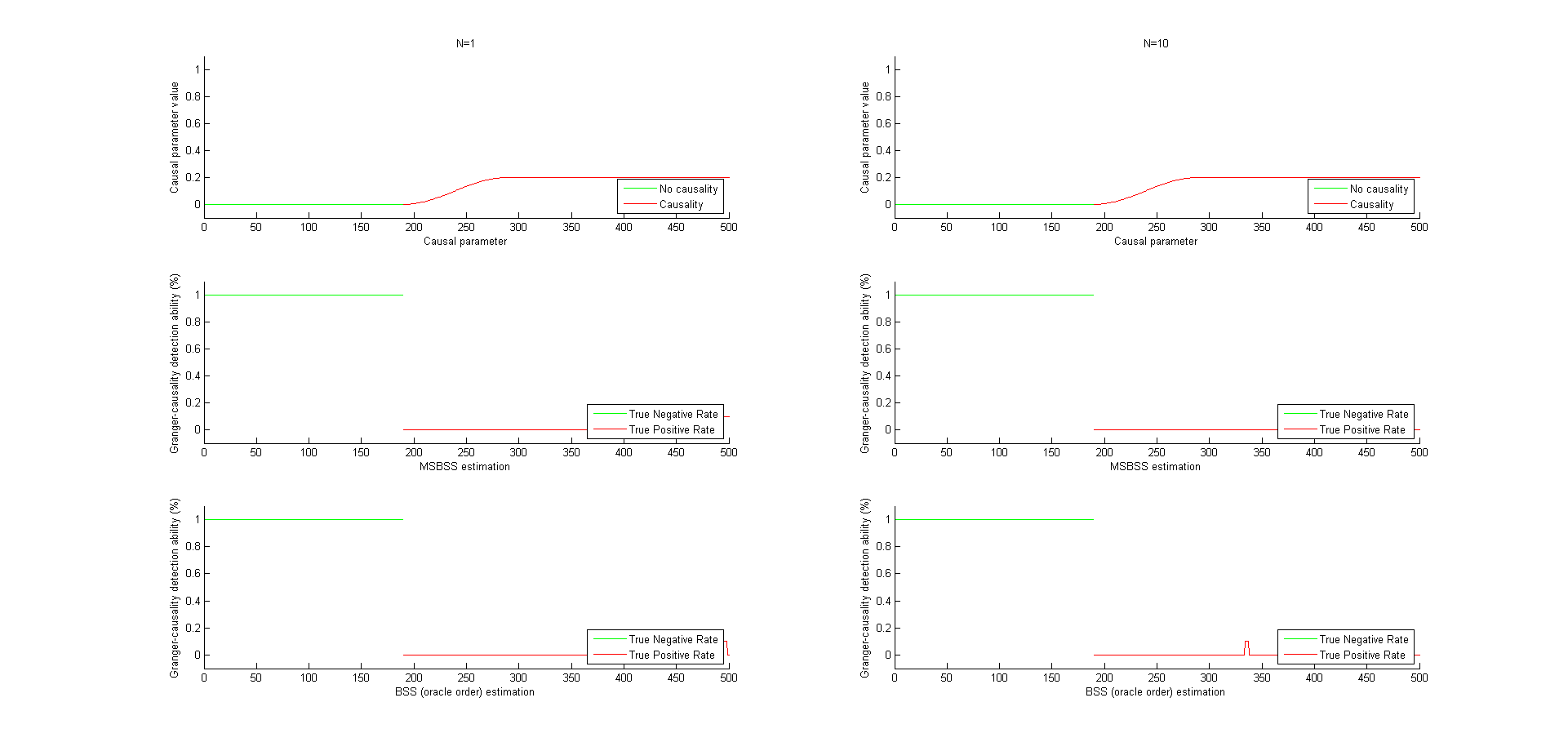}
\caption{Granger-causality detection ability for a non-normal error simulation of order $8$ and series length $500$ and causal parameter $0.2$.}
\label{T_4_1_5}
\end{figure}

\clearpage

\subsection{Windowing estimation procedure}
\label{Windowing estimation}

\begin{figure}[!htb]
\centering
\hspace*{-1in}
\includegraphics[width=6.4in]{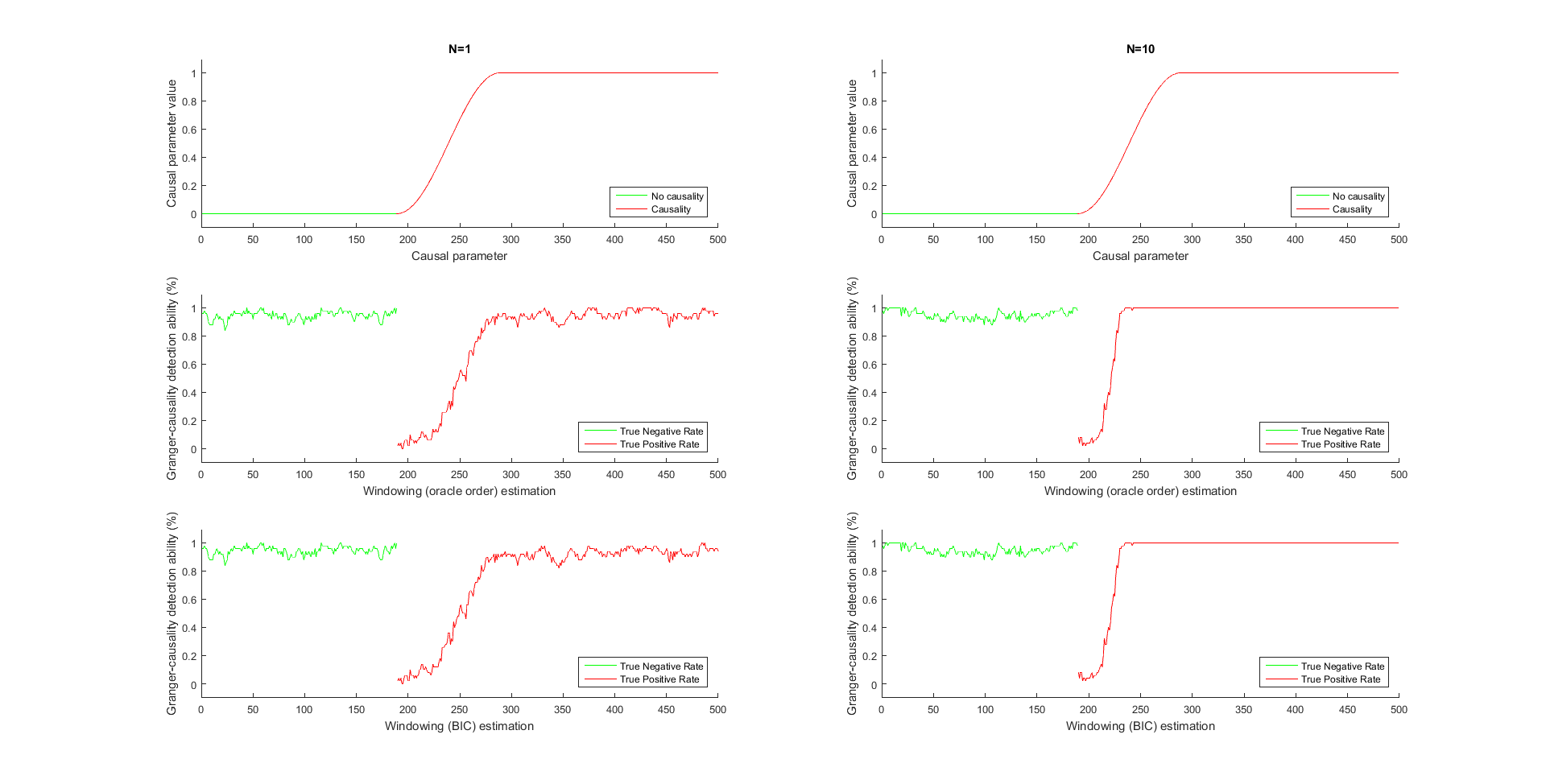}
\caption{Granger-causality detection ability for a windowing estimation procedure with a model order $1$, series length $500$ and causal parameter $1$. The middle graphs show the results for the true model order estimate (oracle) and the bottom graphs those for the model order selected based on the BIC criterion.}
\label{W_1_1_1}
\end{figure}

\begin{figure}[!htb]
\centering
\hspace*{-1in}
\includegraphics[width=6.4in]{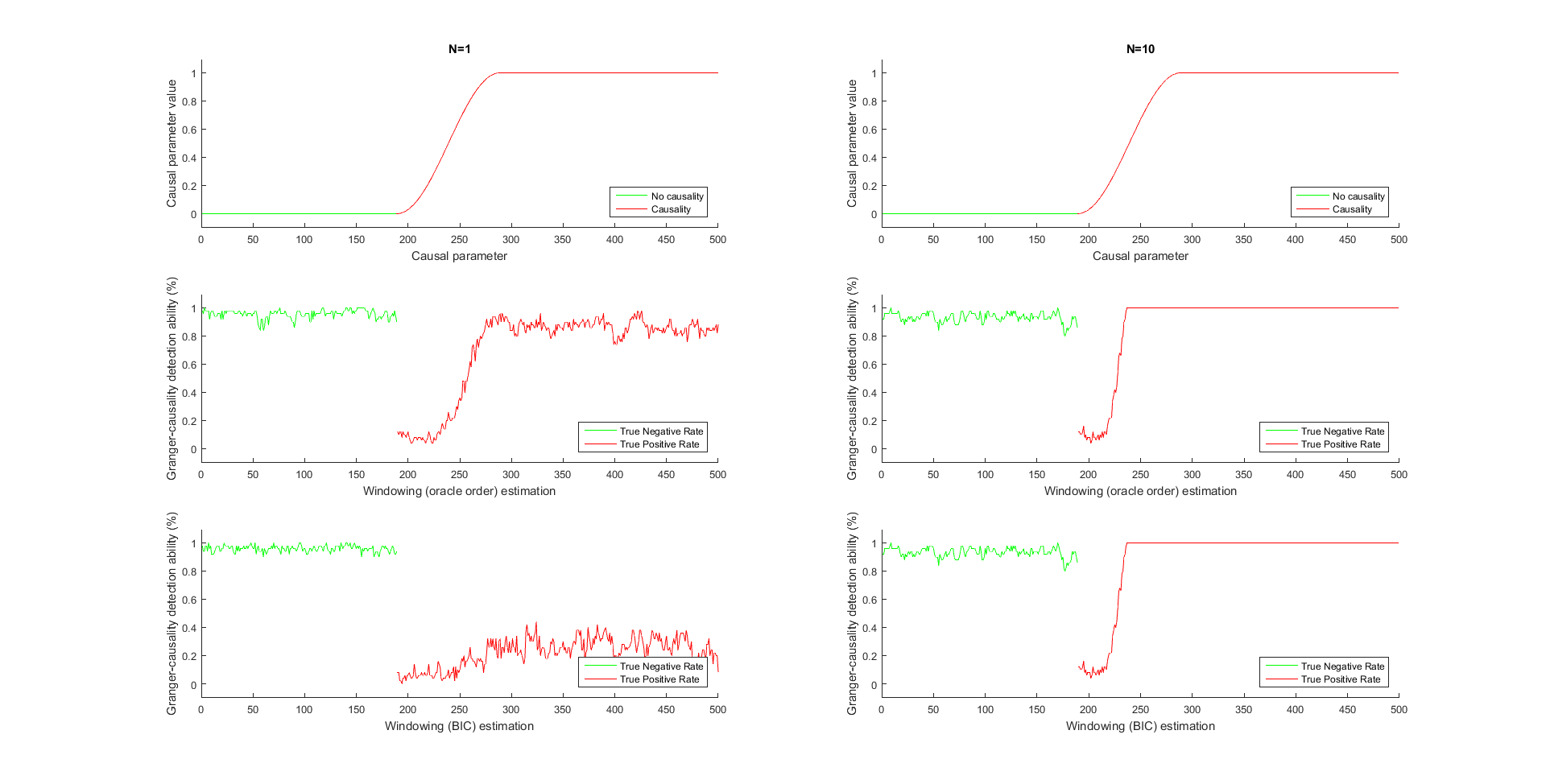}
\caption{Granger-causality detection ability for a windowing estimation procedure with a model order $2$ and series length $500$ and causal parameter $1$.}
\label{W_2_1_1}
\end{figure}


\begin{figure}[!htb]
\centering
\hspace*{-1in}
\includegraphics[width=6.4in]{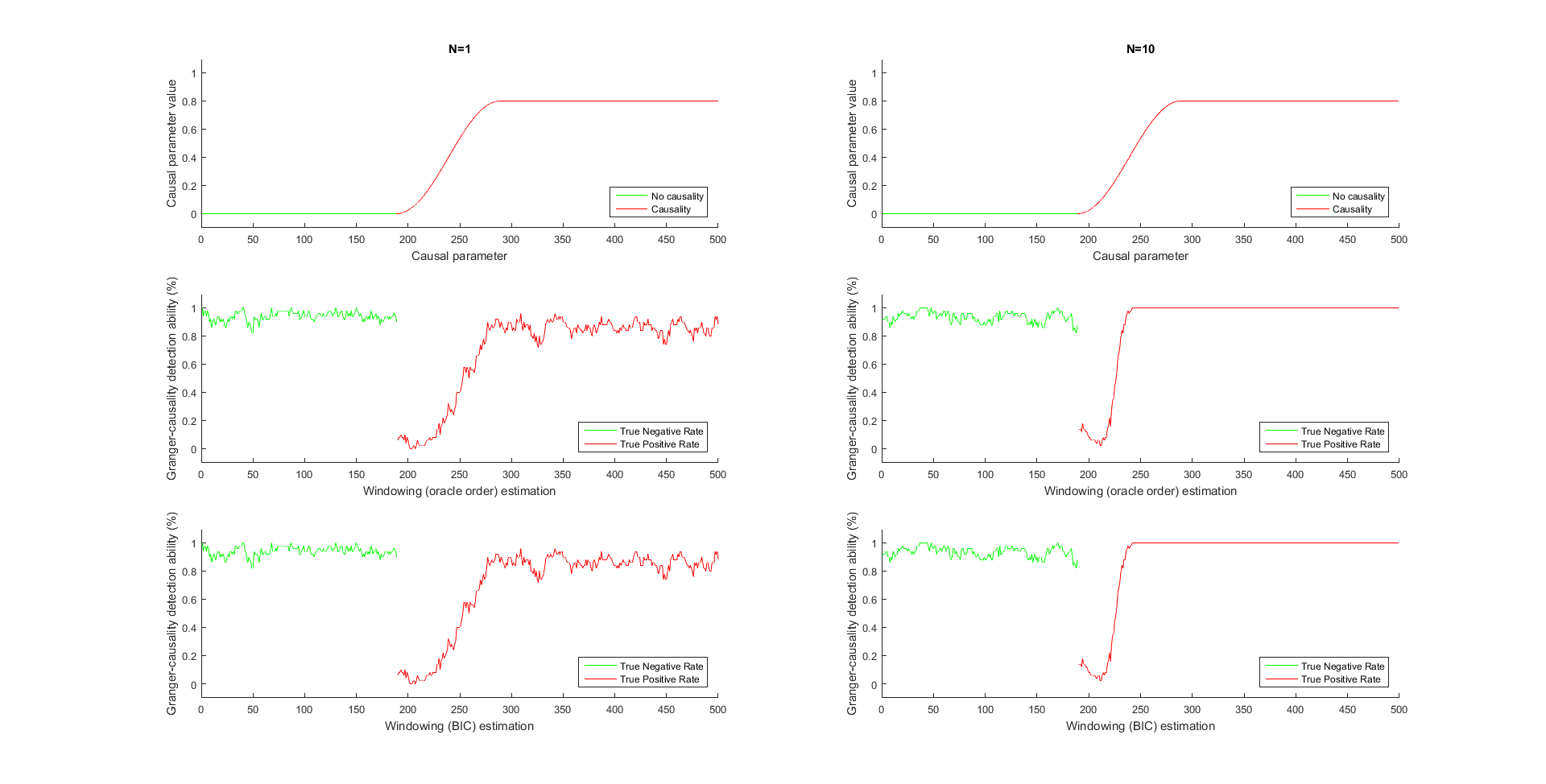}
\caption{Granger-causality detection ability for a windowing estimation procedure with a model order $1$ and series length $500$ and causal parameter $0.8$.}
\label{W_1_1_2}
\end{figure}

\begin{figure}[!htb]
\centering
\hspace*{-1in}
\includegraphics[width=6.4in]{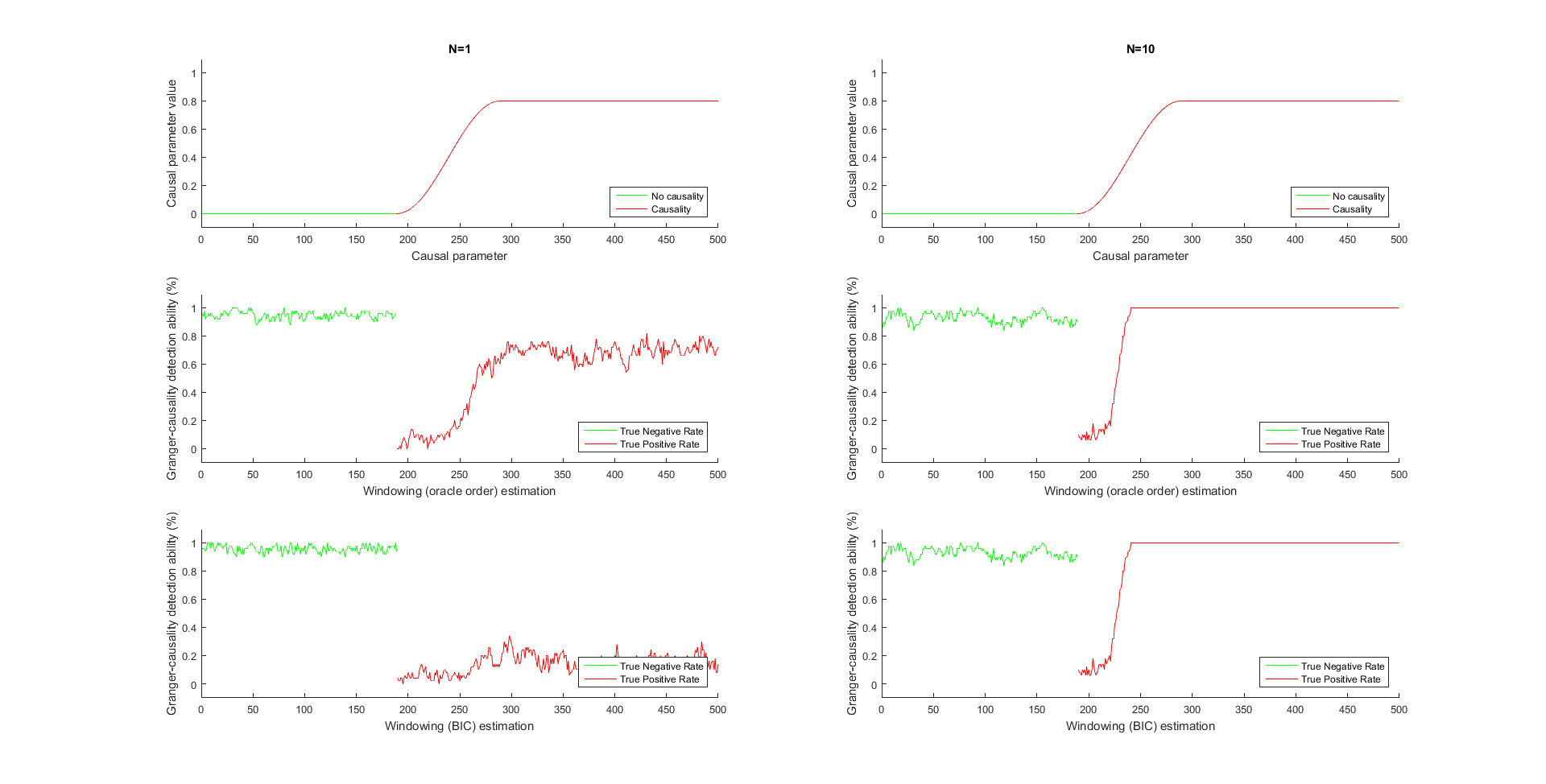}
\caption{Granger-causality detection ability for a windowing estimation procedure with a model order $2$ and series length $500$ and causal parameter $0.8$.}
\label{W_2_1_2}
\end{figure}

\begin{figure}[!htb]
\centering
\hspace*{-1in}
\includegraphics[width=6.4in]{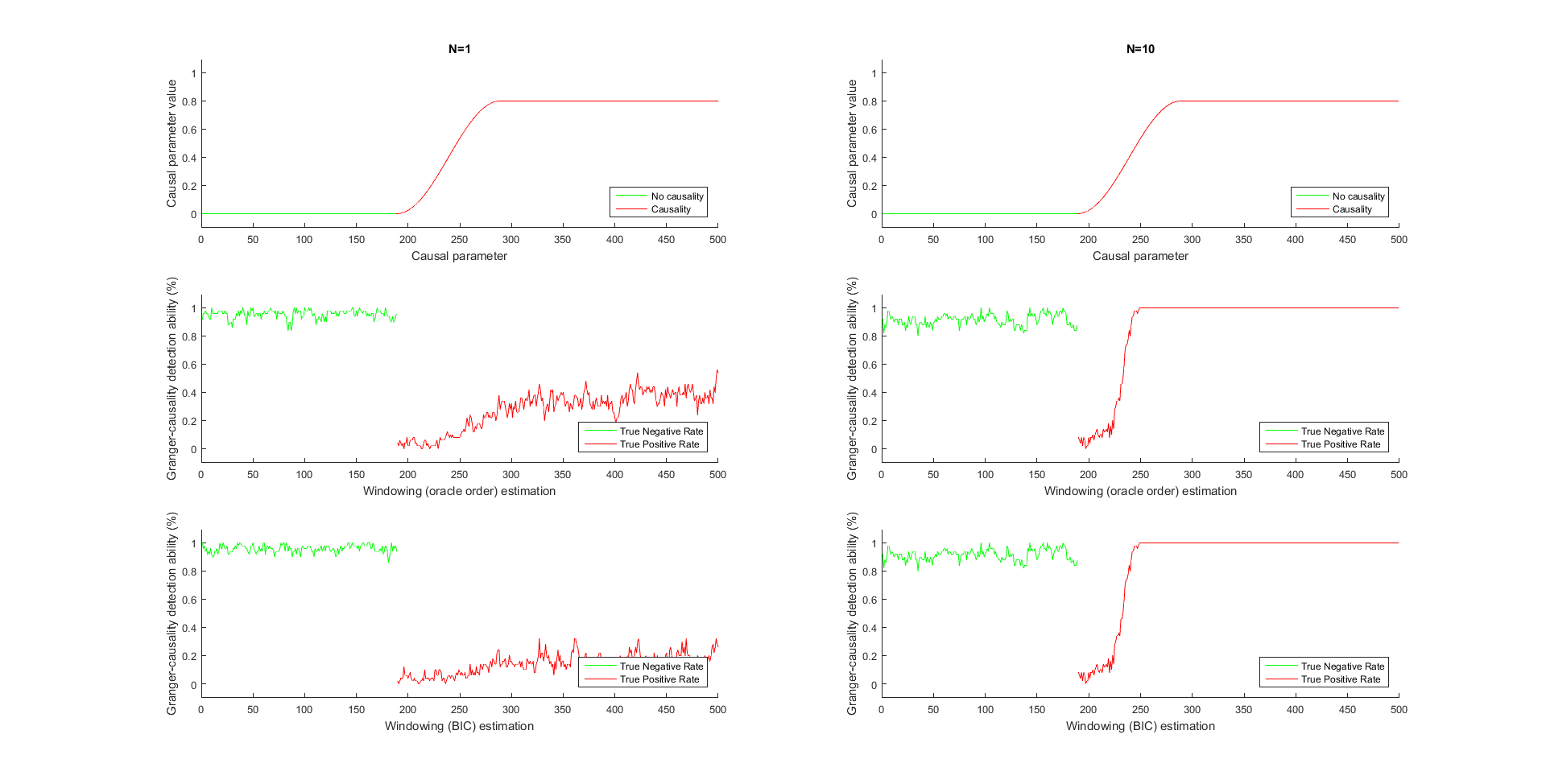}
\caption{Granger-causality detection ability for a windowing estimation procedure with a model order $4$ and series length $500$ and causal parameter $0.8$.}
\label{W_3_1_2}
\end{figure}

\begin{figure}[!htb]
\centering
\hspace*{-1in}
\includegraphics[width=6.4in]{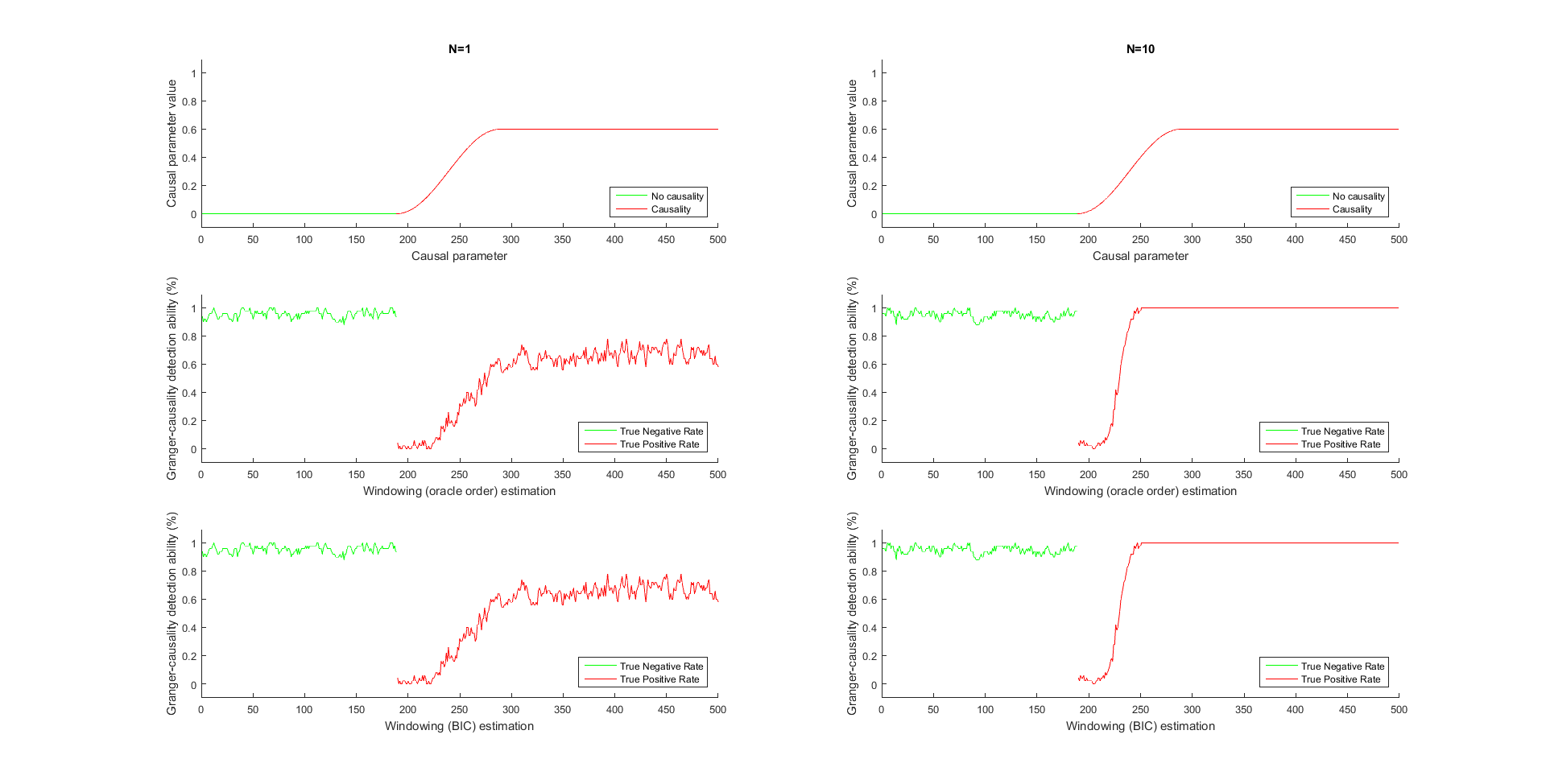}
\caption{Granger-causality detection ability for a windowing estimation procedure with a model order $1$ and series length $500$ and causal parameter $0.6$.}
\label{W_1_1_3}
\end{figure}

\begin{figure}[!htb]
\centering
\hspace*{-1in}
\includegraphics[width=6.4in]{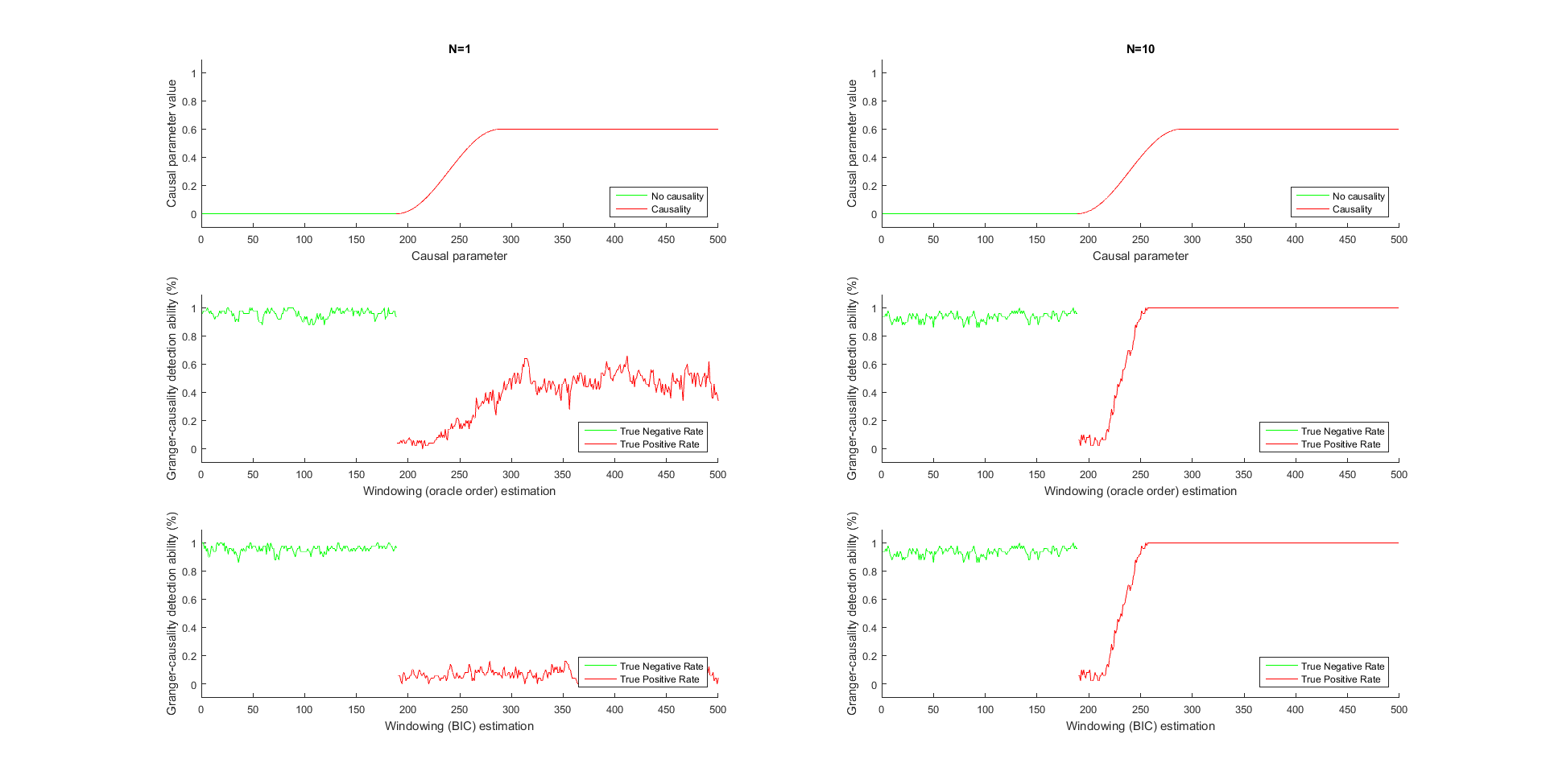}
\caption{Granger-causality detection ability for a windowing estimation procedure with a model order $2$ and series length $500$ and causal parameter $0.6$.}
\label{W_2_1_3}
\end{figure}


\begin{figure}[!htb]
\centering
\hspace*{-1in}
\includegraphics[width=6.4in]{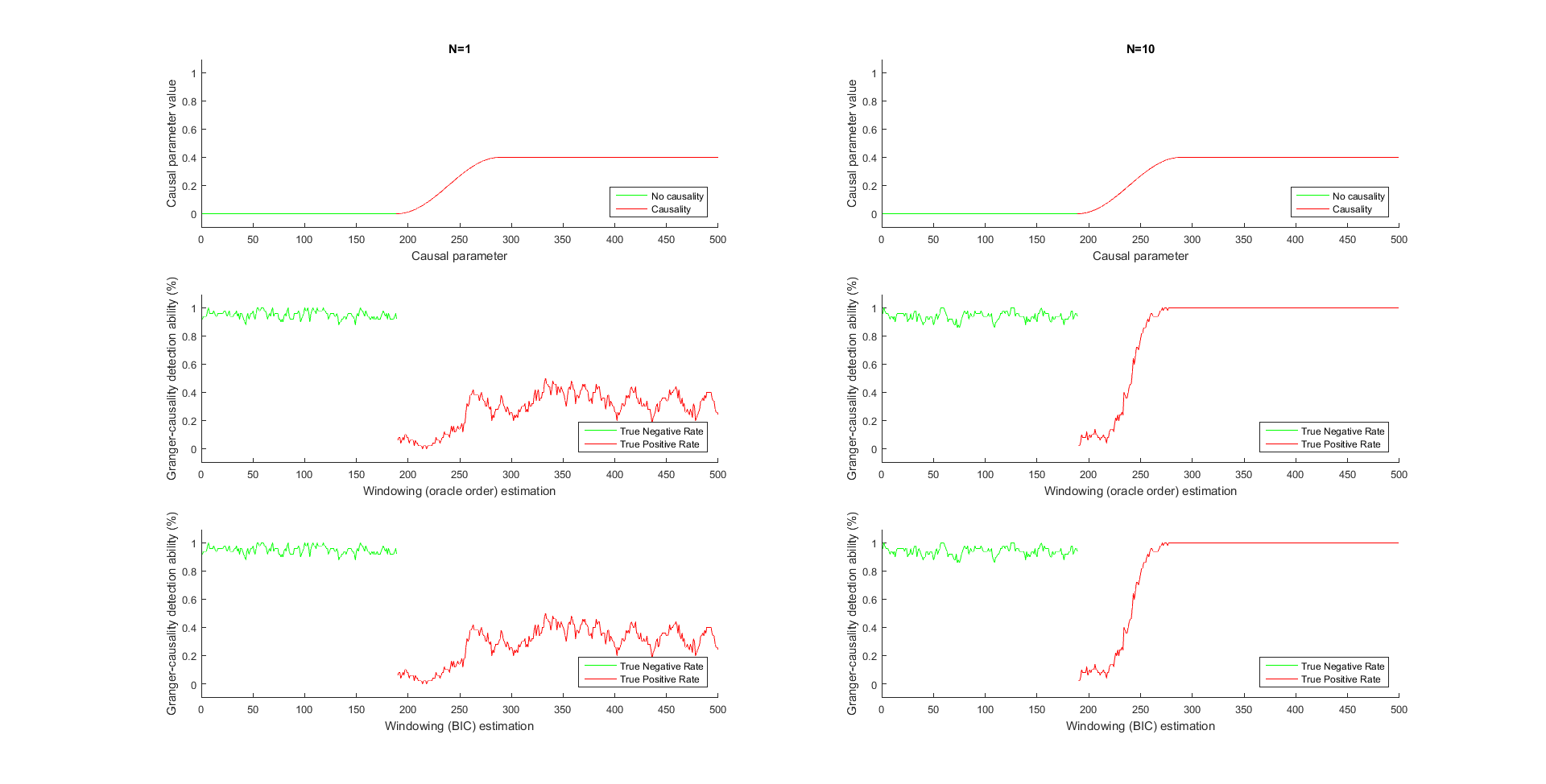}
\caption{Granger-causality detection ability for a windowing estimation procedure with a model order $1$ and series length $500$ and causal parameter $0.4$.}
\label{W_1_1_4}
\end{figure}

\begin{figure}[!htb]
\centering
\hspace*{-1in}
\includegraphics[width=6.4in]{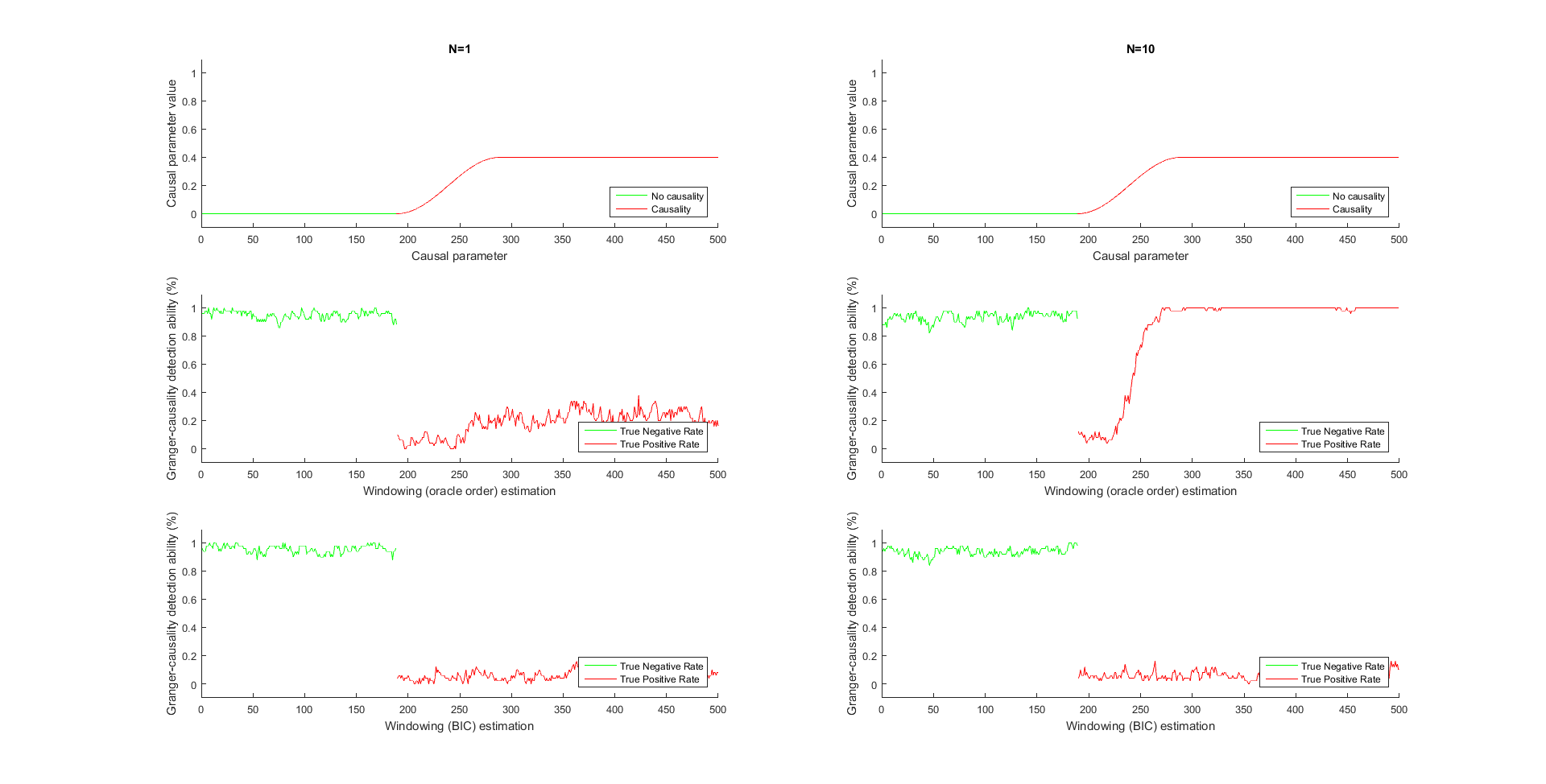}
\caption{Granger-causality detection ability for a windowing estimation procedure with a model order $2$ and series length $500$ and causal parameter $0.4$.}
\label{W_2_1_4}
\end{figure}

\begin{figure}[!htb]
\centering
\hspace*{-1in}
\includegraphics[width=6.4in]{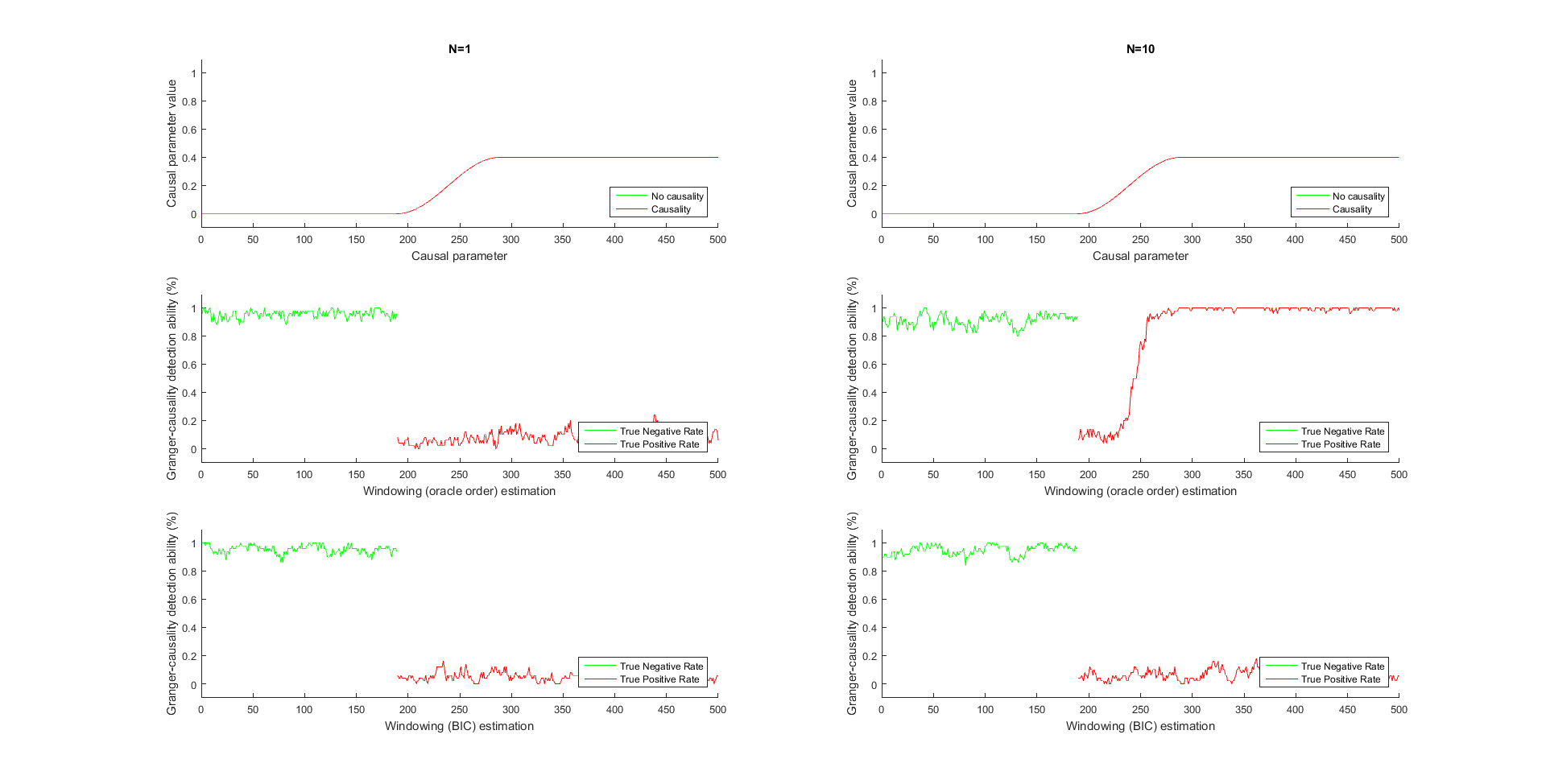}
\caption{Granger-causality detection ability for a windowing estimation procedure with a model order $4$ and series length $500$ and causal parameter $0.4$.}
\label{W_2_1_4}
\end{figure}

\begin{figure}[!htb]
\centering
\hspace*{-1in}
\includegraphics[width=6.4in]{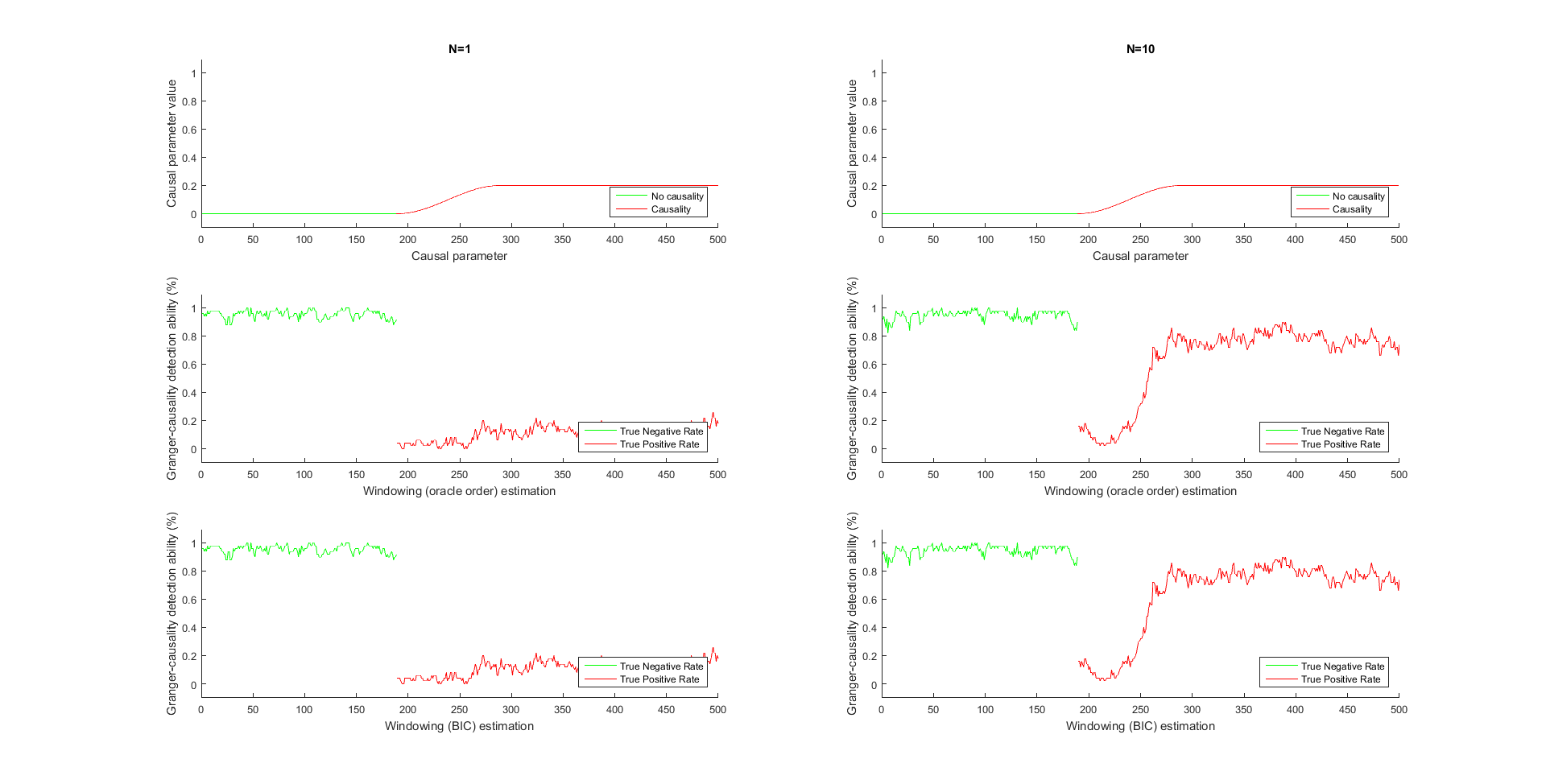}
\caption{Granger-causality detection ability for a windowing estimation procedure with a model order $1$ and series length $500$ and causal parameter $0.2$.}
\label{W_1_1_5}
\end{figure}

\begin{figure}[!htb]
\centering
\hspace*{-1in}
\includegraphics[width=6.4in]{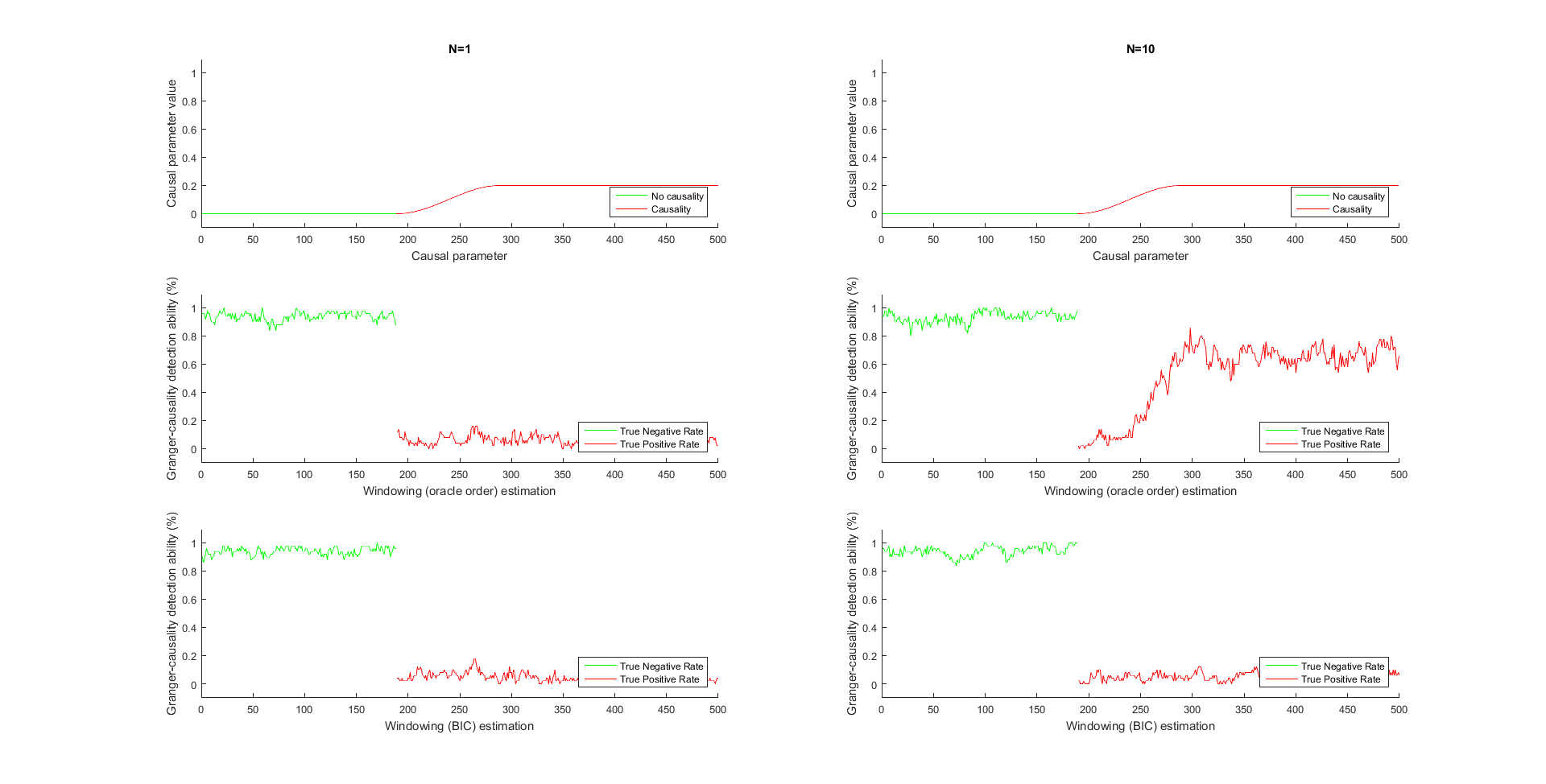}
\caption{Granger-causality detection ability for a windowing estimation procedure with a model order $2$ and series length $500$ and causal parameter $0.2$.}
\label{W_2_1_5}
\end{figure}

\begin{figure}[!htb]
\centering
\hspace*{-1in}
\includegraphics[width=6.4in]{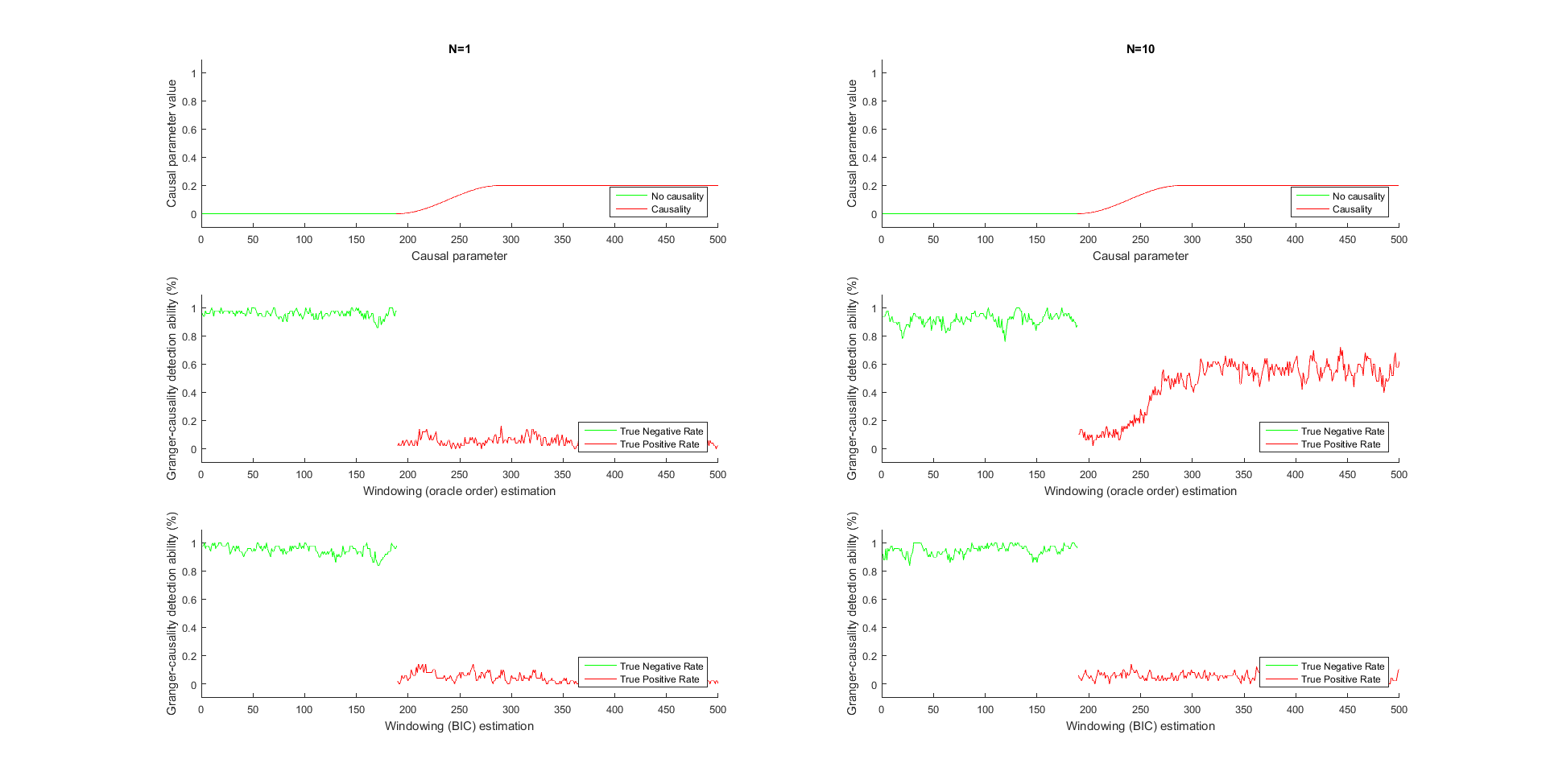}
\caption{Granger-causality detection ability for a windowing estimation procedure with a model order $4$ and series length $500$ and causal parameter $0.2$.}
\label{W_3_1_5}
\end{figure}

\end{document}